\documentclass[usenatbib,usegraphicx,twocolumn]{mn2e}
\usepackage[latin1]{inputenc}
\usepackage{graphicx,amsmath}
\usepackage{times}
\usepackage{amssymb}
\usepackage{natbib}
\usepackage{rotating}
\usepackage{lscape}
\usepackage{multirow}

\title{Imprints of Dark Energy on Cosmic Structure Formation:\\
II) Non-Universality of the halo mass function}

\author[J. Courtin, Y. Rasera, J.-M. Alimi, P.-S. Corasaniti, V. Boucher, A. F\"uzfa ]{J. Courtin$^{1}$\thanks{email:jerome.courtin@ac-caen.fr},Y. Rasera$^{1}$\thanks{email: yann.rasera@obspm.fr},J.-M. Alimi$^{1,2}$\thanks{email: jean-michel.alimi@obspm.fr}, P.-S. Corasaniti$^{1}$\thanks{email: pier-stefano.corasaniti@obspm.fr},  V. Boucher$^{3}$\thanks{email: vincent.boucher@uclouvain.be}, A. F\"uzfa$^{1,2,3}$\thanks{email: andre.fuzfa@fundp.ac.be}  \\
$^{1}$CNRS, Laboratoire Univers et Th\'eories (LUTh), UMR 8102 CNRS, Observatoire de Paris, \\
Universit\'e Paris Diderot ; 5 Place Jules Janssen, 92190 Meudon, France\\
$^{2}$Groupe d'Application des MAth\'ematiques aux Sciences du COsmos (GAMASCO), 
University of Namur (FUNDP), Belgium\\
$^{3}$ Center for Particle Physics and Phenomenology (CP3), Universit\'e catholique de Louvain, \\
Chemin du Cyclotron, 2, B-1348 Louvain-la-Neuve, Belgium} 

\begin{document}
\date{Sent. Accepted. Received. In original form.}
\pagerange{\pageref{firstpage}--\pageref{lastpage}} \pubyear{2010}  

\maketitle
\label{firstpage}

\begin{abstract} 
The universality of the halo mass
function is investigated in the context of dark energy cosmologies. This widely used
approximation assumes that the mass function can be expressed as a
function of the matter density $\Omega_m$ and the root-mean-square 
linear density fluctuation $\sigma$ only, with no explicit 
dependence on the properties of dark energy or redshift. In order
to test this hypothesis we run a series of 15 high-resolution N-body
simulations for different cosmological models. These consist of three 
$\Lambda$CDM cosmologies best fitting WMAP-1, 3 and 5 years data,
which are used for model comparison, and three toy-models
characterized by a Ratra-Peebles quintessence potential with different slopes
and amounts of dark energy density. These toy models have very different evolutionary
histories at the background and linear level, but share the same $\sigma_8$ value. For each of these models
we measure the mass function from catalogues of halos identified in
the simulations using the Friend-of-Friend (FoF) algorithm. We
find redshift-dependent deviations from a universal behaviour, well above numerical
uncertainties and of non-stochastic origin, which are correlated with the
linear growth factor of the investigated cosmologies. Using the spherical
collapse as guidance, we show that such deviations
are caused by the cosmology dependence of the non-linear collapse and virialization
process. For practical applications, we provide a fitting
formula of the mass function accurate to 5 percents over the all range of investigated cosmologies. 
We also derive an empirical relation
between the FoF linking parameter and the virial overdensity which can account for most of the deviations from an exact universal behavior. Overall these results suggest that measurements of the halo mass function at $z=0$ can provide 
additional constraints on dark energy since it carries a fossil
record of the past cosmic evolution.
\end{abstract}
  

\section{Introduction}
In this series of articles we have investigated the imprint of
dark energy on the non-linear structure formation. In a previous
paper \citep{alimi09} we focused on the non-linear matter power spectrum at $z=0$
and showed that dark energy leaves distinctive signatures through a
number of effects. On the one hand the clustering of dark energy 
 modifies the shape and
amplitude of the linear matter power spectrum, on the other hand the
values of the cosmological parameters, such as the matter density $\Omega_m$ and the linear 
root-mean-square (rms) density fluctuations on the $8$~h$^{-1}$Mpc scale, $\sigma_8$,
differ from one model to another such as to satisfy the constraints from 
Supernova Ia and Cosmic Microwave Background (CMB) observations. Because of this, the shape and
amplitude as well as the evolution of the linear power spectrum are affected, with
the non-linear phase of collapse mixing and amplifying these model dependent
features. This is a direct consequence of the fact that a record of the past history 
of forming structures is kept throughout the non-linear regime
\citep[see e.g.][]{ma07}. Although current measurements of the
clustering of matter at small scales are unable to detect such imprints (mainly because
of astrophysical systematic uncertainties related to galaxy bias), these effects are present
and may become detectable with future weak lensing observations (see e.g. \citet{Takada04}). Another
consequence is that any estimate of the non-linear
power spectrum based on parametrized fitting functions written in
terms of the cosmological parameters (e.g. $\Omega_m$) and
instantaneous linear quantities (such as the linear matter power
spectrum, $P(k)$, at $z=0$ and the linear growth factor)
are of limited precision on the non-linear scales. For instance 
this is the case of the \citet{smith03,peacock96} formula.
Deviations with respect to these fitting functions 
depend on the past evolutionary history
of a given cosmology, hence it is not surprising that such discrepancies have been found to
be manifestly accentuated in the context of dark energy cosmologies
\citep{mcdonald06,ma07,francis07,casarini09,jennings09,alimi09}. 

The halo density profile
is another observable for which similar effects
occur. For example \citet{wechsler02} have shown that the concentration parameters
depend on the halo assembly history, and it has been shown that such
dependencies are strengthened in the case of dark energy models \citep[see for instance][]{dolag04}. 
Paradoxically, as a result of state-of-the-art numerical simulations
\citep{jenkins01,warren06}, a universal form of the halo mass function
entirely specified by $\Omega_m$ and linear root-mean-square
(rms) density fluctuations $\sigma$ is usually assumed. Furthermore it has been claimed
that such a universal form holds for dark energy cosmologies as well
\citep{linder03}. If universality is rigorously exact, then in the light of the previous
results on the non-linear matter power spectrum and halo profile, it would
imply that there must exist an unknown gravitational mechanism capable of erasing the
influence of the past evolution of forming structures on the halo mass 
function. Only in such a case a dependence on cosmology (e.g. the
properties of dark energy) and redshift would be absent. 

The universality of the mass function at very high-redshift ($z>5$)
has been long debated in the literature. As an example deviations
from a universal behaviour up to $50\%$ have been found by
\citet{reed03,reed07}. However \citet{lukic07} have shown that most of
these deviations might be caused by numerical
artifacts (finite volume effects or initial conditions set at very low
redshift). In fact results inferred from high redshift simulations are
very sensitive to numerical errors, consequently several works have focused on the 
low-redshift mass function. As an example \cite{tinker08} have shown
deviations from universality up to $30\%$ in the low redshift mass function
($z<3$). Such deviations are thought to be associated with the Spherical
Overdensity (SO) halo finder which tends to underestimate the mass function relative to the
Friend-of-Friend (FoF) algorithm when considering higher redshifts
\citep{lukic09}. Since the SO detection is similar to the observational 
procedure of measuring the mass of galaxy clusters \citep{tinker08}, 
such an effect is more relevant from an observational stand point, but
less informative for a better understanding of the non-linear structure formation. 
To our knowledge, $10\%$ deviations from universality (as a function of redshift) using 
FoF halo finder have been detected by \cite{lukic07,tinker08} at low redshift and 
very recently confirmed by \citet{crocce09}. Nonetheless the physical
origin of these deviations has yet to be understood, especially in the
context of dark energy cosmologies. 

Is the halo mass function 
really universal? To what extent does the universality approximation hold? 
For which cosmologies and for which
redshifts? If there are deviations from an universal behaviour are
these of stochastic nature or do they correlate with physical
effects? What can we learn from deviations to a universal behaviour and
how to model them? These are the questions which we will address in
this work. 

The paper is organized as follows. In Section~\ref{DE} we 
introduce the halo mass function, describe the main features of the 
cosmological models for which we have run a series of N-body
simulations, and discuss the spherical collapse model. In Section~\ref{nbody} 
we describe the characteristics of the N-body simulations, and the halo finder algorithm, while in Section~\ref{subsec_numerics}, we discuss various numerical tests which we have performed to identify potential sources of systematics errors. In Section~\ref{universality}, we present the results of the non-universality of the mass function, and discuss the mechanisms responsible for the measured deviations in Section~\ref{insight}. We finally discuss our conclusions in Section~\ref{conclu}.

\section{Dark energy and structure formation }\label{DE}
\subsection{The halo mass function}\label{theory}

Current analytical predictions of the halo mass function are
based on the original work by \citet{press74}. The basic idea is that
virialized objects of mass $M$ correspond to regions where the
linear density fluctuation field smoothed on the scale $R$ lies
above a critical density contrast threshold $\delta_c$. Then the halo mass
function is simply proportional to the fraction of volume occupied by 
the collapsed objects with mass greater than $M$. Assuming a Gaussian
distribution of density fluctuations, this is given by:
\begin{equation}
F(>M)=\frac{2}{\sqrt{2\pi}\sigma}\int_{\delta_c}^\infty e^{-\frac{\delta^2}{2\sigma^2}} d\delta,\label{fofm}
\end{equation}
where
\begin{equation}
  \sigma^2(R)=\frac{1}{2 \pi^2}\int_0^\infty k^2P(k)W^2(k,R)dk,
\end{equation}
is the variance of the density field smoothed on the scale $R$,
with $P(k)$ being the linear matter power spectrum today and $W(k,R)$
is the window function. For a spherical top-hat filter in real space
of radius $R$ containing a mass $M\approx4\pi\bar{\rho}_0/3 R^3$ where $\bar{\rho}_0$
is the present matter density, 
we have $W(k,R)=3\times(sin(kR)/(kR)^3-cos(kR)/(kR)^3)$. 
The only additional ingredient needed to solve the model is the overdensity
threshold $\delta_c$, which is assumed
to be given by the spherical collapse model prediction of the linearly 
extrapolated density fluctuation at the time of collapse.
 
Subsequent studies of the mass function have largely improved this simple modelling
\citep{bond91,lacey93}, and corrections have been included to account for the
ellipsoidal collapse \citep{audit97,sheth99,sheth01}. Recently a modelling of the halo
mass function in the context of the excursion set formalism as well as its
extension to the case of non-gaussian initial conditions has been
presented in \citep{maggiore09}.

A clear understanding of what universality of the mass function
implies can be gathered by writing Eq.~(\ref{fofm}) as
\begin{equation}
 F(>M)=\int^\infty_0 S\left(\delta,\frac{\delta}{\sigma}\right)\mathcal{F}\left(\frac{\delta}{\sigma}\right) d\left(\frac{\delta}{\sigma}\right),
\end{equation}
where $S(\delta,\delta/\sigma)$ is a selection function in
$\delta$-space and $\mathcal{F}$ is the probability distribution of
the primordial density fluctuations smoothed on the scale $R$ 
\citep[for more details about the following discussion see][]{blanchard92}. 
In such a case the halo mass function reads
\begin{equation}
\frac{dn}{dM}=-\frac{\bar{\rho}_0}{M}\frac{1}{\sigma^2}\frac{d\sigma}{dM} \int^\infty_0 \delta \frac{\partial S} {\partial \delta }\mathcal{F}\left(\frac{\delta}{\sigma}\right) d\delta \label{eq_theory}.
\end{equation}
This very general formulation should be a good approximation for
slowly evolving primordial power spectra (such as the one used in this
paper). A crucial point resulting from Eq.~(\ref{eq_theory}) is that all effects
associated with the non-linear collapse are encoded in the form of the
selection function $S$. Although the precise shape of $S$ is difficult to
compute, one may expect as a general trend that $S$ varies from zero
to one near the non-linear density threshold $\delta_S$. In the end it
is this threshold that determines the precise form of the mass function. 
For instance, by assuming $S$ to be a Heaviside in
$\delta_S$, one recovers the Press-Schechter halo mass function,
\begin{equation}
\frac{dn}{dM}=-\frac{\bar{\rho}_0}{M}\frac{1}{\sigma^2}\frac{d\sigma}{dM} \mathcal{F}\left(\frac{\delta_S}{\sigma}\right) \delta_S.
\end{equation}
The dependence on $\delta_S$ accounts for the effect of the
non-linear gravitational collapse and the virialization process. The
former is estimated in terms of the extrapolated linear density at the
time of collapse $\delta_c^{S}$, and the latter by the virial overdensity at the same
time $\Delta_{vir}^S$\footnote{These quantities do not refer to any
specific non-linear collapse model, thus we distinguish them 
from $\delta_c$ and $\Delta_{vir}$ of the spherical collapse.}. Since  
the collapse and virialization processes are specific
to each mass, redshift and cosmology, we can expect
$\delta_S$ to be cosmology and redshift dependent, and consequently, for a given $\sigma$
the mass function as well. This is explicitly
manifest in the Press-Schechter (PS) mass function formula, with $\delta_S\equiv\delta_c$ 
\begin{equation}
\frac{dn}{dM}=-\frac{\bar{\rho}}{M}\frac{1}{\sigma^2}\frac{d\sigma}{dM}f\left(\frac{\delta_c}{\sigma}\right)\label{dndm}
\end{equation}
where the functional form of $f$ is given by:
\begin{equation}
f_{PS}\left(\frac{\delta_c}{\sigma}\right)\equiv\mathcal{F}\left(\frac{\delta_c}{\sigma}\right) \delta_c=\left(\frac{2}{\pi}\right)^{1/2}\frac{\delta_c}{\sigma}e^{-\delta_c^2/(2\sigma^2)}\cdot
\end{equation}
While, in the case of the Sheth-Tormen formula (ST) \citep{sheth99,sheth01}, the functional form of $f$
is given by:
\begin{equation}
f_{ST}\left(\frac{\delta_c}{\sigma}\right)=A\left(\frac{2a}{\pi}\right)^{1/2}\frac{\delta_c}{\sigma}\left[1+\left(\frac{\delta_c}{\sigma\sqrt{a}}\right)^{-2p}\right]e^{-\delta_c^2
  a/(2\sigma^2)},\label{ST}
\end{equation}
with $A=0.322$, $a=0.707$ and $p=0.3$. Again in these formula $\delta_c$ is the
spherical collapse prediction of a given cosmology, nevertheless the cosmology dependence encoded
in $\delta_c$ is often neglected in the literature (see however \citet{percival05,francis09a,francis09b,grossi09}). This is because the spherical collapse model in the SCDM
scenario, which has been for long time the reference cosmology to study structure formation, 
predicts $\delta_c=1.686$ and $\Delta_{vir}=178$ constant in redshift. Therefore it has become common to set
$\delta_c$ to such a value. Alternatively the functional form of 
$f$ in Eq.~(\ref{dndm}) has been directly fitted against the
mass function measured in numerical simulations as a function of $\sigma$ only \citep{jenkins01,linder03,warren06}. 
For example, \citet{jenkins01} found
\begin{equation}
f(\sigma)=0.315 \exp{(-|\ln{\sigma^{-1}}+0.61|^{3.8})},\label{JK}
\end{equation}
over the range $-1.2\le\ln{\sigma^{-1}}\le 1.05$, with deviations
for different cosmologies within $20\% $ level. In such a case
Eq.~(\ref{dndm}) depends on the matter density $\Omega_m$ and the rms
linear density fluctuation $\sigma$ only, thus manifestly independent
  of the specificities related to the cosmological and redshift
  evolution of the non-linear collapse and
  virialization process. This is what is commonly
understood as ``{\it universality}'' of the mass function. Another way
  of rephrasing this idea is to say that the function $f$ in Eq.~(\ref{dndm})
\begin{equation}
f(\sigma)\equiv\int_0^\infty \delta\frac{\partial S}{\partial\delta}\mathcal{F}(\frac{\delta}{\sigma})d\delta,\label{fsig}
\end{equation}
is universal if the selection function $S$ is independent of cosmology
and redshift.
How exact is this statement, and to what extent it remains valid 
especially in the context of dark energy cosmologies?

\begin{figure} 
\begin{center}
\includegraphics[width=\hsize]{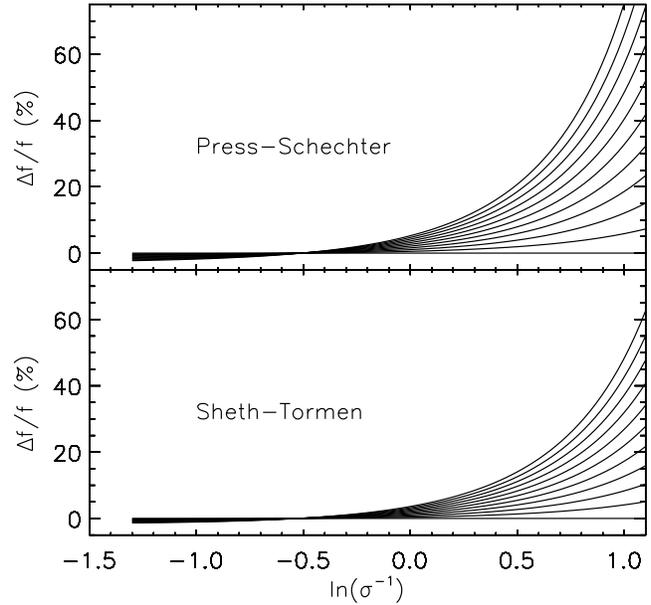}
\end{center}
\caption{Relative variation of $f(\delta_c/\sigma)$ for different values of the
spherical collapse derived density threshold $\delta_c$ as a function of $\ln(\sigma^{-1})$.
 Top panel shows variation for the Press-Schechter
formula, while we plot the variation for the Sheth-Tormen formula in
the bottom panel. Curves from top to bottom correspond to regularly spaced values of
$\delta_c$ in the range $1.638$ to $1.686$ (covering the interval of
values inferred from the spherical collapse for the cosmological models considered in this
  paper).}  
\label{deltafsurf}
\end{figure}

We can have an estimate of the influence of varying $\delta_S$
on $f(\sigma)$ by evaluating the relative error
\begin{equation}
\frac{\Delta f}{f}=\frac{\int^\infty_0 \delta \frac{\partial^2S}{\partial \delta^2}\mathcal{F}(\frac{\delta}{\sigma})d\delta}{\int^\infty_0 \delta \frac{\partial S}{\partial \delta}\mathcal{F}\left(\frac{\delta}{\sigma}\right)d\delta}\Delta{\delta_S},
\end{equation}
we may notice that deviations from a universal behaviour are proportional 
to $\Delta \delta_S$. In Fig.~\ref{deltafsurf}, we illustrate this for
the (extended) Press-Schechter (PS) formula and the Sheth-Tormen (ST) parametrization,
where we let $\delta_S$ ($=\delta_c$) varying in the range 1.638 to 1.686 
(corresponding to the typical range of $\delta_c$ values covered by the
cosmological models studied in this paper). We can see that the
effect of varying $\delta_S$ is more important on the high mass tail of
the mass function (i.e. $\ln{(\sigma^{-1})}>0$). This is because the mass
function is exponentially sensitive to the value of $\delta_S$. It is for
this very reason that we will specifically focus on the high mass tail,
nevertheless this does not imply that there are no effects
on smaller masses. From Fig.~\ref{deltafsurf} we can also notice that the
variations induced on $f(\sigma)$ are similar
for the PS and ST prescriptions, thus suggesting that we can study deviations from universality independently of the specific form fo the mass function. To this purpose, we focus on models which exhibit the same rms fluctuations $\sigma$, while being characterized by different background expansion histories and evolutions of the linear perturbations.


\subsection{Cosmological toy-models: background and linear evolution}
\label{cosmo}

We consider as reference cosmology a $\Lambda$CDM model best fit to WMAP-5 years data
\citep{komatsu09} which hereafter we refer to as $\Lambda$CDM-W5. For
comparison we also consider two additional $\Lambda$CDM models
calibrated to WMAP-1 and 3 years data \citep{spergel03,spergel07}, which we dub
as $\Lambda$CDM-W3 and $\Lambda$CDM-W1 respectively.
The model parameter values of these $\Lambda$CDM-WMAP cosmologies are 
listed in Table~\ref{wmap_cosmo}, the most noticeable difference between these models concerns the $\sigma_8$ value,
nevertheless their expansion and linear growth histories are almost identical as we will discuss later in this Section. 
In order to investigate the imprint of dark energy on the halo mass function
and test the universality hypothesis, we confront the $\Lambda$CDM-W5 cosmology with a set of
``toy-models''. These are flat cosmological models with different
background expansion and linear growth of the density perturbations.
Following the discussion of the previous paragraph we additionally require the models to have
the same distribution of linear density fluctuations at $z=0$, hence
the same $\sigma_8$ value. 

\begin{table} 
\begin{center}
\begin{tabular}{cccccc}
\hline \hline
Model & $\Omega_{DE}$ &  $h$ & $\sigma_8$ & $n_s$ & $\Omega_b$ \\
\hline\hline
$\Lambda$CDM-W5&0.74&0.72 & 0.79&0.963&0.044\\
\hline
$\Lambda$CDM-W3&0.76&0.73&0.74&0.951&0.042\\
\hline
$\Lambda$CDM-W1&0.71&0.72&0.90&0.99&0.047\\
\hline\hline
\end{tabular}
\caption{\label{wmap_cosmo} Cosmological parameter values of the WMAP
 calibrated cosmologies. Our reference model is the $\Lambda$CDM-W5.}  
\end{center}
\end{table}

We focus on a quintessence model with Ratra-Peebles potential
\citep{ratra88}:
\begin{equation}
\label{rp}
V(\phi)=\frac{\lambda^{4+\alpha}}{m_{Pl}^\alpha\phi^\alpha},
\end{equation}
where $\alpha$ and $\lambda$ are the slope and amplitude of the scalar
self-interaction respectively, and $\phi$ is the quintessence field
evolving according to the Klein-Gordon equation. For $\alpha=0$ the quintessence cosmology resembles a standard
$\Lambda$CDM, provided that the initial
field velocity vanishes. We choose the RP
model since it corresponds to a dark energy component
whose equation of state can vary from $w=-1$ (cosmological constant
value) to an evolving function of the redshift ($w(z)\ge-1$) 
by changing the slope of the
potential through the parameter $\alpha\ge0$. As we are specifically interested 
in cosmological 
evolutions which largely differs from that of $\Lambda$CDM, 
we consider a quintessence model with $\alpha=10$, which we dub as L-RPCDM
(the letter L in the acronym means large $\alpha$ value). 

We also construct two models with different amount of
dark energy density. In particular we consider a $\Lambda$CDM 
model characterized by a large value of the dark energy
density, $\Omega_{DE}=0.9$, which we refer to as L-$\Lambda$CDM (here
L meaning large $\Omega_{DE}$ value), 
and a cold-dark matter dominated cosmology, SCDM$^*$, with $\Omega_{DE}=0$
or equivalently $\Omega_m=1$ (the $*$ symbol is to remind that
the other model parameter values differ from the 
SCDM usually considered in the literature \footnote{SCDM with $\sigma_8=0.51$, $h=0.5$ and shape parameter $\Gamma=0.5$ as in \citet{jenkins98}}). 
In Table~\ref{para_toy1} we quote the toy-model parameters, while all the other
cosmological parameters ($h,n_s,\sigma_8,..$) are set to the $\Lambda$CDM-W5 values.
An important point is that given the same initial
conditions at some early time (e.g. at recombination) these models predict different 
matter power spectra at $z=0$. However, we would like to investigate
deviations from a universal behaviour of the mass function independently of the
present form of the matter power spectrum. 
Therefore we artificially force these
models to have the matter power spectrum at $z=0$ of the 
$\Lambda$CDM-W5 reference cosmology. From a practical point of view this means that when
running the N-body simulations, for each model we generate the initial
conditions such that the linearly extrapolated matter power
spectrum at $z=0$ (obtained by using the growth factor $D^+(z)$
specific to each model) 
coincides with that of the $\Lambda$CDM-W5. Since $D^+(z)$ is
model dependent, it implies that the various models will have the same
initial power spectrum, but a different initial redshift (see Sect.\ref{nbody} and Sect.\ref{subsec_numerics} for technical details about this point). We want to
stress that such toy-models are not intended to be
compatible with observations, contrary to the ``realistic models''
considered in the previous paper \citep{alimi09} which were calibrated
against CMB and SN Ia data. Again, our aim here is to perform a
physical study of the cosmological dependence of the mass function. 
Nonetheless the conclusions drawn from this
study will be extended to more realistic cosmological models as well in a forthcoming paper.

\begin{table} 
\begin{center}
\begin{tabular}{cccc}
\hline \hline
Model & $\Omega_{DE}$ & $\alpha$ & P(k) \\
\hline\hline
L-RPCDM & 0.74 &10 & $\Lambda$CDM-W5\\
\hline
L-$\Lambda$CDM &0.9 &- &$\Lambda$CDM-W5\\
\hline
SCDM$^*$ &0 &-  & $\Lambda$CDM-W5\\
\hline\hline
\end{tabular}
\caption{\label{para_toy1} Toy-models characteristics; the values of the other parameters
  $h$, $\sigma_8$ and $n_s$ are set to the $\Lambda$CDM-W5 values quoted
  in Table~\ref{wmap_cosmo}.}  
\end{center}
\end{table}

\begin{figure*} 
\begin{tabular}{ccc}
\includegraphics[width=0.3\hsize]{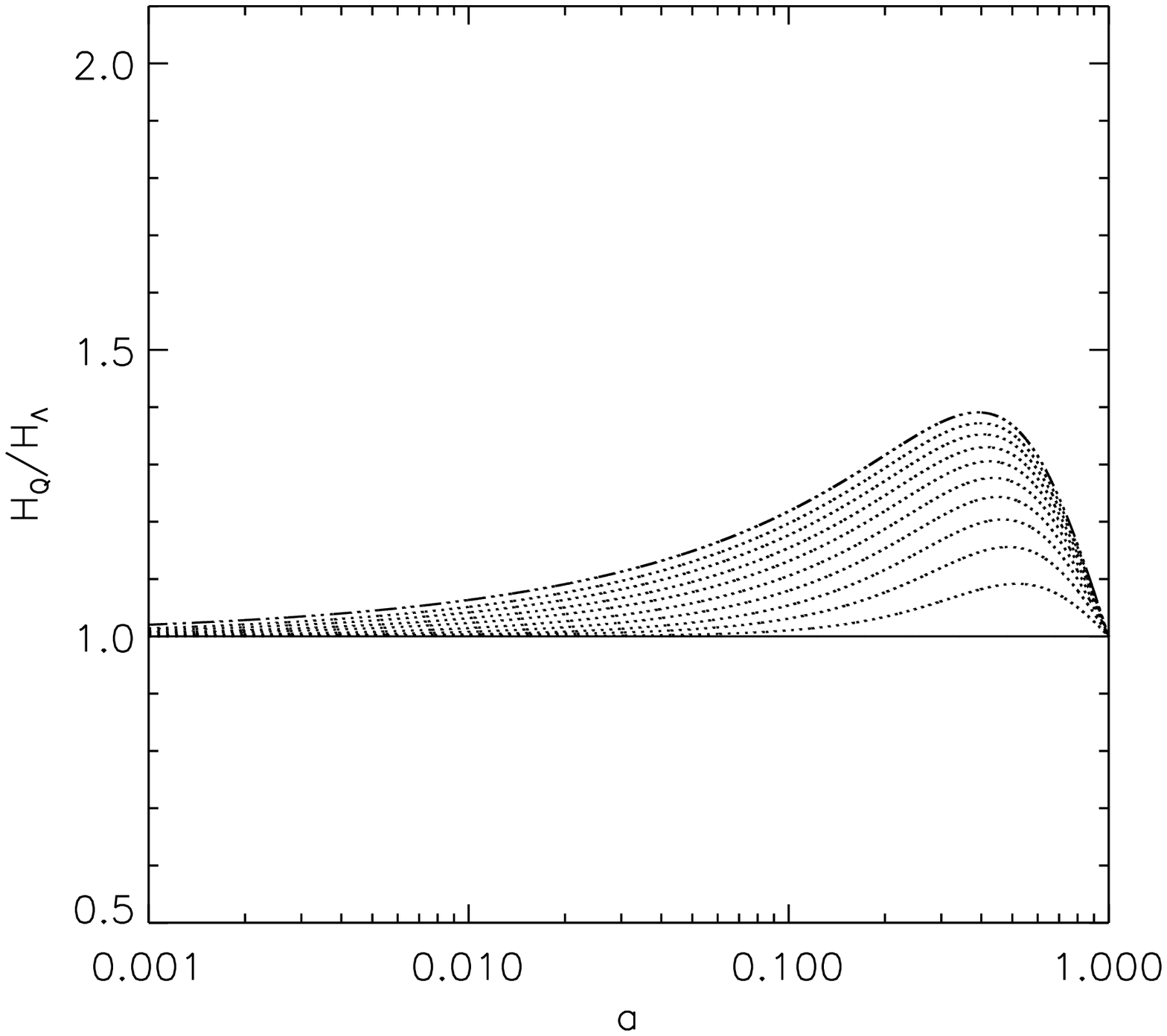}&
\includegraphics[width=0.3\hsize]{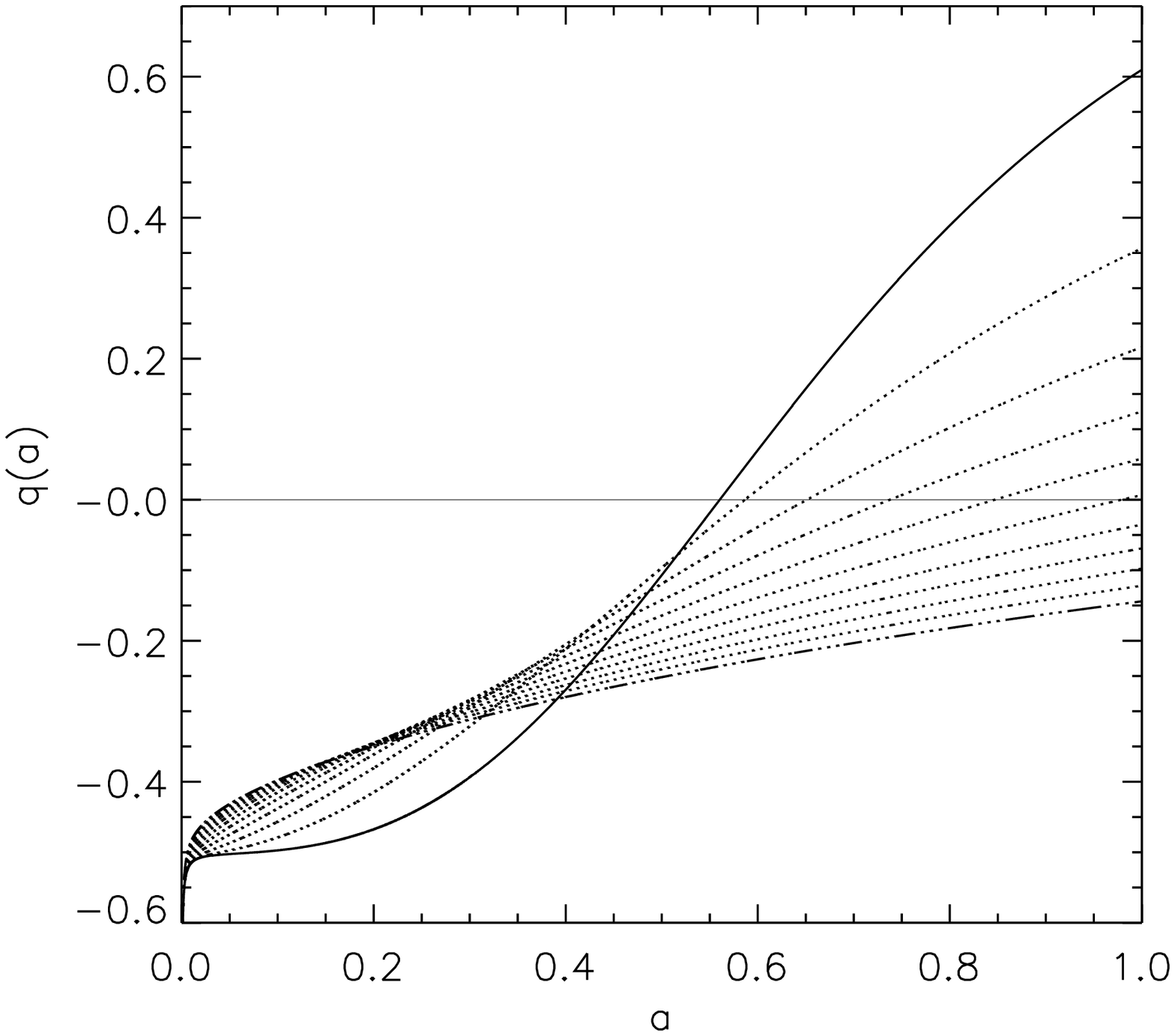}&
\includegraphics[width=0.3\hsize]{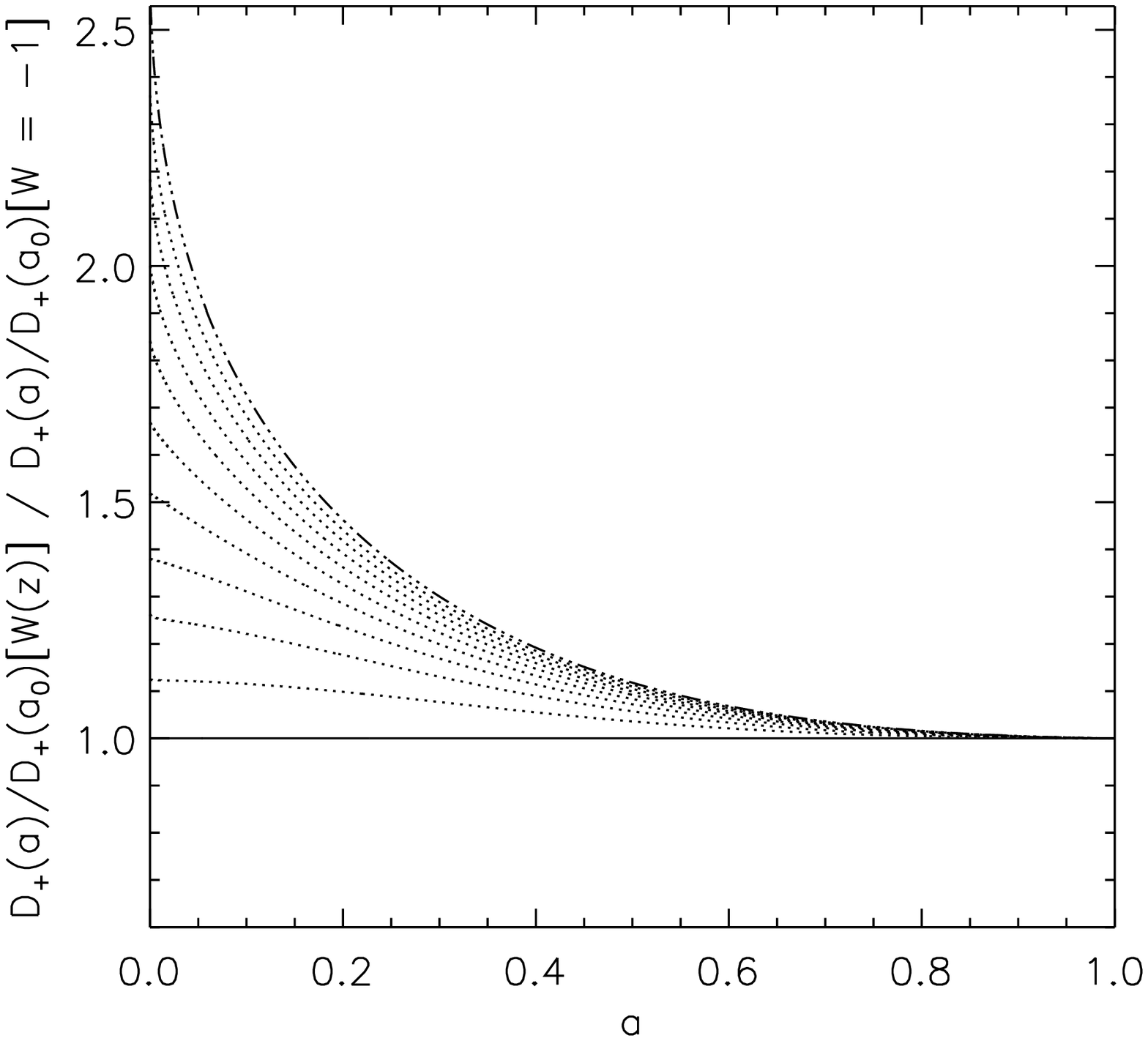}\\
\includegraphics[width=0.3\hsize]{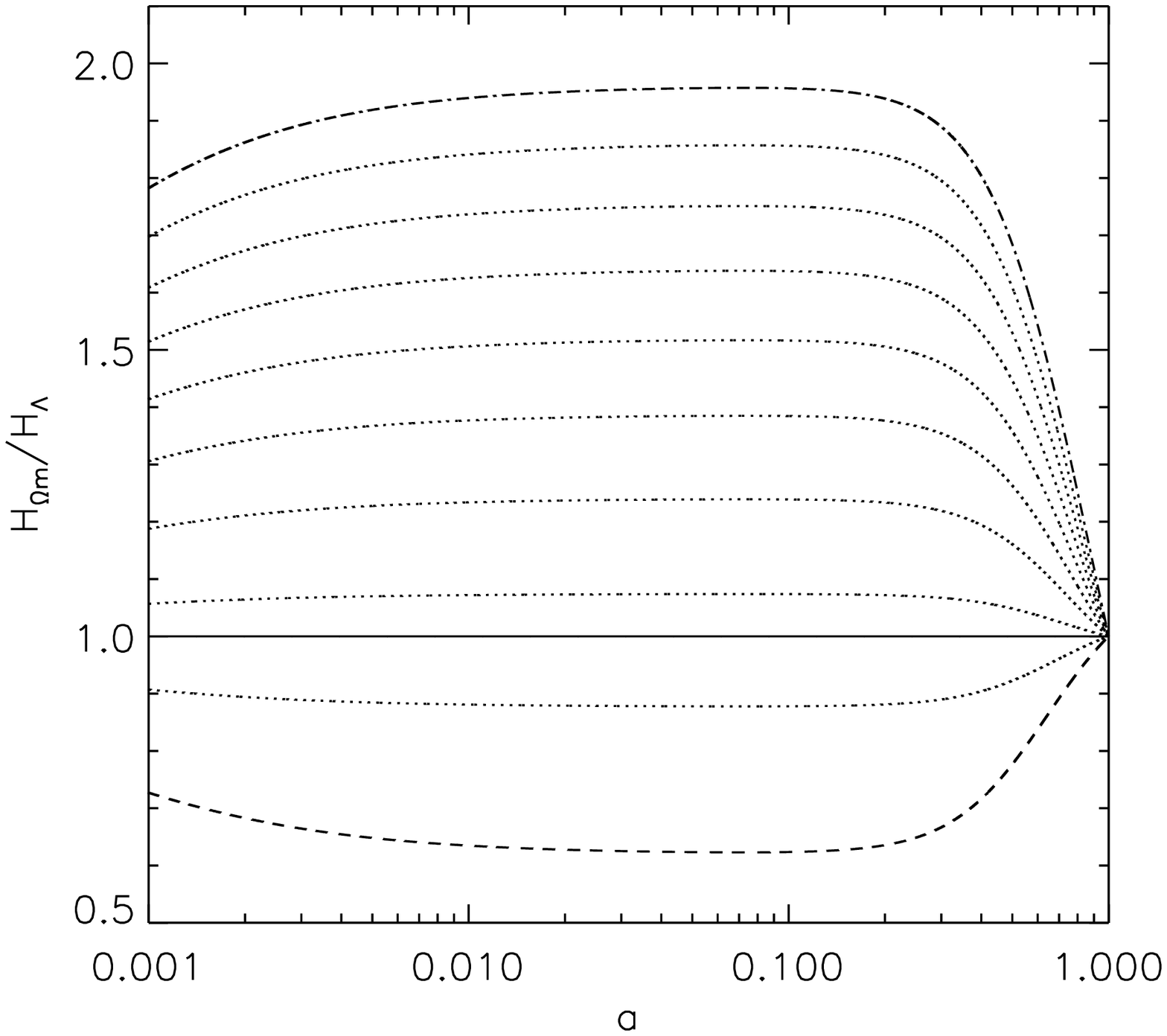}&
\includegraphics[width=0.3\hsize]{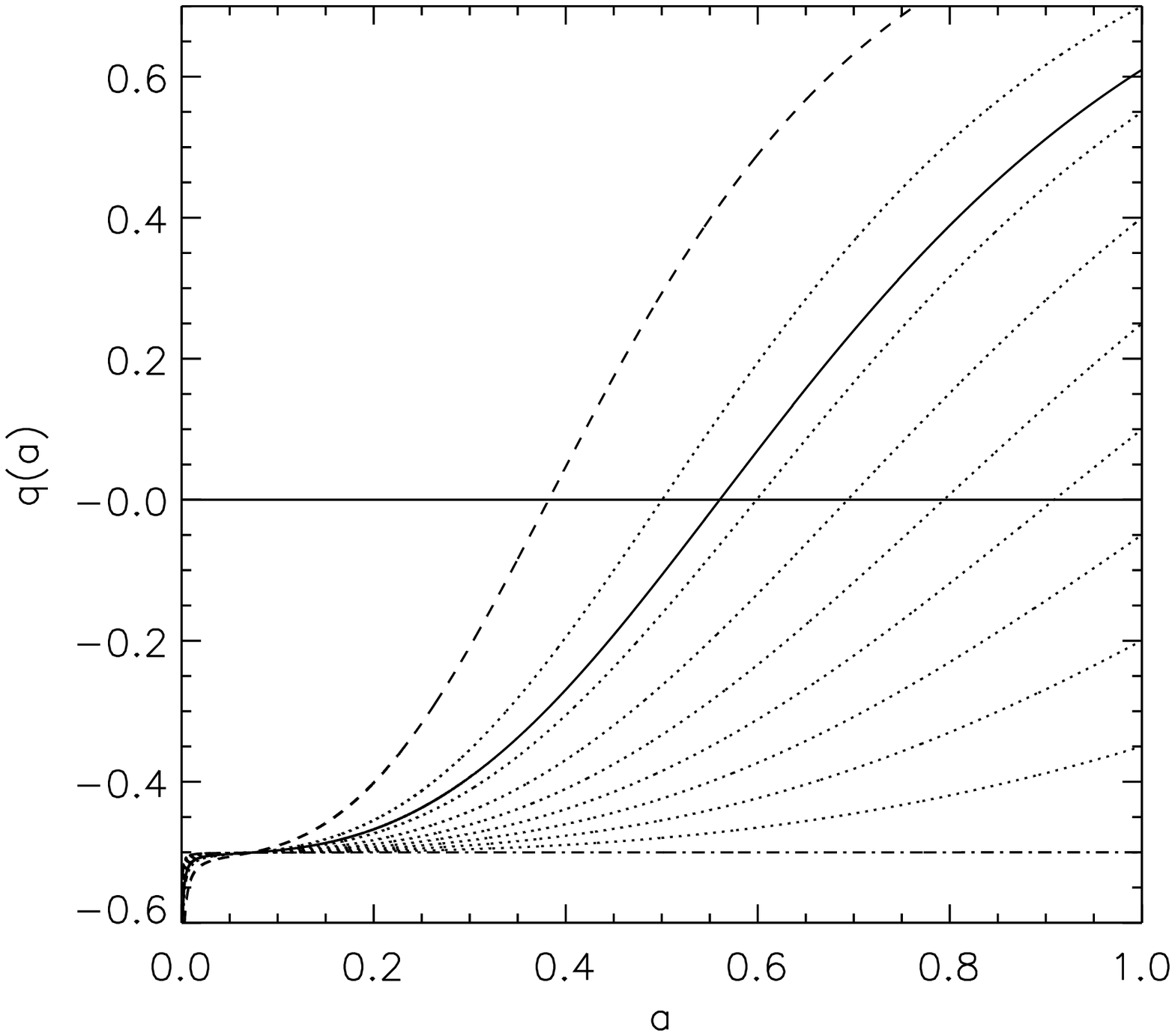}&
\includegraphics[width=0.3\hsize]{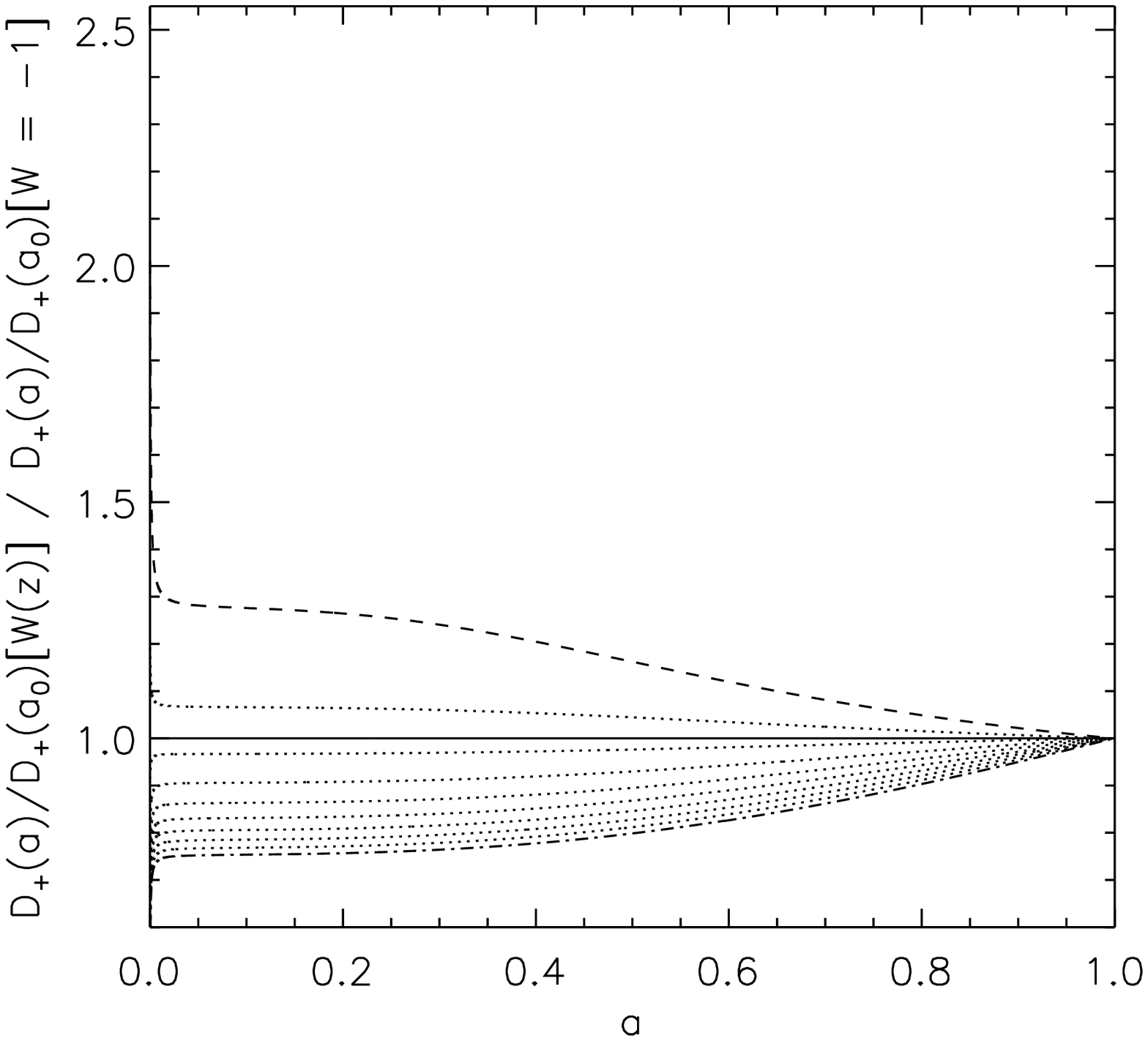}\\
\includegraphics[width=0.3\hsize]{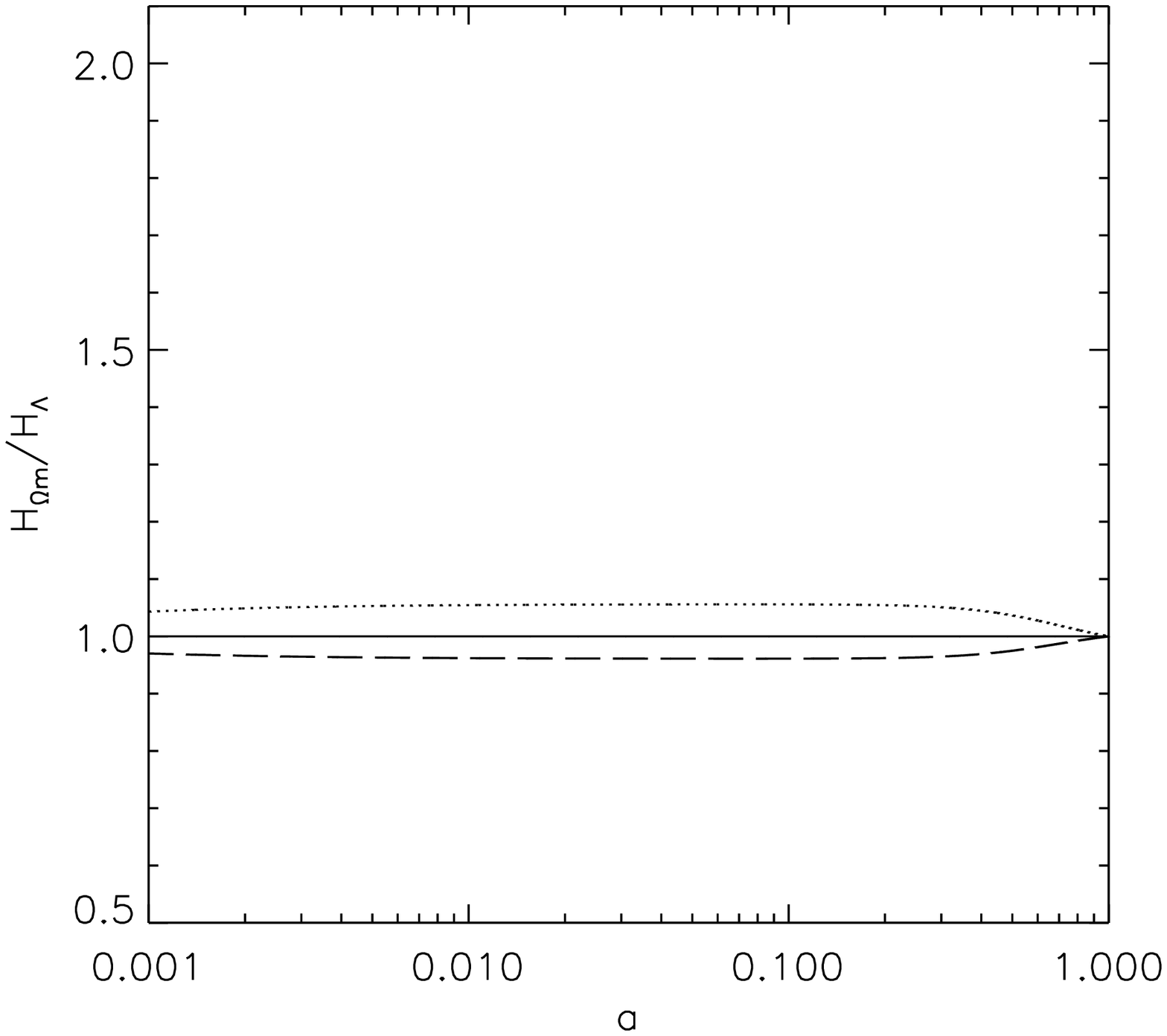}&
\includegraphics[width=0.3\hsize]{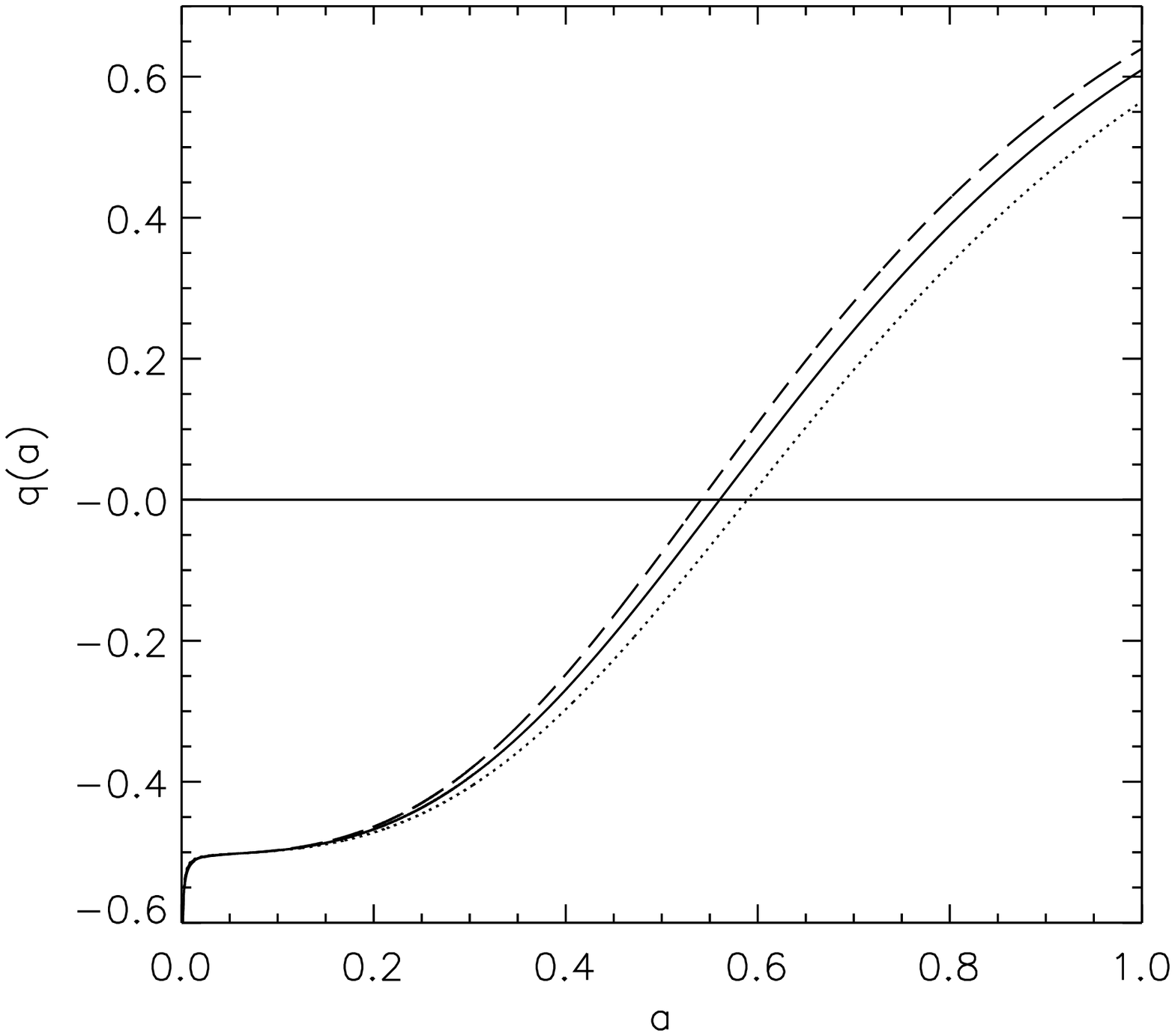}&
\includegraphics[width=0.3\hsize]{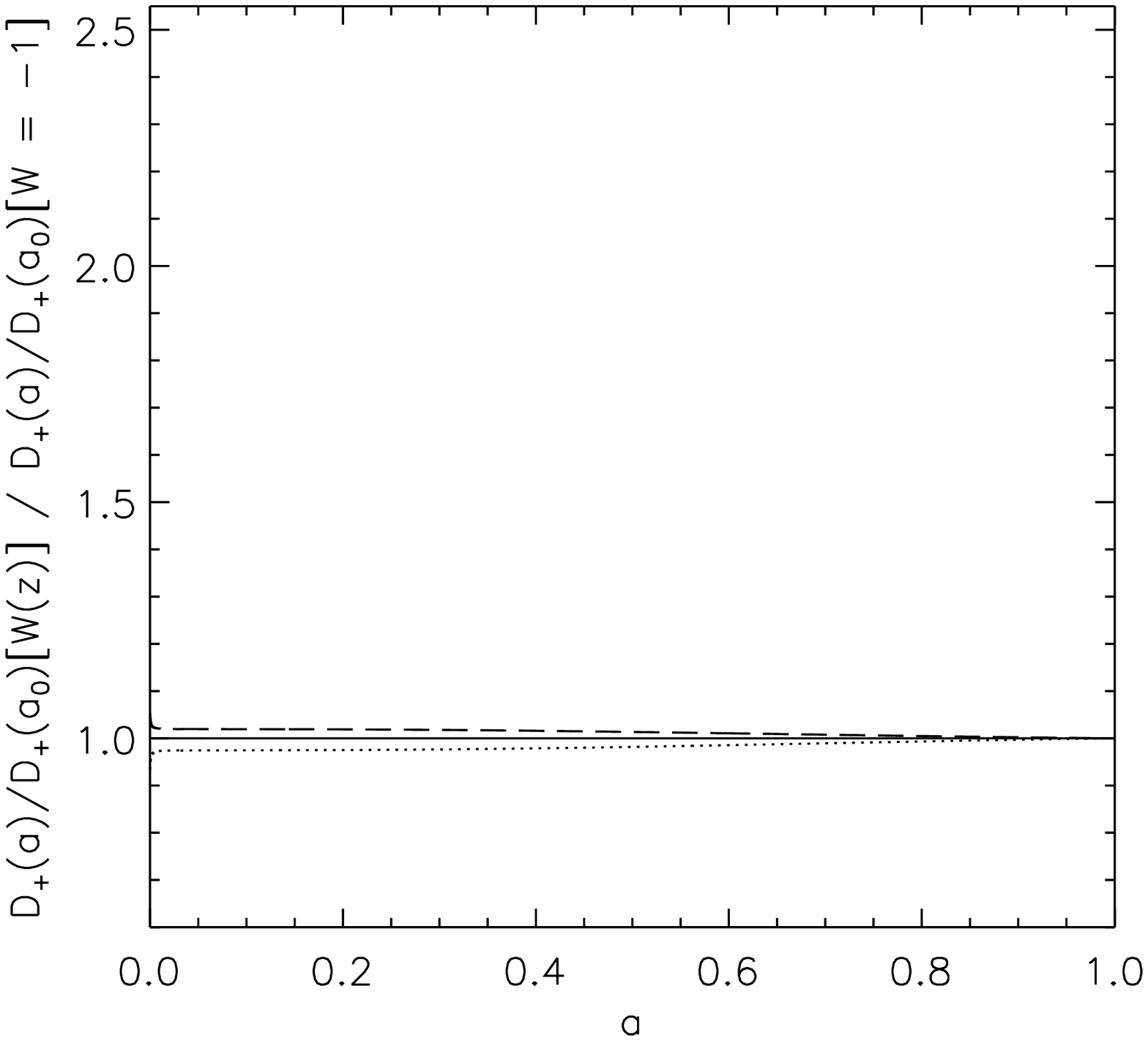}
\end{tabular}
\caption{\label{phen} Scale factor dependence of the Hubble rate (left
  column panels), acceleration parameter (middle column panels) and
  linear growth factor (right column panels) for the RP models with
  different values of $\alpha=0,1,..,10$ (upper row panels), for $\Lambda$CDM with
  different values of $\Omega_{DE}=0,0.1,..,1$ (central row panels) and for the
  WMAP models (lower row panels). In all panels the solid line
  corresponds to the $\Lambda$CDM-W5; in the upper panels the L-RPCDM
  model corresponds to the triple-dot-dashed line, while in the
  central panels the L-$\Lambda$CDM and SCDM$^*$ models correspond to
  the short-dashed and dot-dashed lines respectively. In the bottom
  panels $\Lambda$CDM-W1 and $\Lambda$CDM-W3 models corresponds to the
  dotted and long-dashed lines respectively.}
\end{figure*}

In Fig.~\ref{phen} we plot the Hubble rate $H(a)$, 
the acceleration parameter $q(a)$ and the linear growth factor
$D^+(a)/D_0^+$ for different cosmologies including our toy-models.
The Hubble rate and the
growth factor are normalized to that of the $\Lambda$CDM-W5 cosmology.
In the upper panels we plot RP models with
$\alpha=0,1,..,10$; the L-RPCDM model is shown as
triple-dot-dashed line. We may notice that the Hubble rate increasingly 
deviates at higher redshifts from that of the reference cosmology for larger
values of $\alpha$ (curves bottom to top in the upper left panel). 
Similarly, in such models the evolution of the acceleration
parameter shows that the expansion is less decelerated during the
matter dominated era and less accelerated during the dark energy
dominated era (see upper central panel). Besides the acceleration starts later for
larger values of $\alpha$, such that above
some values (corresponding to dark energy models with equation of state $w>-1/(3\Omega_{DE})$) 
the acceleration never takes place. As dark energy dominates earlier (for increasing
values of $\alpha$) the deviation of the linear growth rate with respect to
the $\Lambda$CDM-W5 case is larger at high
redshift, with differences up to a factor $2$ at $a=0.1$ (see curves
bottom to top in the upper right panel). 
In the middle panels of Fig.~\ref{phen} are shown $\Lambda$CDM models (equivalent to RP models with $\alpha=0$)
with different amount of dark energy density. The various
curves correspond to $\Omega_{DE}=0,0.1,..,1$ (curves bottom to top). The
L-$\Lambda$CDM and SCDM$^*$ models are plotted as short-dashed and 
dot-dashed lines respectively. We can see here a trend which is similar to that
of varying $\alpha$ in the RP models. However when increasing $\Omega_{DE}$
the deviations on the Hubble rate are much larger compared to the previous
case. Moreover, since the dark energy equation of state here is fixed
to the cosmological constant
value ($\alpha=0$ or $w=-1$), for decreasing values of $\Omega_{DE}$ the acceleration 
occurs later, and eventually never takes place for sufficiently low values of $\Omega_{DE}$ (corresponding to matter dominated cosmologies). 
On the other hand we may notice that the linear growth
factor deviates from that of the reference cosmology both at low and
high redshifts (middle right panel). Finally in the lower panels are shown the $\Lambda$CDM-WMAP models.
As we can see these have very similar behaviours both at the
background and linear level. Henceforth our toy-models are characterized by a cosmic evolution
which differs from that of standard vanilla $\Lambda$CDM
cosmology. Deviations on the Hubble rate are up to $40\%$
for the L-RPCDM model, $50\%$ for the L-$\Lambda$CDM and $95\%$ for
SCDM$^*$ respectively. Also the redshift dependence of the
linear growth factor differs from that of the reference cosmology with
larger deviations at higher redshifts, up to a factor $2$ at $a=0.1$
for the L-RPCDM model. In the case of L-$\Lambda$CDM
and SCDM$^*$ such deviations are present both at high and low redshifts and up to
$30\%$ level respectively. If the universality hypothesis holds then such 
differences should not leave any imprint in the halo mass function.

\subsection{Cosmological toy-models: spherical collapse}
\label{spherical}

\begin{figure*} 
\begin{tabular}{cc}
\includegraphics[width=0.45\hsize]{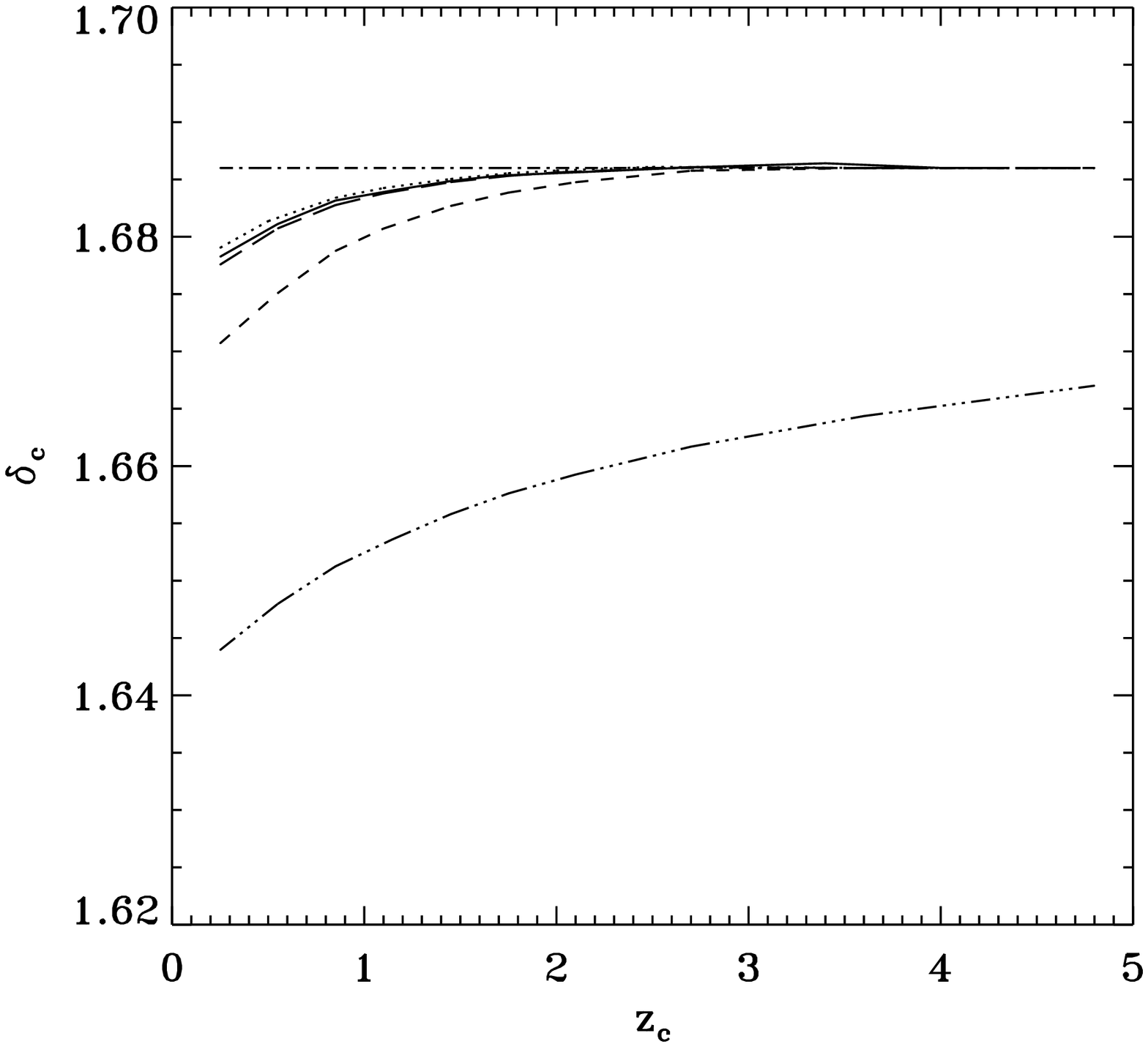}&
\includegraphics[width=0.45\hsize]{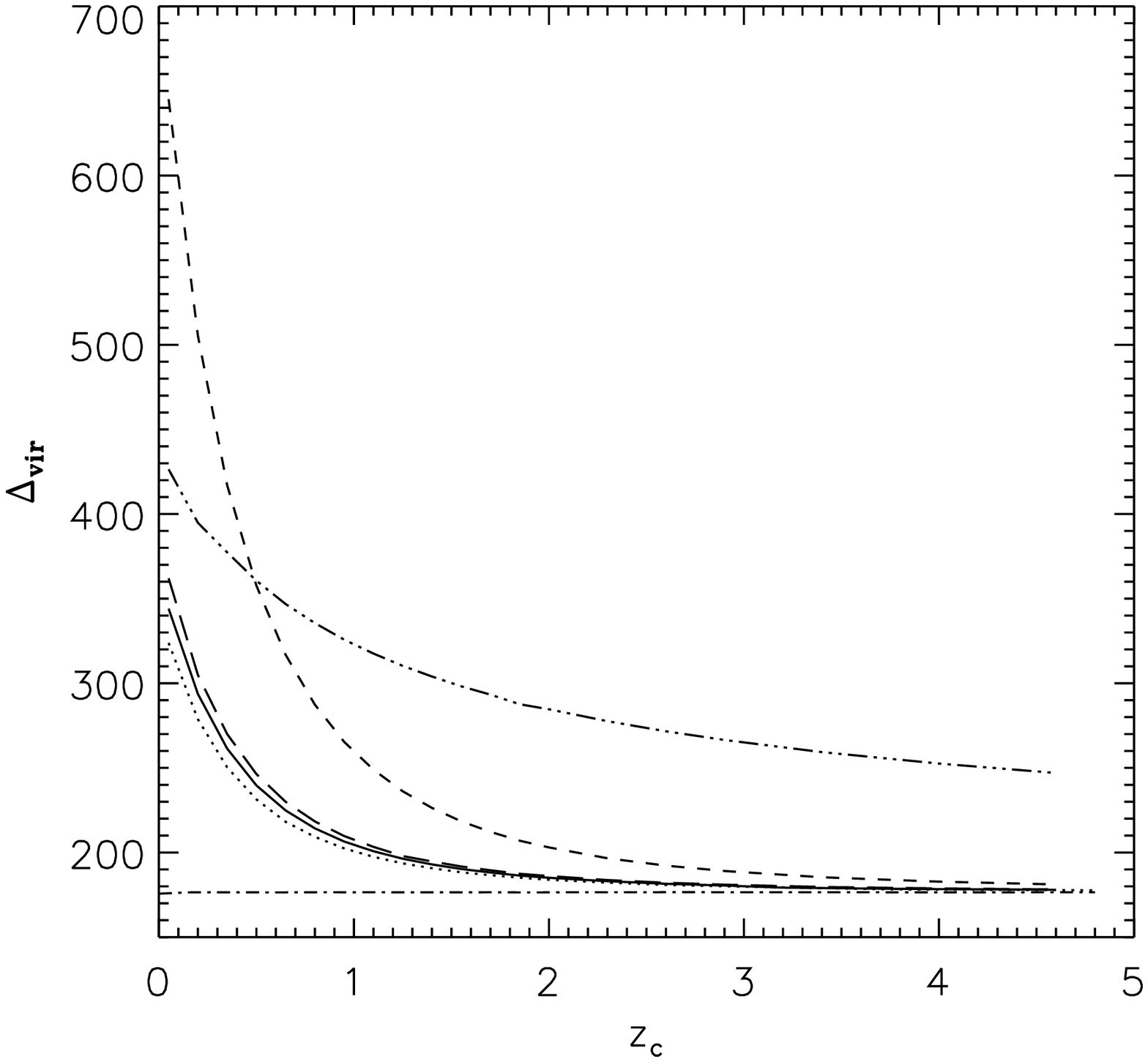}
\end{tabular}
\caption{\label{delta_sc_evol} Evolution of $\delta_c$ (left panel)
  and $\Delta_{vir}$ as a function of the redshift of collapse for the
  different cosmological models, the line coding is the same as in Fig.~\ref{phen}. Note that $\delta_c$ and $Delta_{vir}$ for L-RPCDM does tend towards 1.686 and 178 but at very high redshift.}
\end{figure*}

The collapse of a spherical matter overdensity embedded in an
expanding Friedmann-Lema\^itre-Robertson-Walker (FLRW) background is the simplest
model to describe the evolution of a density perturbation throughout the
non-linear phase of gravitational collapse. For a given cosmology it
allows us to estimate the time of collapse of an initial density
perturbation, as well as the value of the overdensity at the time of 
virialization (i.e. assuming that the perturbation virializes at some time before collapse).
Formally, the spherical collapse equation describes the evolution of a matter overdensity 
$\delta_{m}$ with a top-hat profile in a sphere of radius $R$ as derived from the Raychaudury
equation and which reads as\footnote{For a recent study of the spherical
  collapse model in dark energy cosmologies and a fully relativistic
  derivation of the Raychaudury equation we refer to \citet{creminelli09}}:
\begin{equation}
\frac{\ddot{R}}{R}=-\frac{4\pi G}{3}(\rho+3p),\label{R}
\end{equation}
where $\rho$ and $p$ are the total energy density and pressure within the sphere. Here we
treat dark energy as a homogeneous component presents only at the
background level. This is required for consistency with the fact that
our N-body simulations do not explicitly account for the presence of
dark energy perturbations, except for the linear regime of the realistic model
simulations discussed in \citep{alimi09}. 
Hence the halos detected in the simulations
are effectively dark matter overdensities embedded in a FLRW
background in the presence of a homogeneous dark energy. Since we aim
to interpret the properties of these halos using the spherical collapse model,
we solve the model equations in a similar cosmological setup.

In such a case the energy density and pressure in the spherical region are given by
$\rho=\rho_m^c+\bar{\rho}_{DE}$ and $p=w_{DE}\bar{\rho}_{DE}$, 
where $\rho_m^c\equiv\bar{\rho}_m(1+\delta_m)$
is the matter density in the sphere of radius $R$, while $\bar{\rho}_m$ 
and $\bar{\rho}_{DE}$ are the background matter and dark energy density
respectively, with $w$ being the dark energy equation of state. The background densities evolve according to 
\begin{equation}
\dot{\bar{\rho}}_m+3\frac{\dot{a}}{a}\bar{\rho}_m=0,\label{rhombar}
\end{equation}
and
\begin{equation}
\dot{\bar{\rho}}_{DE}+3(1+w)\frac{\dot{a}}{a}\bar{\rho}_{DE}=0.
\end{equation}
For the quintessence models considered here the dark energy density 
and equation of state depends are given in terms of the scalar field 
potential and kinetic energy
$\bar{\rho}_{DE}=\dot{\phi}^2/2+V(\phi)$ and $w=\frac{\dot{\phi}^2/2-V(\phi)}{\dot{\phi}^2/2+V(\phi)}$, 
with the scalar field $\phi$ evolving according to the Klein-Gordon equation which we solve numerically. 
The evolution of the local matter density inside the sphere is given by
\begin{equation}
\dot{{\rho}}_m^c+3\frac{\dot{R}}{R}{\rho}_m^c=0.\label{rhoc}
\end{equation}
Integrating Eq.~(\ref{rhoc}) and substituting the definition of 
$\rho_m^c$ in terms of the background matter density, we find the
relation between $\delta_m$ and $R$
at time $t$, 
\begin{equation}
1+\delta_m=(1+\delta_m^i)\left(\frac{a}{a_i}\right)^3\left(\frac{R_i}{R}\right)^3,\label{deltam}
\end{equation}
where $\delta_m^i$, $a_i$ and $R_i$ are the matter overdensity, 
scale factor and radius at the initial time $t_i$. Finally using
Eq.~(\ref{deltam}) we can 
rewrite Eq.~(\ref{R}) in terms of a second order differential equation 
for the spherical matter overdensity \citep[see also]{Nunes06,Abramo07}:
\begin{equation}
\ddot{\delta}_m+2H\dot{\delta}_m=4\pi G\bar{\rho}_m\delta_m(1+\delta_m)+\frac{4}{3}\frac{\dot{\delta}^2_m}{1+\delta_m}.\label{dmnl}
\end{equation}
Equation~(\ref{dmnl}) is a non-linear ODE, which at the linear order reduces to 
the standard equation for spherical linear matter perturbations in the presence of a 
homogeneous dark energy component \citep[see e.g.][in the inhomogeous case]{creminelli09}. Notice that 
dark energy affects the matter spherical collapse only because of its influence on the background dynamics through the friction
term appearing in Eq.~(\ref{dmnl}). 

Past works have solved the spherical collapse using Eq.~(\ref{R}), however we
prefer to work with Eq.~(\ref{dmnl}) which we solve numerically as an initial
conditions problem rather than a boundary value one as in the case of 
Eq.~(\ref{R}). This allows us to relax some the assumptions used in 
the literature, such as estimating the time of collapse $t_c$ assuming that $t_c=2t_{ta}$, where
$t_{ta}$ is the time of turn-around, i.e. when $\dot{R}=0$ after an initial phase of radial expansion. 
Besides by solving the linearized form of Eq.~(\ref{dmnl}) for the same set of initial conditions
that lead to collapse at $t_c$ (or equivalently $z_c$), we are able to directly infer
the value of the linear density contrast at the time of collapse, $\delta_c$, without 
the need of using semi-analytical
formula for the linear growth factor. 
In order to numerically solve Eq.~(\ref{dmnl}), we first rewrite it in terms of the redshift variable $z$ 
(i.e. $\frac{d}{dt}=\frac{d}{dz}\frac{dz}{dt}$, hereafter $'=d/dz$) 
and set initial conditions with the perturbation $\delta_m^{i}$ initially starting in the Hubble flow 
($\dot{R}_i/R_i=H_i$, i.e. ${\delta_m^{i}}'=0$). 
At the beginning the system evolves linearly, but as 
the quadratic terms in Eq.~(\ref{dmnl}) become non-negligible the collapse enters in the non-linear evolution 
which ultimately leads to a diverging non-linear matter density contrast, $\delta_m\rightarrow\infty$,
at the redshift of collapse $z_c$ (corresponding to the radius of the spherical region $R\rightarrow 0$).
Since our goal is to determine the linear density contrast for a perturbation
collapsing at $z_c=0$ and more generally at a given $z_c$, we have implemented 
a numerical algorithm which iteratively searches for the initial density perturbation value $\delta_m^i$ 
for which $\delta_m$ diverges at the input redshift value $z_c$. Formally $\delta_m(z_c)=\infty$, but
numerically it is not possible to verify such a condition exactly, hence we use an additional 
algorithm, similar to that
developed in \citet{London82}, to determine for a given level of accuracy when $\delta_m$ reaches the singularity.
We fix the accuracy to $\Delta{z_c}=10^{-5}$, then the algorithm alts the search for $\delta_m^i$
as soon as a divergent solution for $\delta_m$ is found within $\Delta{z_c}$ from the specific $z_c$ 
value of interest. At the same time we calculate on flight the turn-around condition $\delta_m'=0$ and determine
the corresponding redshift $z_{ta}$. Furthermore we numerically solve
the linearized form of Eq.~(\ref{dmnl}) using ${\delta_m^i}'=0$ and the $\delta_m^i$ value previously determined 
for which collapse occurs at $z_c$ to finally infer
the value of the linear density contrast at the time of collapse, $\delta_c$.
  
From the spherical collapse dynamics we can also calculate the
redshift and the non-linear overdensity value at the time of virialization. 
In the spherical collapse model, this can be implemented
only as an external condition by requiring that at the time of
virialization the kinetic energy of the system $T$ and the gravitational 
potential energy $U$ satisfy the virial condition
$T_{vir}=(R_{vir}/2)(\partial{U}/\partial{R})_{vir}$ \citep[see e.g.][]{Peebles93}. Then using energy conservation, 
$T_{vir}+U(z_{vir})={\rm const.}$, one can infer the redshift of virialization by solving the algebraic
equation
\begin{equation}
\frac{R}{2}\frac{\partial{U}}{\partial{R}}(z_{vir})+U(z_{vir})=U(z_{ta}),\label{vir}
\end{equation}
where we have used the fact that at turn-around the kinetic energy of
the system vanishes, $T_{ta}=0$. We refer to \citet{maor05} for an explicit form of
the gravitational potential energy in terms of the matter overdensity and dark energy
density for different cosmological set up,
including SCDM, $\Lambda$CDM and homogeneous dark energy
models. Using the numerical computation previously described 
we numerically solve Eq.~(\ref{vir}) to derive $z_{vir}$ and compute the virial overdensity 
$\Delta_{vir}\equiv\bar{\rho}_m(z_{vir})(1+\delta_m^{vir})/\bar{\rho}_m(z_c)$.
However we would like to remind the reader that the use of
Eq.~(\ref{vir}) is rigorously justified only in
the SCDM and $\Lambda$CDM models. This is because in the presence of a
homogeneous dark energy component for which the background density
evolves in time at a rate different from that of the background matter
component, energy is not conserved within the spherical overdensity
region. Despite several attempts
to account for such a loss of energy \citep[see e.g.][]{maor05}, we still
lack a full relativistic calculation\footnote{We thank Jorge Norena and
  Filippo Vernizzi for pointing this to us.}
\citep[see discussion in][]{creminelli09}. Thus similar to previous
analysis the virial overdensity value which we infer for the L-RPCDM
model should be considered only as approximative. Given this state-of-art
computation we cannot do better.

\begin{table} 
\begin{center}
\begin{tabular}{ccc}
\hline \hline
Model&$\delta_c$& $\Delta_{vir}$\\
\hline\hline
$\Lambda$CDM-W5&1.673&368\\
\hline
$\Lambda$CDM-W3&1.672&387\\
\hline
$\Lambda$CDM-W1&1.674&344\\
\hline
L-RPCDM&1.638&436\\
\hline
L-$\Lambda$CDM&1.665&708\\
\hline
SCDM*&1.686&178\\
\hline\hline
\end{tabular}
\caption{\label{deltac_deltavir} Values of the linearly extrapolated
  critical density threshold $\delta_c$ and virial overdensity
  $\Delta_{vir}$ at $z_c=0$ predicted by the 
  spherical collapse model. Note that $\Lambda$CDM-WMAP
  cosmologies have very similar values for the collapse and
  virialization parameters. The L-RPCDM model differs mainly in the
  value of $\delta_c$, while the effect of varying the dark energy
  density as described by the SCDM$^*$ and L-$\Lambda$CDM models
  primarily affects the virial overdensity value.}  
\end{center}
\end{table}

In Table~\ref{deltac_deltavir} we list the values of $\delta_c$ and
$\Delta_{vir}$ at $z_c=0$ for the different cosmological models. 
We recover the standard SCDM$^*$ values $\delta_c=1.686$ and $\Delta_{vir}=178$; 
the values corresponding to the various $\Lambda$CDM models are
consistent with those found in \citet{eke96}, similarly the values of the
L-RPCDM model are compatible with those quoted in \citet{mainini03}.
As expected, the $\Lambda$CDM-WMAP cosmologies have very similar
collapse and virialization parameters, this is coherent with the fact
that their expansion history and linear growth evolution are very similar as well.
In contrast the L-RPCDM model predicts the lowest value, $\delta_c=1.638$, 
which in the framework of the Press-Schechter formalism means that in such a 
model structures form earlier. The spherical collapse prediction of $\delta_c$ for 
L-$\Lambda$CDM and SCDM$^*$ gives $\delta_c=1.665$ and $1.686$ respectively, only a few percent
lower and higher than those of $\Lambda$CDM-WMAP models. In contrast the former models predict
very different values for the virial overdensity ($708$ and $178$ respectively). 
Since a larger virial overdensity implies 
a more compact object, we expect that virialized objects tend to be more compact 
as the amount of dark energy increases.

In Fig.~\ref{delta_sc_evol} we plot the evolution of $\delta_c$ (left
panel) and $\Delta_{vir}$ (right panel) as a function of the redshift of collapse. We can see
that both the collapse threshold and the virial overdensity is closed
to the SCDM values with large $z_c$. This is expected
since at higher redshift the collapse occurs in a matter
dominated universe. The redshift evolution of $\delta_c$ in L-RPCDM shows the largest deviation with
respect to the trend of the other models, with a much slower convergence toward SCDM. 
This is because dark energy
starts dominating earlier in L-RPCDM than the other cosmological
models. In contrast the evolution
$\Delta_{vir}(z_c)$ has the largest deviation for the
L-$\Lambda$CDM model, with $\Delta_{vir}$ rapidly decreasing towards high
redshifts due to the fact that in such a model the acceleration occurs
earlier (see Fig.~\ref{phen}).

The spherical collapse model remains a very simplistic description of
the gravitational collapse of structures in the universe. Typically 
these are not isolated, or spherical, and their velocity
dispersion and angular momentum are certainly not
negligible. Nevertheless the model captures some features of the non-linear
phase of collapse (see for instance \citet{valageas09} for analytical results in the rare events limit). It can therefore guide us toward a better understanding of the formation and evolution of dark matter halos as detected in N-body simulations.

\section{N-Body simulations}
\label{nbody}
\subsection{Simulation sets}
The numerical set up is the same as the one presented in
\citet{alimi09} and
we refer the interested reader to the previous article for a detailed description
of the numerical codes used in this series of papers. 
The N-body simulations are performed using the RAMSES-AMR
(Adaptative Mesh Refinement) code \citep{teyssier02,rasera06} based on
a multigrid Poisson solver. The matter power spectra are computed
using the CAMB code \citep{lewis00}, then Gaussian initial conditions 
using the Zel'dovich approximation are
generated with MPGRAFIC (\citet{prunet08}). For the toy-models we use the LCDM-W5 power spectrum (see previous sections). For each model the various cosmological variables ($a(t), H(t)$,
etc...) are pre-computed with an independent code and stored into
tables, these are subsequently interpolated to input the various time dependent
quantities into appropriately modified version of MPGRAFIC and RAMSES.
We have used the same phase for the initial conditions of the various simulations,
more specifically the white noise from the Horizon Project. We have 
performed a total of 15 simulations at the ``Centre de Calcul
Recherche et Technologie'' (CCRT\footnote{www-ccrt.cea.fr}). Our
longest simulation run for 350 hours of elapsed time on 64 cores. 
In Table~\ref{simu_set} we list the characteristics of the 15 cosmological simulations covering 6 cosmological models at various scales. 
\begin{table} 
\begin{center}
\begin{tabular}{cccc}
\hline \hline
Model&162~h$^{-1}$Mpc&648~h$^{-1}$Mpc&1296~h$^{-1}$Mpc\\
\hline\hline
\textbf{$\Lambda$CDM-W5}&&&\\
$z_i$&$93$&$56$&$41$\\
$m_p$~(h$^{-1}$M$_\odot$)&$2.28\times 10^9$&$1.46\times10^{11}$&$1.17\times10^{12}$\\
$\Delta$x~(h$^{-1}$kpc)&$2.47$&$19.78$&$39.55$\\
$\ell_{\textrm{max}}$&$7$&$6$&$6$\\
\hline\hline
\textbf{L-RPCDM}&&&\\
$z_i$&-&$136$&$97$\\
$m_p$~(h$^{-1}$M$_\odot$)&-&$1.46\times10^{11}$&$1.17\times10^{12}$\\
$\Delta$x~(h$^{-1}$kpc)&-&$9.89$&$39.55$\\
$\ell_{\textrm{max}}$&-&$7$&$6$\\
\hline\hline
\textbf{L-$\Lambda$CDM}&&&\\
$z_i$&-&$71$&$52$\\
$m_p$~(h$^{-1}$M$_\odot$)&-&$5.61\times10^{10}$&$4.50\times10^{11}$\\
$\Delta$x~(h$^{-1}$kpc)&-&$9.89$&$39.55$\\
$\ell_{\textrm{max}}$&-&$7$&$6$\\
\hline\hline
\textbf{SCDM*}&&&\\
$z_i$&-&$42$&$30$\\
$m_p$~(h$^{-1}$M$_\odot$)&-&$5.61\times10^{11}$&$4.50\times10^{12}$\\
$\Delta$x~(h$^{-1}$kpc)&-&$19.78$&$79.10$\\
$\ell_{\textrm{max}}$&-&$6$&$5$\\
\hline\hline
\textbf{$\Lambda$CDM-W3}&&&\\
$z_i$&84&$52$&$38$\\
$m_p$~(h$^{-1}$M$_\odot$)&$2.11\times 10^9$&$1.35\times10^{11}$&$1.08\times10^{12}$\\
$\Delta$x~(h$^{-1}$kpc)&2.47&$79.10$&$158.20$\\
$\ell_{\textrm{max}}$&7&$4$&$4$\\
\hline\hline
\textbf{$\Lambda$CDM-W1}&&&\\
$z_i$&110&$65$&$47$\\
$m_p$~(h$^{-1}$M$_\odot$)&$2.55\times 10^9$&$1.63\times10^{11}$&$1.31\times10^{12}$\\
$\Delta$x~(h$^{-1}$kpc)&2.47&$19.78$&$39.55$\\
$\ell_{\textrm{max}}$&7&$6$&$6$\\
\hline\hline
\end{tabular}
\caption{\label{simu_set} Parameters of the 15 N-body simulations
  for the various cosmological models: $z_i$ is
  the initial redshift, $m_p$ is the mass of the particle, $\Delta$x the
  comoving resolution and $\ell_{\textrm{max}}$ is the maximum refinement
  level. Each simulation has $512^3$ particles, with a $512^3$ grid on the coarse
  level and evolved down to $z=0$. All the simulations share the same realization
  of the initial conditions (namely the Horizon Project white noise), and
  start at the same level of rms density fluctuation ($\simeq 0.05$) at the scale of the
  resolution of the coarse grid. Our refinement strategy consist in refining
  when the number of particles in one cell is greater than 8.}  
\end{center}
\end{table}

Since we expect deviations from universality to manifest in the
high-mass tail of the mass function, our box lengths of interest are
$648~{\rm h^{-1}Mpc}$ and $1296~{\rm h^{-1}Mpc}$. In the case of the 
WMAP $\Lambda$CDM cosmologies we have also performed three simulations with 
a box length of $162{\rm h^{-1}Mpc}$ for comparison with the mass
functions at intermediate mass ranges which have been 
estimated in previous studies.
\begin{figure} 
\begin{tabular}{cc}
\includegraphics[width=0.48\hsize]{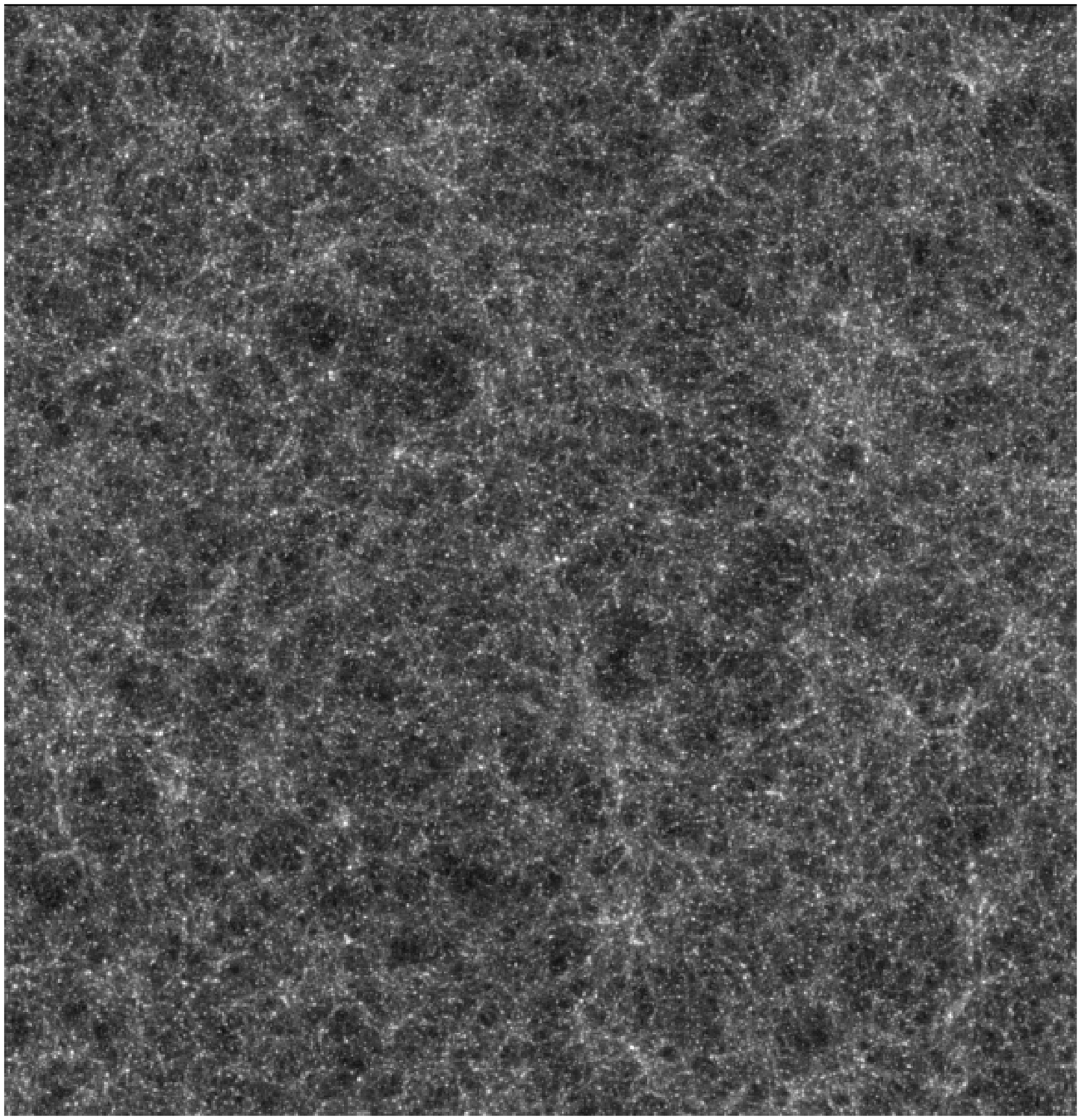}&\includegraphics[width=0.48\hsize]{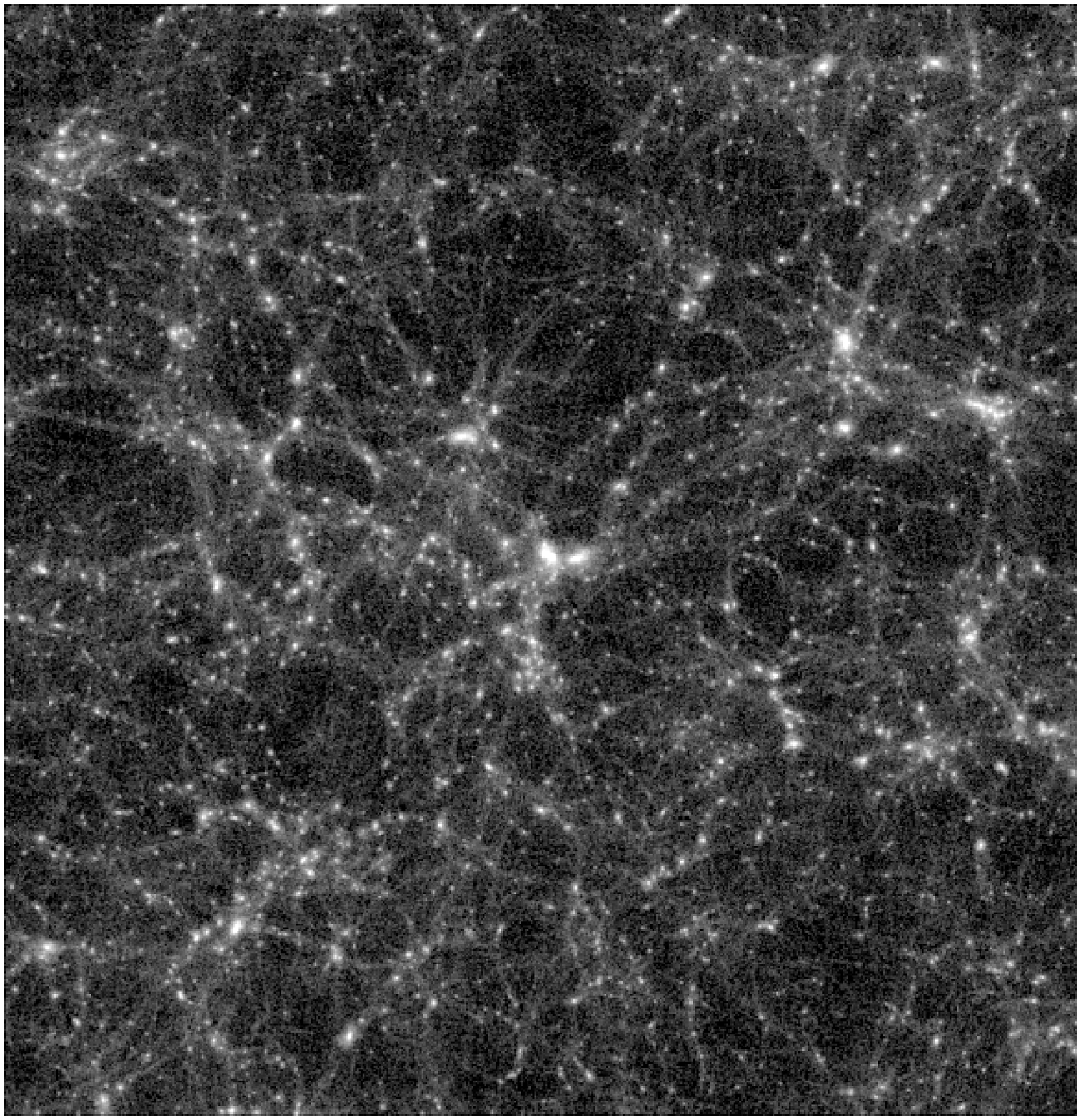}\\
\includegraphics[width=0.48\hsize]{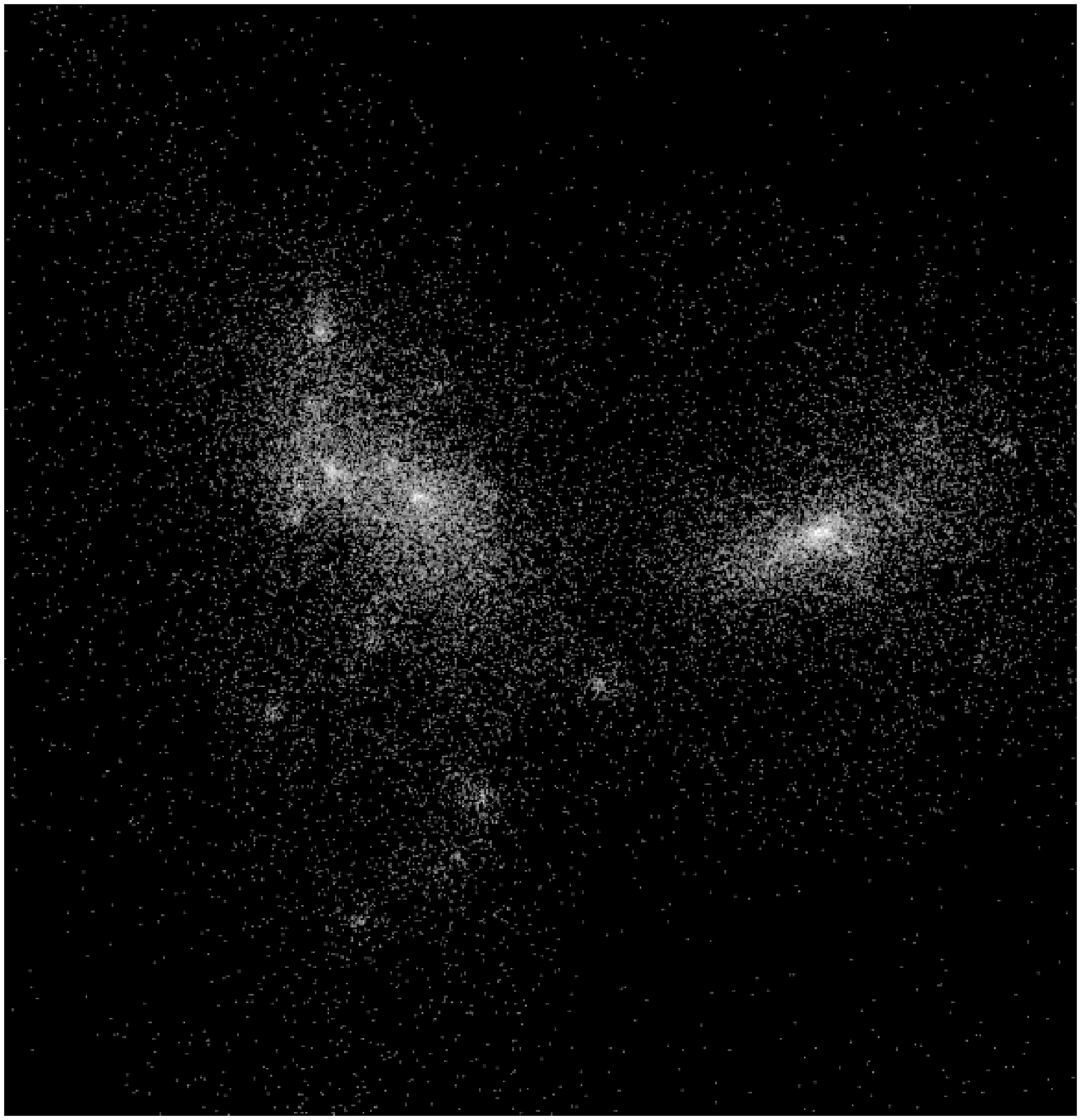}&\includegraphics[width=0.48\hsize]{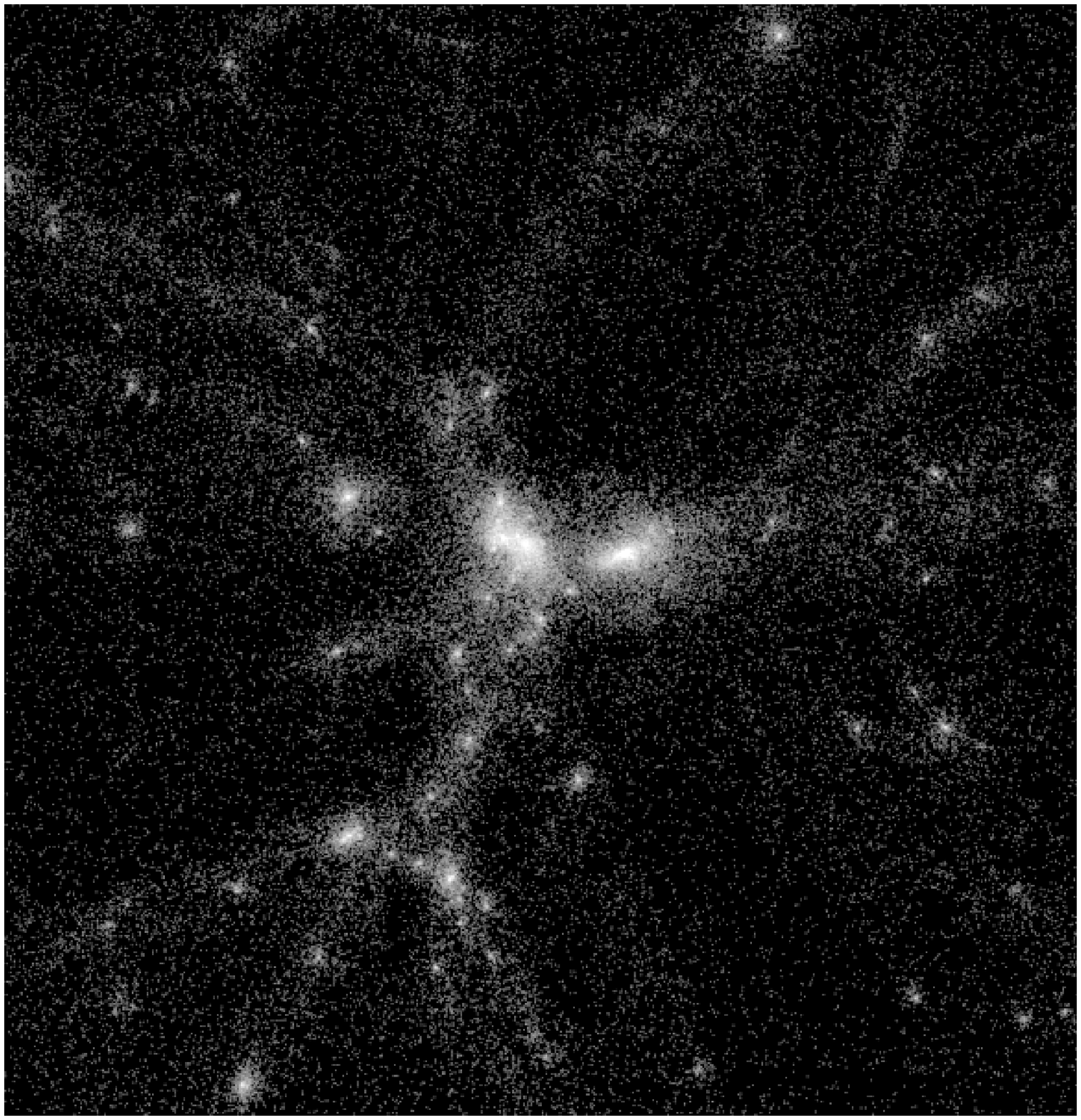}
\end{tabular}
\caption{\label{view_32BW} 2-dimensional projection of the dark matter density at z=0 from the
  $\Lambda$CDM-W5 simulation in the 648~h$^{-1}$Mpc box. The four
  images represent a zooming sequence (clockwise from top to bottom)
  of a factor four for each image from the full box size (upper
  left) to a 10~h$^{-1}$Mpc scale (bottom left).}
\end{figure}
In Fig.~\ref{view_32BW}, we illustrate the dynamics of the RAMSES AMR
code with a zooming sequence of images of the dark matter particle distribution
in the $\Lambda$CDM-W5 at $z=0$ from cosmological to galaxy group scale 
(clockwise from top to bottom $648~{\rm h^{-1}Mpc}$, $162{\rm
  h^{-1}Mpc}$, $40{\rm h^{-1}Mpc}$ and  $10{\rm h^{-1}Mpc}$). Because of the
large simulated volume and the high spatial and mass resolutions, 
these simulations provide a robust statistics of well resolved halos.


\subsection{Halo finder and mass function}
\label{subsec_halodef}
The halos in the simulation boxes are identified using the
Friend-of-Friend halo finder (FoF) \citep{davis85}.
This algorithm detects halos as group of particles characterized by
an intra-particle distance smaller than a given linking length
parameter $b$. The algorithm runs over
a list of particle coordinates in the box, firstly
it regroups all those which are within distance $b$ of an initial
particle. Then it moves to the next particle 
of the group and reiterates the same
selection until no new neighbours are found. At this point the
algorithm moves onto the next untagged
particle repeating the above procedure to detect a new group (halo).

Halos can also be detected using the Spherical Overdensity
algorithm \citep[SO,][]{lacey94}, which provides halo mass
estimations similar to that performed observationally, moreover
density profiles are directly given by the algorithm. However
the halo detection is restricted to spherical geometry, 
halos can overlap, and the last particles of the list could 
merge into some unphysical halos.
In contrast under the FoF algorithm all particles belong
to halos as in the case of the halo model, and it is applicable to
non-spherical particle groups as well, though it has tendency to overlink
bridged-halos. Therefore FoF and SO are complementary algorithms.
Here we use FoF since by relaxing the assumption 
on spherical symmetry this allows for a better physical study of the 
collapsed structures in the simulations (as we shall see hereafter). 

As can be seen in Fig.~\ref{view_32BW}, halos are much more complicated than spherical
isolated objects with a clearly defined radius. 
Halos in numerical simulations are non-spherical, interacting with
their environment, contain lots of subhalos and, moreover the density
can vary rapidly. This makes the definition of their boundary
somewhat arbitrary. 
The SO halo definition is based on the enclosed overdensity $\Delta$ in a given spherical region. This implies that it is rather straightforward to run a SO halo finder with threshold given by $\Delta_{vir}$ from the spherical collapse. Similarly, it is possible to run a
FoF halo finder with a linking length parameter $b=b_{vir}$, where
$b_{vir}$ corresponds to the value of a spherical overdensity $\Delta_{vir}$, provided that a relation between the two is assumed. 
A practical conversion formula is given by $\Delta/178\approx(0.2/b)^3$ (\citet{cole96}), nevertheless such a conversion is very approximative. A thorough comparison between the FoF and SO algorithms and the accuracy of conversions has been performed in \citet{lukic09} \citep[see also][]{audit98}. Furthermore since $b_{vir}$ depends on an approximate treatment of the non-linear collapse, its use as a linking length parameter, introduces built-in assumptions on the mass function obtained with FoF($b_{vir}$), which may hide the non-linear contributions associated with the halo formation and virialization process. For these reasons we start by considering the FoF
halo finder with a
constant $b$, which we set to $b=0.2$. As we will show, our conclusion about universality
will be independent of the exact value of $b$, remaining valid
in the range $0.1$ to $0.3$ (corresponding to $\Delta_{vir}$ varying roughly
from $1424$ to $53$).

One last point concerns the numerical definition of the halo mass
function, which reads as
\begin{eqnarray}
  \frac{dn(M)}{d\textrm{log}(M)}=\frac{1}{L^3}\frac{N}{\Delta \textrm{log}(M)},
\end{eqnarray}
where N is the number of halos in a logarithmic mass bin between 
$\textrm{log}(M)-\Delta \textrm{log}(M)/2$ and $\textrm{log}(M)+\Delta
\textrm{log}(M)/2$, with $\Delta \textrm{log}(M)$ the bin width and $L$ the comoving box length.
Our default choice of the bin width is very conservative, $\Delta
M/M\simeq0.2$. Moreover when measuring the mass function we focus
on mass ranges which corresponds to halos containing at least $350$ particles and Poisson shot
noise below the $10\%$ level. As we will discuss in the next
Section, such conservative assumptions are necessary to limit the
effect of mass resolution, bin size and Poisson noise, which can be
important source of systematic errors in the high-mass tail.

\section{Numerical Consistency Checks}\label{subsec_numerics}

In this Section we present a detailed study of various source of
numerical systematic effects. As result of this analysis we are able to 
control numerical errors within few percent level, thus validating 
our final results at the same percentage level.

\subsection{Initial conditions}
We have performed a series of tests on the initial conditions
generated with MPGRAFIC. Firstly, we have checked that the power
spectrum of the initial particle distribution is consistent within a
few thousandth with the input spectrum from CAMB
near the Nyquist frequency (at this high frequencies the Poisson noise is minimal). 
In addition, we have compared the MPGRAFIC power
spectrum with that extracted from the GRAFIC code for the same white
noise. After turning off the Hanning filter subroutine in GRAFIC we have 
found good agreement even at low $k$ \citep{bertschinger95,bertschinger01}.
The Hanning filter suppress the high $k$ modes, hence potentially causing
unphysical numerical artifacts on the generated particle distribution 
even in the linear regime. Therefore previous studies that have inadvertently used the Hanning filter may have underestimated the mass function, which is known to be sensitive to the gravitational dynamics on all scales. In our case, the use of the Hanning filter, led to a roughly $10\%$ suppression of the mass function over the full range of masses tested in our simulations ($10^{14}-10^{15}{\rm h^{-1}M_\odot}$). Therefore, we preferred not to use this filter.

A further test concerns the choice of the initial redshift of the
simulations. This can be responsible for spurious effects due to transients
from initial conditions when using the Zel'dovich approximation
\citep{reed03,reed07,crocce06,lukic07}. Several works 
(\citet{crocce06}, \citet{tinker08} and \citet{crocce09}) suspect that a very low initial
redshift combined with the Zel'dovich approximation might be
responsible for the discrepancies between their estimated mass
functions and those inferred by \citet{jenkins01} and \citet{warren06}.
To be as conservative as possible we have chosen higher initial redshifts $z_i$ 
compared to previous studies \citep{sheth01,jenkins01,warren06}. This
is realized by imposing that the standard deviation of the density 
fluctuations smoothed at the scale of the coarse grid
$\Delta^{coarse}_x$ is $\sigma (\Delta_x^{coarse})=0.05$. With this
choice our $z_i$ for the different cosmologies and simulations boxes 
vary in the range $30-110$, which are much higher then for instance
the initial redshifts considered in \citet{warren06} which range from
$24$ to $34$ depending on the box lengths, and for which \citet{crocce06} have estimated 
the suppression of the high-mass tail of the mass function to be of order $10-15\%$.


\subsection{N-body code}
We have run a series of tests to check the accuracy of the RAMSES code
as well as the modifications that have been implemented to account for the
quintessence scenario. Firstly, we have verified that for the various
models the ratio of the measured power spectrum to the linear
prediction from CAMB is constant at the percent level on 
the large linear scales (k$\simeq$0.01-0.1~h/Mpc) of the simulations at all redshifts \citep[see][]{alimi09}.
Furthermore, we have checked that by reducing the integration time steps
the results remain unchanged, thus validating our implementation of
quintessence models in RAMSES.
Secondly, we have compared the mass function from the RAMSES
simulation of the $\Lambda$CDM-W3 cosmology with box length $648~{\rm h^{-1}Mpc}$
and $256^3$ particles against a simulation with identical
characteristics and same seed for the initial conditions obtained
using the GADGET code \citep{springel01,springel05}. We find
that the two mass functions are in very good agreement (few percent
level) over the range of masses corresponding
to halos with more than $350$ particles and less than $10\%$ Poisson
noise. Using the \citet{warren06} 
correction does not improve noticeably the inferred mass function below
350 particles. In principle the halo mass correction by \citet{warren06}
depend on code and/or cosmology. 
Another point concerns the results of \citet{heitmann05,heitmann08}, the authors have computed the mass function with various codes using the same set of initial conditions. They found a 
scatter of order 10\% in the high mass range, the HOT code 
used in \citet{warren06} seems to give lower mass functions than other 
codes such as GADGET by a factor of about ten percent at $\simeq 10^{15}$~h$^{-1}$M$_\odot$.

\begin{figure*} 
\begin{tabular}{cc}
\includegraphics[width=0.48\hsize]{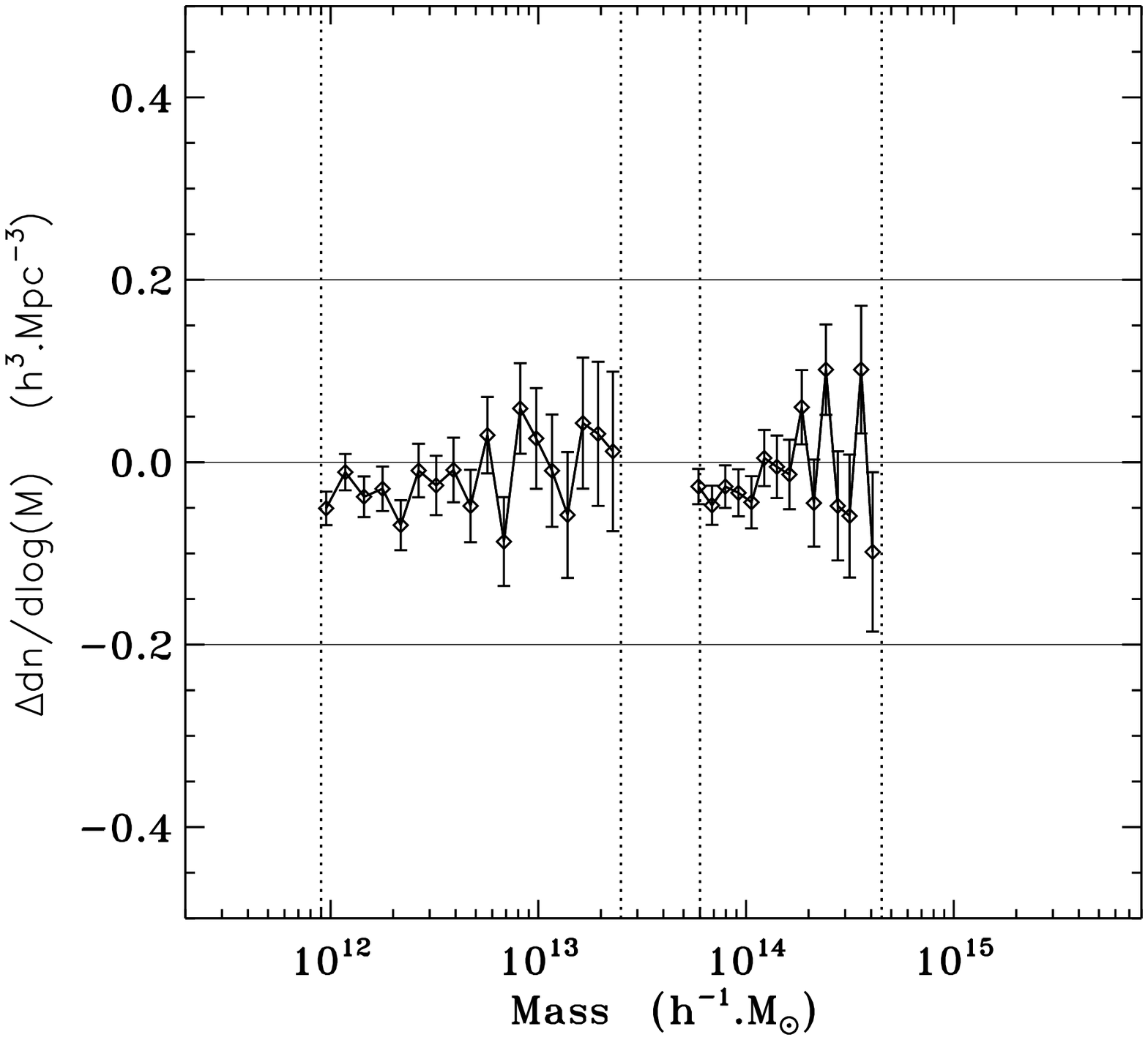}&
\includegraphics[width=0.48\hsize]{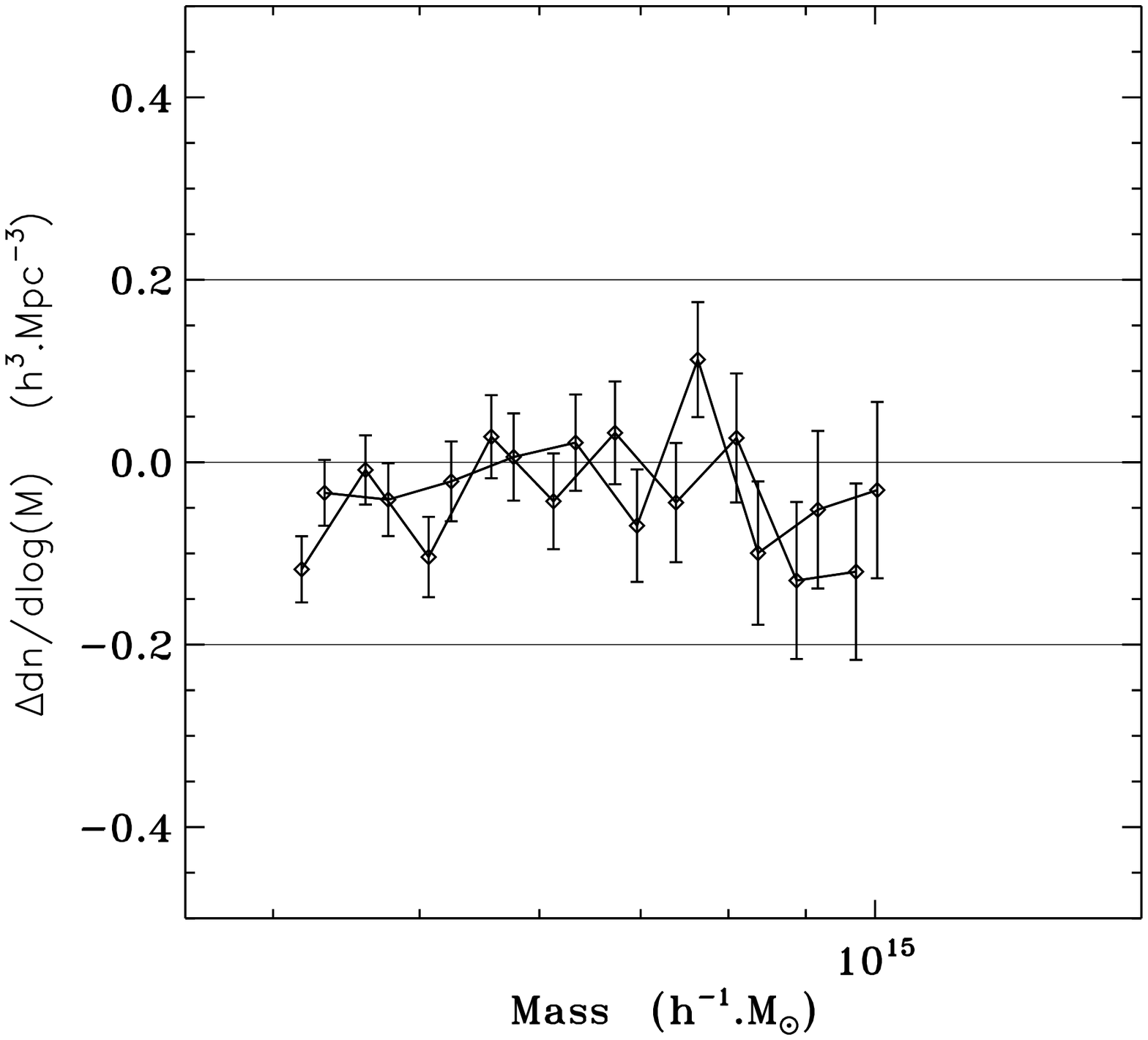}
\end{tabular}
\caption{\label{coherence_phase} Left: residual between the measured mass function and our fitting formula for two $\Lambda$CDM-W5 simulations with box lengths 162~h$^{-1}$Mpc and 648~h$^{-1}$Mpc at z=0. The number of particles is 512$^3$, the minimal number of particles per halo is 350 and the Poisson noise is less than $10$\% (vertical lines). This illustrates the coherence between the simulations. Right: As in the previous figure for two simulations with different realisations of the initial conditions. The box length is 1296~h$^{-1}$Mpc and the number of particles 512$^3$.}
\end{figure*}

\subsection{Particles, box length, mass and force resolution}
Another set of tests concerns the sensitivity of the results to various
simulation characteristics. First we have checked the stability of
the results when varying the number of particles from $256^3$ up to
$1024^3$, while keeping the same phase of the initial
conditions\footnote{The results of these simulations with 1 billion
  particles are part of the Dark Energy Universe Simulation Series (DEUSS) and will be presented in an
  upcoming paper.}. Assuming 350 particles as a minimum for the halo mass, we
find that the mass functions are consistent to better than $5\%$ at
all masses. Secondly, we have evaluated the influence of varying
the maximum level of refinement in RAMSES from $6$ to $4$. This parameter
changes the spatial resolution of the simulations, however we found no
effect on the mass functions. Besides in our simulations the
refinement level evolves freely, and never reaches the maximum allowed level.
   
Another important consistency check concerns the coherence
of the mass functions as measured from simulations with different
box lengths. Discrepancies may indicate effects due to either
mass resolution or finite volume. With our conservative choice of a
minimum of $350$ particles per halo and $10\%$ level of Poisson noise,
the coherence in the mass range of interest is better than
$5\%$, as it can be seen in the left panel of
Fig.~\ref{coherence_phase}, where  
we plot the difference between the mass function measured in
$\Lambda$CDM-W5 simulations at $z=0$ with $512^3$ particles and box lengths of 
$648~{\rm h^{-1}Mpc}$ and $1296~{\rm h^{-1}Mpc}$ respectively, and the
Sheth-Tormen parametrization, Eq.~(\ref{ST}), calibrated over the whole set of $\Lambda$CDM-W5
simulations. The vertical dotted lines identify the mass intervals
where the Poisson noise is below $10\%$ and halos have at least $350$
particles (the right interval corresponding to $1296~{\rm h^{-1}Mpc}$
simulation box and the left one to $648~{\rm h^{-1}Mpc}$). We can see
that the scatter of the points around the zero residual is well within
the $5\%$ level. We did not 
investigate finite volume effects, or cosmic variance specifically. 
In fact in our largest box ($1296~{\rm h^{-1}Mpc}$), 
the cosmic variance error on the largest halo masses is
likely to be dominated by Poisson shot noise as shown in
\citep{hu03,crocce09}, hence it should be within the $10\%$ level
set by our requirement for the Poisson noise.
We have also checked the effect of changing the phase of the initial
conditions. In the right panel of Fig.~\ref{coherence_phase} we plot
the residuals of the measured mass function in $\Lambda$CDM-W5
simulations at $z=0$ with box lengths of $1296~{\rm h^{-1}Mpc}$ and
$512^3$ particles with two different phases, with respect to the 
$\Lambda$CDM-W5 best fitting formula. In the mass range of interests the differences
between the residuals are well within the Poisson errorbars.

\begin{figure*} 
\begin{tabular}{ccc}
\includegraphics[width=0.3\hsize]{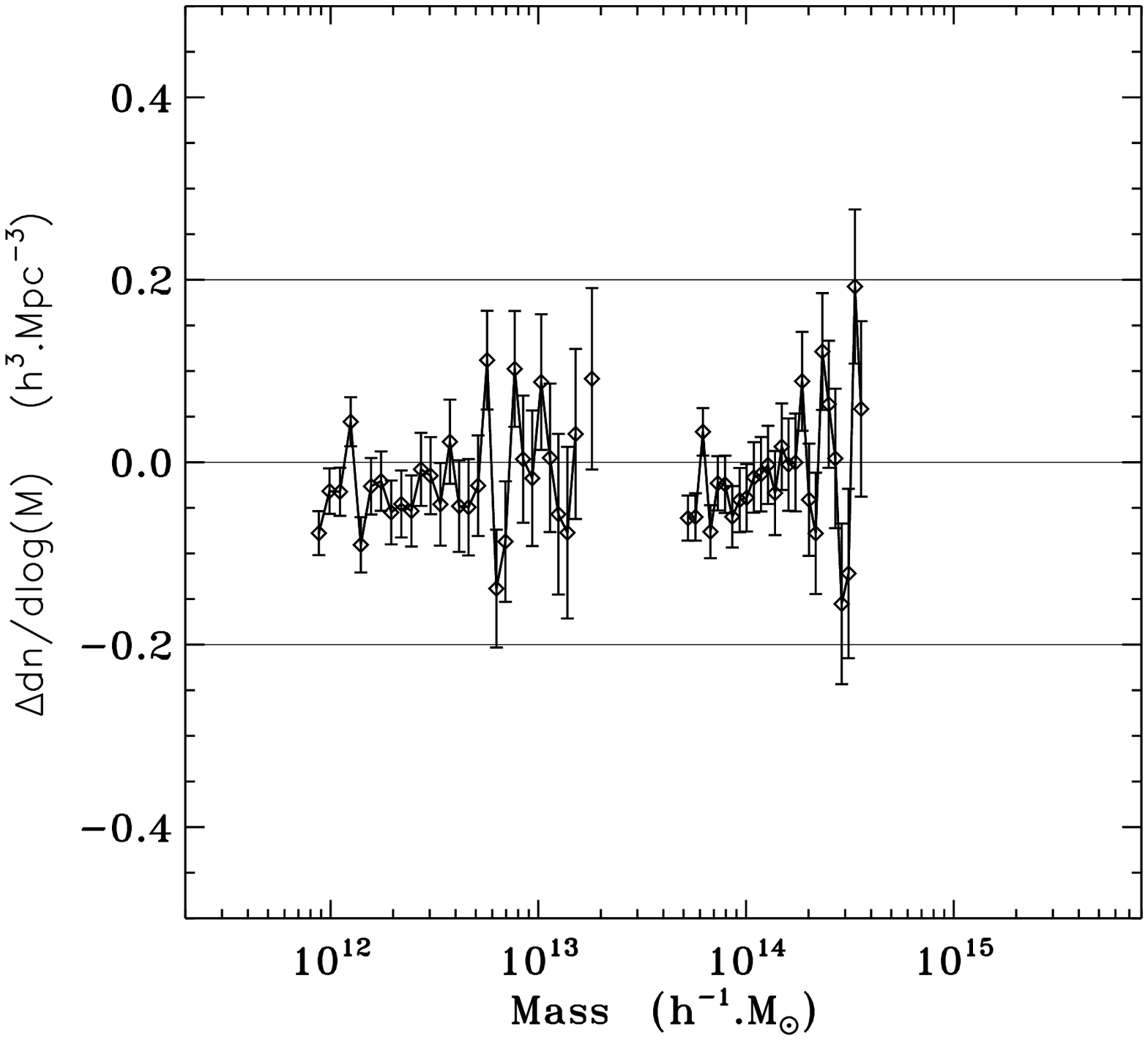} & \includegraphics[width=0.3\hsize]{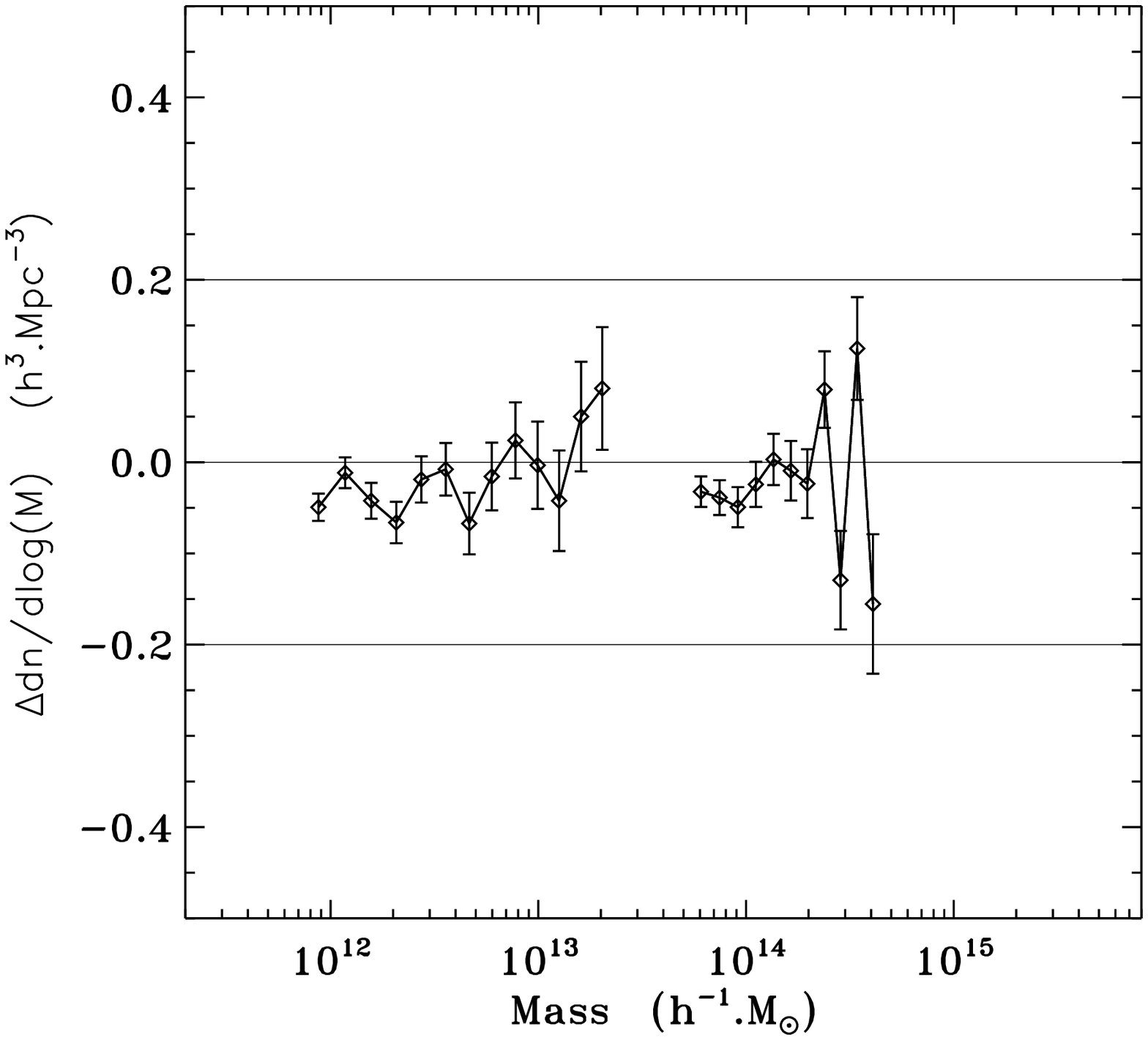} &  \includegraphics[width=0.3\hsize]{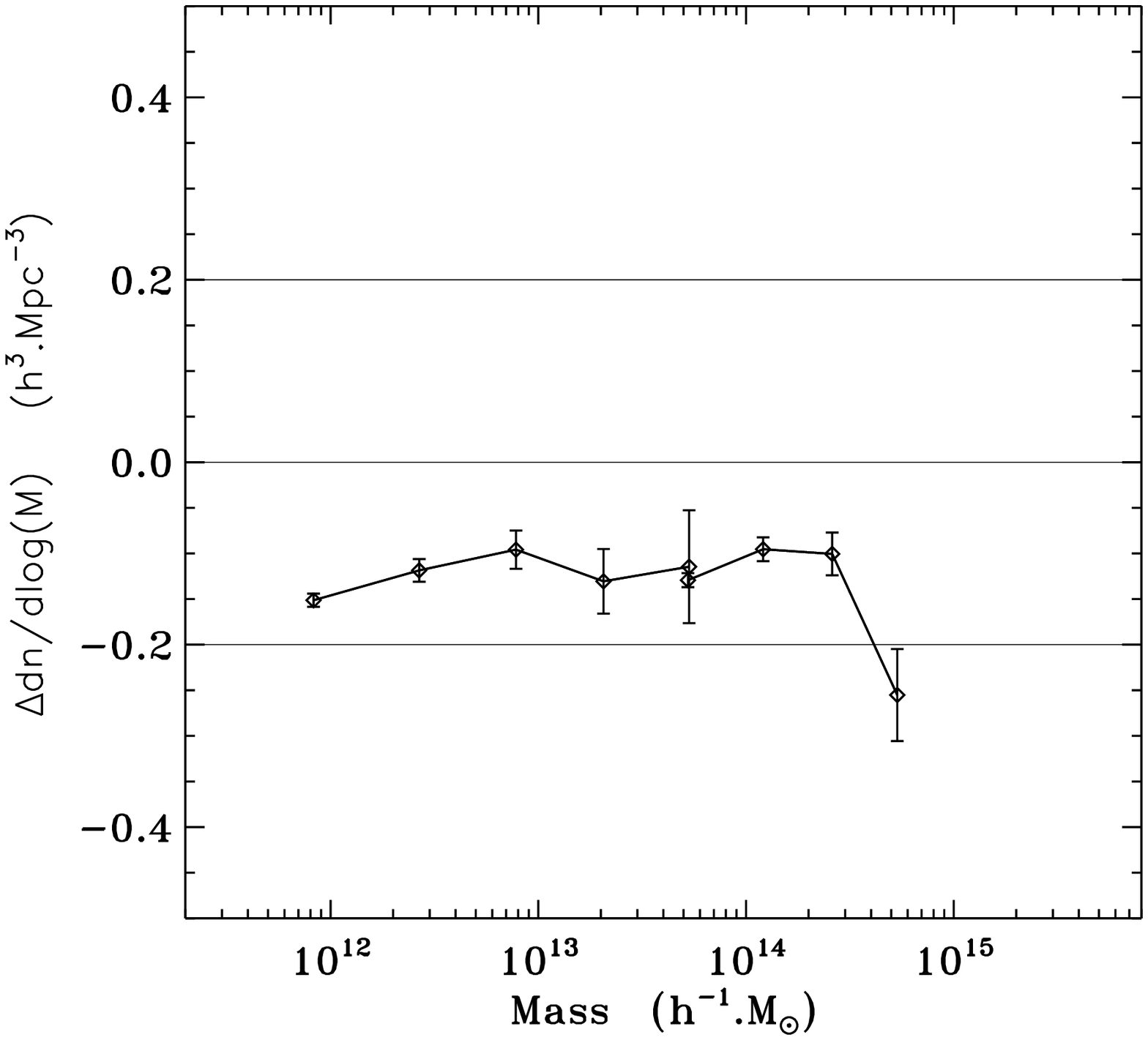}\\
\includegraphics[width=0.3\hsize]{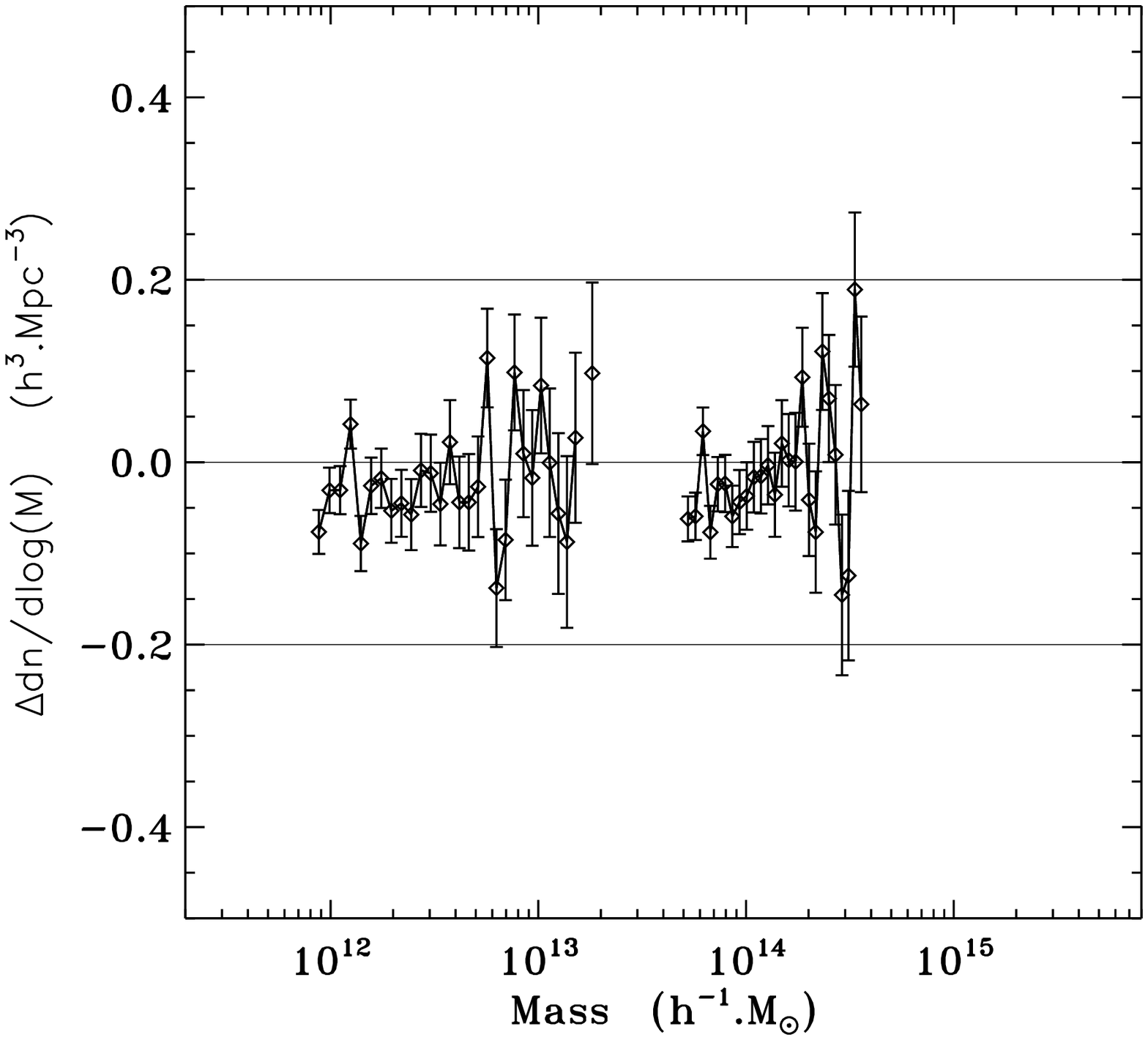} & \includegraphics[width=0.3\hsize]{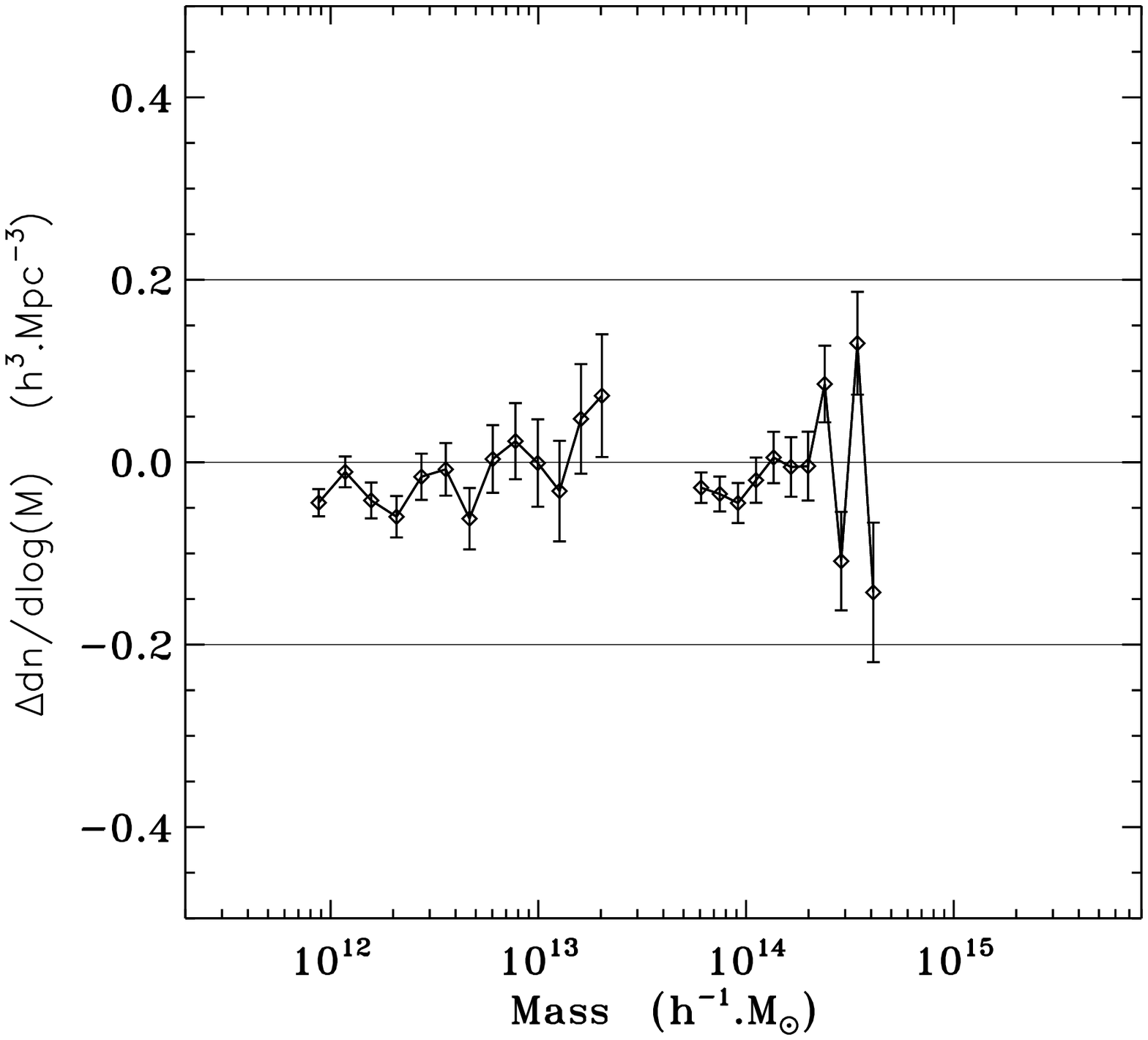} &  \includegraphics[width=0.3\hsize]{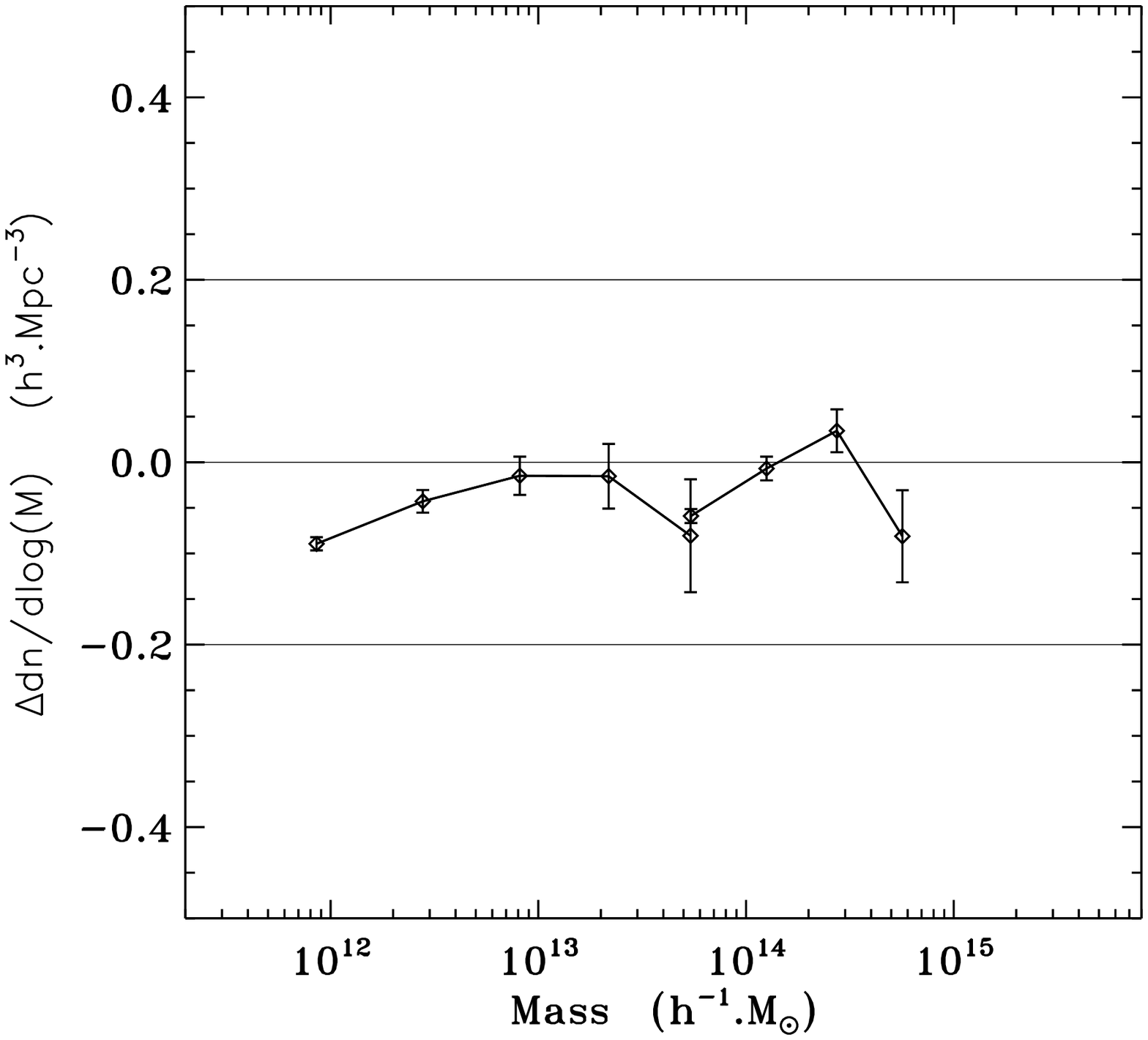}\\
\end{tabular}
\caption{\label{bins} Effect on the mass function of varying the bin width and binning strategy. The plots show the residuals between the measured mass functions and our fitting formula Eq.~(\ref{bestfit}) as a function of $\ln(\sigma^{-1})$ for two $\Lambda$CDM-W5 simulations with box lengths 648~h$^{-1}$Mpc and 1296~h$^{-1}$Mpc at z=0. The minimal number of particles per halo is 350 and the Poisson noise is less than $10$\%. {\bf Left:} bins width $\Delta M/M=0.1$. {\bf Middle:} $0.25$. {\bf Right:}  $1$. {\bf Top raw:} ``average-mass'' bins. {\bf Bottom raw:} ``Centered-mass'' bins. Using too large bins decrease the mass function, especially for the ``average-mass'' bins. To be as conservative as possible we use ``centered-mass'' bins of width $\Delta M/M \simeq 0.2$.}
\end{figure*}

\subsection{FoF code \& Mass Binning}
We have compared the results from our halo finder code to those obtained using ``Halomaker'' FoF
\citep{tweed09} and found no difference in the measured mass
function. We have also performed a study of the influence 
of varying the bin-width and the binning strategy on the mass
function. These tests are
particularly important, since some authors \citep[see
  e.g.][]{jenkins01} have used analytical corrections to remove 
effects caused by large mass-bins and smoothing functions. 

In Fig.~\ref{bins} we plot the residual mass function of 
$\Lambda$CDM-W5 simulations at $z=0$ with $512^3$ particles and box lengths of 
$648~{\rm h^{-1}Mpc}$ and $1296~{\rm h^{-1}Mpc}$ respectively with
respect to the Sheth-Tormen best fitting formula, for different bin-widths: 
$\Delta{M}/M=0.1$ (left panels), $\Delta{M}/M=0.25$ (central panels) and
$\Delta{M}/M=1$ (right panels). In the top panels we plot the case
with averaged-mass bins, i.e. the mass of each point is the average
mass of the halos in the bin, while in the bottom panels we plot the
case with centered-mass bins, i.e. the mass of each point is the
central value of the logarithmic mass bin. We can see that for both
binning strategies increasing the bin-width reduces the Poisson
noise and smooths the curves. However increasing the bin size
is particularly delicate in the high mass tail
especially for the averaged-mass binning. In fact 
we can see that a larger binning ($\Delta{M}/M=1$)
tends to lower the mass function intrinsic scatter. This is because as
the bin-size increases, each bin has a larger number of halos
thus leading to smaller Poisson errors on the other hand in the
high-mass tail the mass function drops steeply and the
use of averaged-mass bins tends to lower its value. As noticeable from Fig.~\ref{bins},
such an effect is much weaker in the case of centered-mass bins which we adopt hereafter,
and for the study of the universality of the mass function
we consider centered-mass bins with width $\Delta{M}/M=0.2$, a
compromise between the Poisson noise and the number of bins. 
It is worth noticing that \citet{jenkins01} use large bins and a gaussian smoothing (of
rms 0.08 in $\ln{(M)}$ corresponding to a width of 0.37 in our units) 
which is supposed to strongly raise the curve. An analytical
correcting factor is then applied to recover a proper estimate of the 
mass function. We find this procedure to be quite uncertain. 
Again as seen in Fig~\ref{bins} using large bins tends to lower the mass 
function unlike expectations, and such an analytical correction might cause an
underestimate in the high mass end at
the $10\%$ level. In such a case, it is safer to use smaller bins such that
deviations remain within $5\%$ (as in our case) and 
not rely on analytical corrections. 
In the light of these various
tests, our precision on the absolute FoF mass function is expected to
be of order $5-10\%$. However, we want to stress that the precision on
the relative mass function is even higher (typically less than $5\%$)
since for all our simulations we use the same phase for the
initial conditions, and
thus systematic errors cancel out. The Poisson error bars are consequently
not relevant for quantifying the error on the relative 
mass functions, since the scatter of the mass functions in two different
cosmologies is strongly correlated. Moreover from the above analysis we can see that
intrinsic fluctuations of the measured mass functions for a given
cosmology are much smaller than the Poisson error bars.

\section{Results}
\label{universality}
\subsection{WMAP-$\Lambda$CDM cosmologies and universality at $z=0$}
\label{UNIV_WMAP}
\begin{figure} 
\includegraphics[width=\hsize]{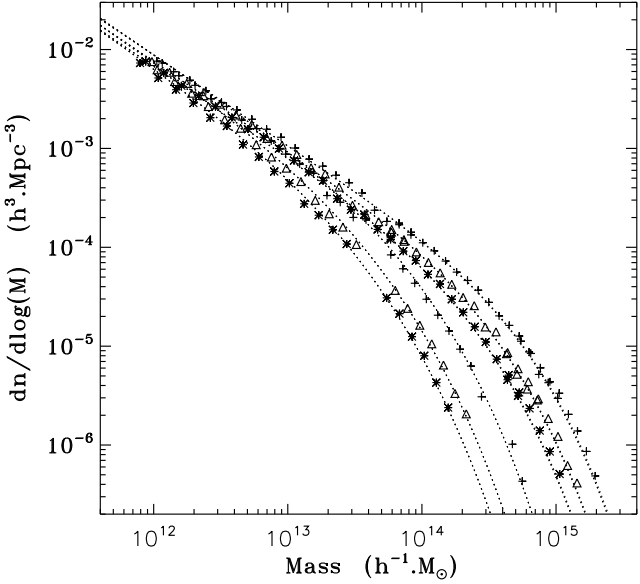}
\caption{\label{ndem} Mass functions $dn(M)/d\textrm{log}(M)$ for the $\Lambda$CDM-WMAP cosmologies at z=0 (upper curves), and z=1 (lower curves) from $162$~h$^{-1}$M$_{\odot}$, $648$~h$^{-1}$M$_{\odot}$ and $1296$~h$^{-1}$M$_{\odot}$ box lengths simulations. Halos are detected with FoF($b=0.2$). The dotted line is our best fit $\Lambda$CDM-W5 
Eq.~(\ref{bestfit}). Crosses correspond to $\Lambda$CDM-W1, stars for $\Lambda$CDM-W3 and triangles to $\Lambda$CDM-W5. 
The mass functions cover a range from Milky-Way size halos to cluster of galaxies.}
\end{figure}

In Fig.~\ref{ndem} we plot the mass function of 
the $\Lambda$CDM-WMAP cosmologies at $z=0$ and $1$ measured in $162{\rm h^{-1}Mpc}$, 
$648{\rm h^{-1}Mpc}$ and $1296~{\rm h^{-1}Mpc}$ simulation boxes
respectively. Notice the large mass range covered by the simulations, which spans
from Milk-Way to cluster size halos ($10^{12}-10^{15}{\rm h^{-1}M_\odot}$).
In order to test the universality of the mass function it is preferable to work with the function  
\begin{equation}
f(\sigma)=\frac{M}{\bar{\rho}_0}\frac{dn}{d\textrm{ln}(\sigma^{-1})},
\end{equation}
where $dn/d\textrm{ln}(\sigma^{-1})$ is the comoving number of halos
per unit of natural logarithm of $\sigma^{-1}$. As mentioned in
Section~\ref{theory}, if the mass function is universal the selection
function accounting for the non-linear collapse in the definition of $f$ 
is independent of cosmology (i.e. independent of the density threshold). In such a case the function $f$
for different cosmological simulations should be identical to
numerical precision. 

\begin{figure} 
\includegraphics[width=\hsize]{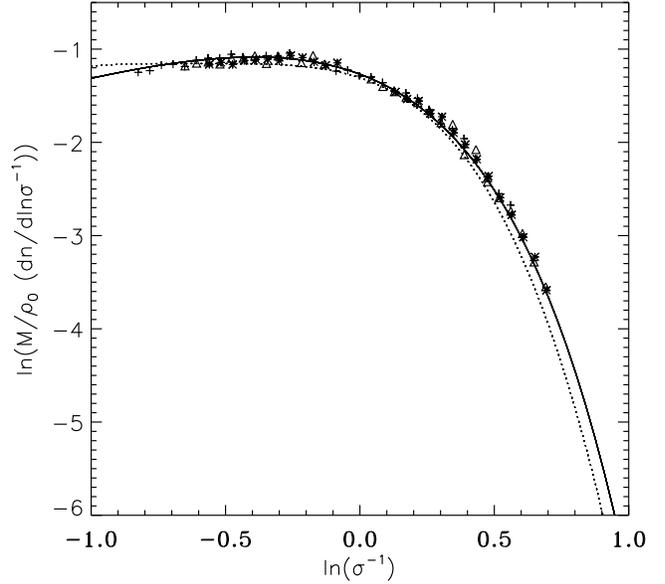}
\caption{\label{jenkinsplot} Mass functions for the $\Lambda$CDM-WMAP cosmologies
  at z=0, in the $f-\ln(\sigma^{-1})$ plane (as defined in the text) from the $162$~h$^{-1}$M$_{\odot}$, $648$~h$^{-1}$M$_{\odot}$ and $1296$~h$^{-1}$M$_{\odot}$ box length simulations. Halos are detected with FoF($b=0.2$). The continuous line is the $\Lambda$CDM-W5 fit described in this section (Eq.~\ref{bestfit}) and the dotted line is the \citet{jenkins01} fit. The conventions are the same than in Fig.~\ref{ndem}. Deviations from universality are below numerical precision. }
\end{figure}

In Fig.~\ref{jenkinsplot} we plot the function $f$ for the three
$\Lambda$CDM-WMAP models at $z=0$ in the range
$-0.8<\ln{(\sigma^{-1}})<0.7$. We can see that the data points
nicely overlap to numerical precision,
this is quite remarkable given the fact that these models
have different initial conditions with very different $\sigma_8$
values. This implies that changing the initial conditions does not break the
universality of the mass function in the $\Lambda$CDM cosmologies, which is
in agreement with the results of \citet{jenkins01,warren06}. Nonetheless
we find that the functional form of $f$ differs from standard fitting
formula inferred in previous analysis \citep{sheth01,jenkins01,warren06}. As an example
in Fig.~\ref{jenkinsplot} we plot against the simulation data-points the 
fitting formula Eq.~(\ref{JK}) given in \citep{jenkins01}. 
Large deviations with respect to this fit occurs in the high-mass end, 
instead we find a better fit by using
the functional form of the ST formula, with different parameter
values calibrated on the $\Lambda$CDM-W5 model simulations. In
particular by fixing $\delta_c$ to the spherical collapse model
prediction of the $\Lambda$CDM-W5 cosmology, $\tilde{\delta}_c=1.673$, our
best fitting function (using b=0.2) reads as:
\begin{equation}
f(\sigma)=\tilde{A}\left(\frac{2\tilde{a}}{\pi}\right)^{1/2}\frac{\tilde{\delta}_c}{\sigma}\left[1+\left(\frac{\tilde{\delta}_c}{\sigma\sqrt{\tilde{a}}}\right)^{-2\tilde{p}}\right]e^{-\tilde{\delta}_c^2
  \tilde{a}/(2\sigma^2)},\label{bestfit}
\end{equation}
with $\tilde{A}=0.348$, $\tilde{a}=0.695$ and $\tilde{p}=0.1$. 

\begin{figure} 
\includegraphics[width=\hsize]{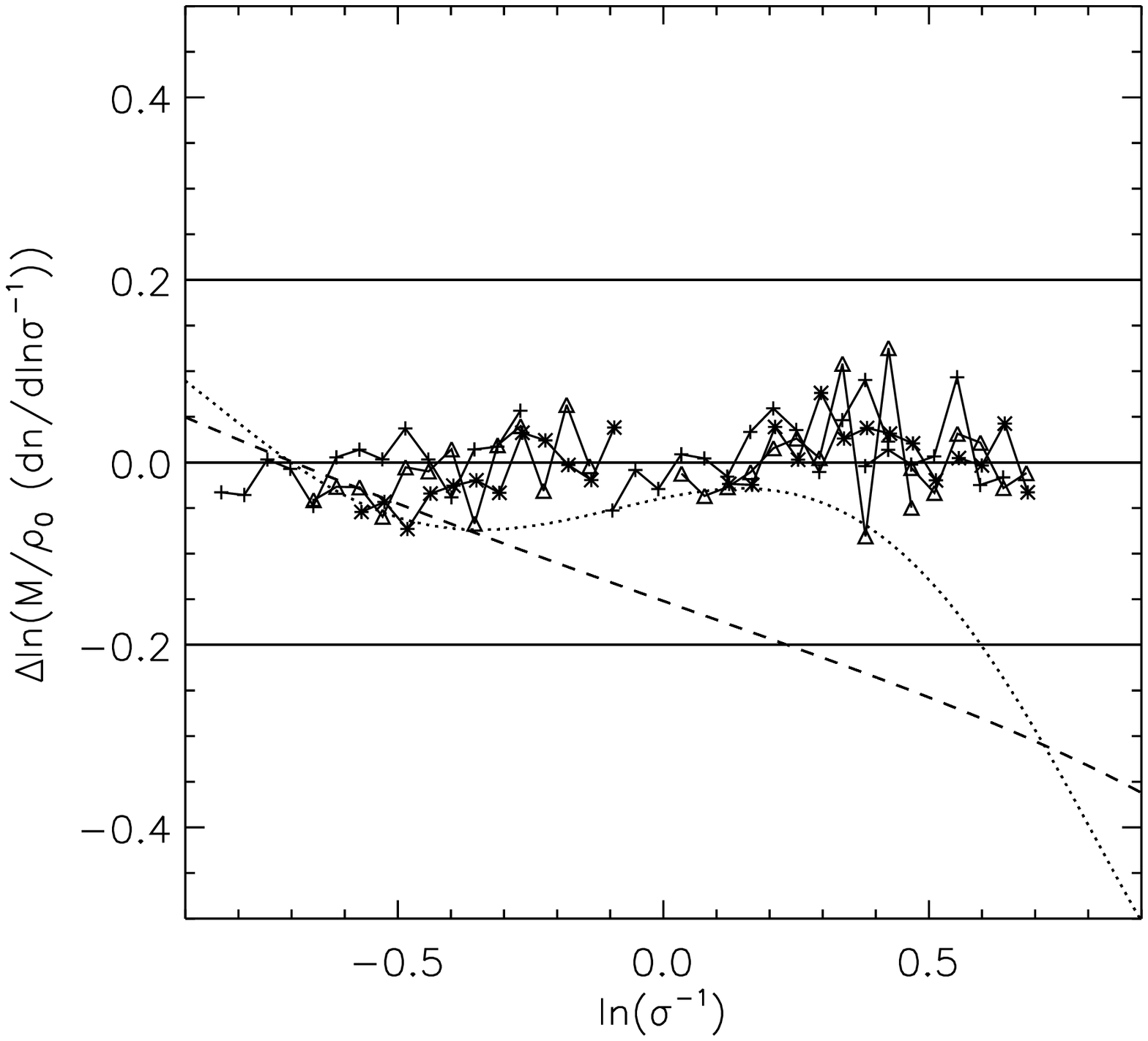}
\caption{\label{univ_jenk} The residual between the fitting formula Eq.~(\ref{bestfit}) and the measured FoF($b=0.2$) mass functions at z=0 for the ``WMAP $\Lambda$CDM cosmologies''. The conventions are the same than in Fig.~\ref{ndem}. Halos come from $162$~h$^{-1}$M$_{\odot}$, $648$~h$^{-1}$M$_{\odot}$ and $1296$~h$^{-1}$M$_{\odot}$ box lengths simulations. For reference the \citet{jenkins01} (dotted line) and the \citet{sheth99,sheth01} (here in dashed line) fitting formula are also plotted. We recover the universality up to numerical precision ($5-10\%$).}
\end{figure}

We plot in Fig.~\ref{univ_jenk} the residual of the measured function
$f$ against Eq.~(\ref{bestfit}), we can see that 
the $\Lambda$CDM-WMAP cosmologies have a universal mass function to
$5-10\%$ accuracy level. For comparison we also plot the original ST
formula Eq.~(\ref{ST}) with $\delta_c=1.686$ (dashed line), and the Jenkins et al. one Eq.~(\ref{JK}) (dotted
line). The former badly reproduce the measured mass functions at ${\rm
ln(\sigma^{-1})>-0.5}$, while
the latter  provide a good description only in the lower
mass range (${\rm
ln(\sigma^{-1})<0.4}$). In contrast
at higher masses there are deviations $>20\%$, thus confirming
previous results by \citet{tinker08, crocce09}.

The fact that in $\Lambda$CDM-WMAP
cosmologies the mass function at $z=0$ shows a universal behaviour is not surprising, since from
the considerations of Section~\ref{theory} these models despite having different
$\sigma_8$ values and small differences in the other parameters, they have
nearly identical expansion histories, linear growth evolution and spherical
collapse model parameters. The spherical collapse
model is only approximative we may expect small deviations from
universality to be present also in these models, however
it is likely that such deviations are within the numerical errors of our simulations.

The universality of the mass function for cosmologies characterized by
the same expansion history but different $\sigma_8$ values (and slightly different values of the other cosmological parameters) 
implies that the position of the mass function in the plane $f-\rm{ln}(\sigma^{-1})$ is independent of $\sigma_8$ 
(within the accuracy of our simulations). This is an important point to keep in mind, as we will show in the next Section
this will allow us to isolate cosmological dependent effects on the halo mass function, which are related to the record 
of the past expansion history, and also extend our conclusions on the limits of universality to models 
with different $\sigma_8$ values.

\subsection{Dark energy models and the limit of universality}
\label{limit}
\begin{figure*} 
\begin{center}
\begin{tabular}{cc}
\includegraphics[width=0.48\hsize]{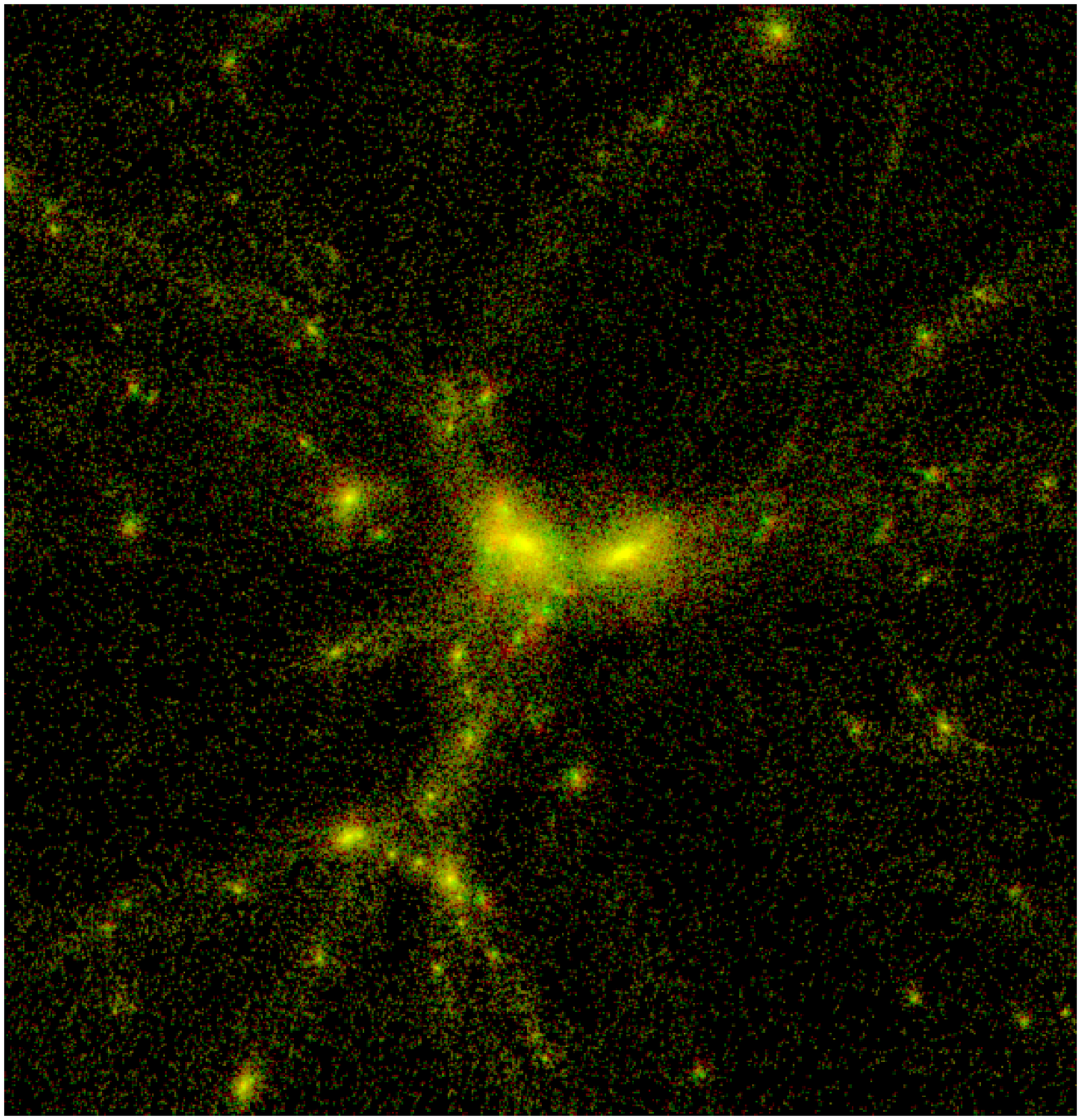}&
\includegraphics[width=0.48\hsize]{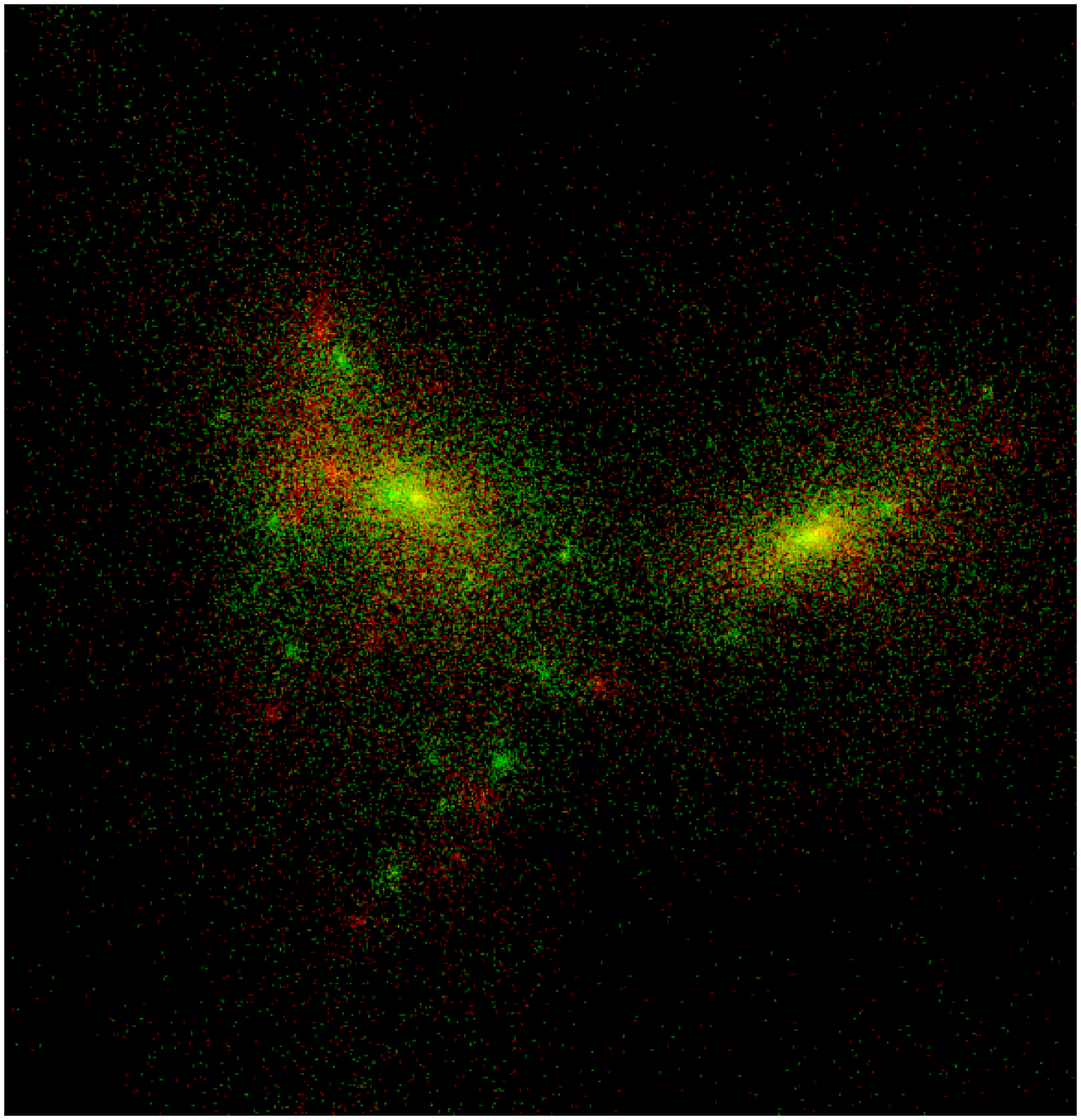}\\
\includegraphics[width=0.48\hsize]{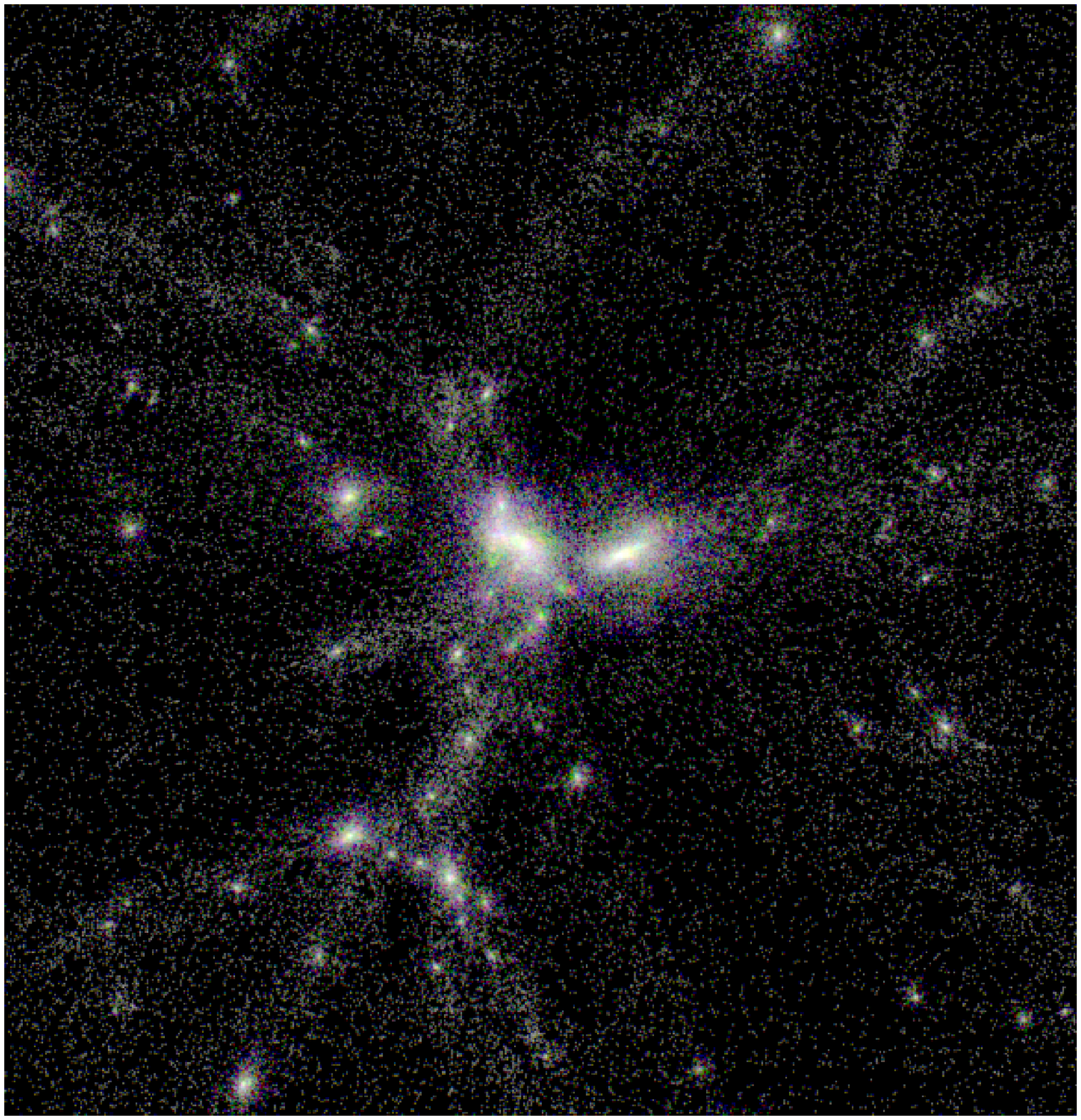}&
\includegraphics[width=0.48\hsize]{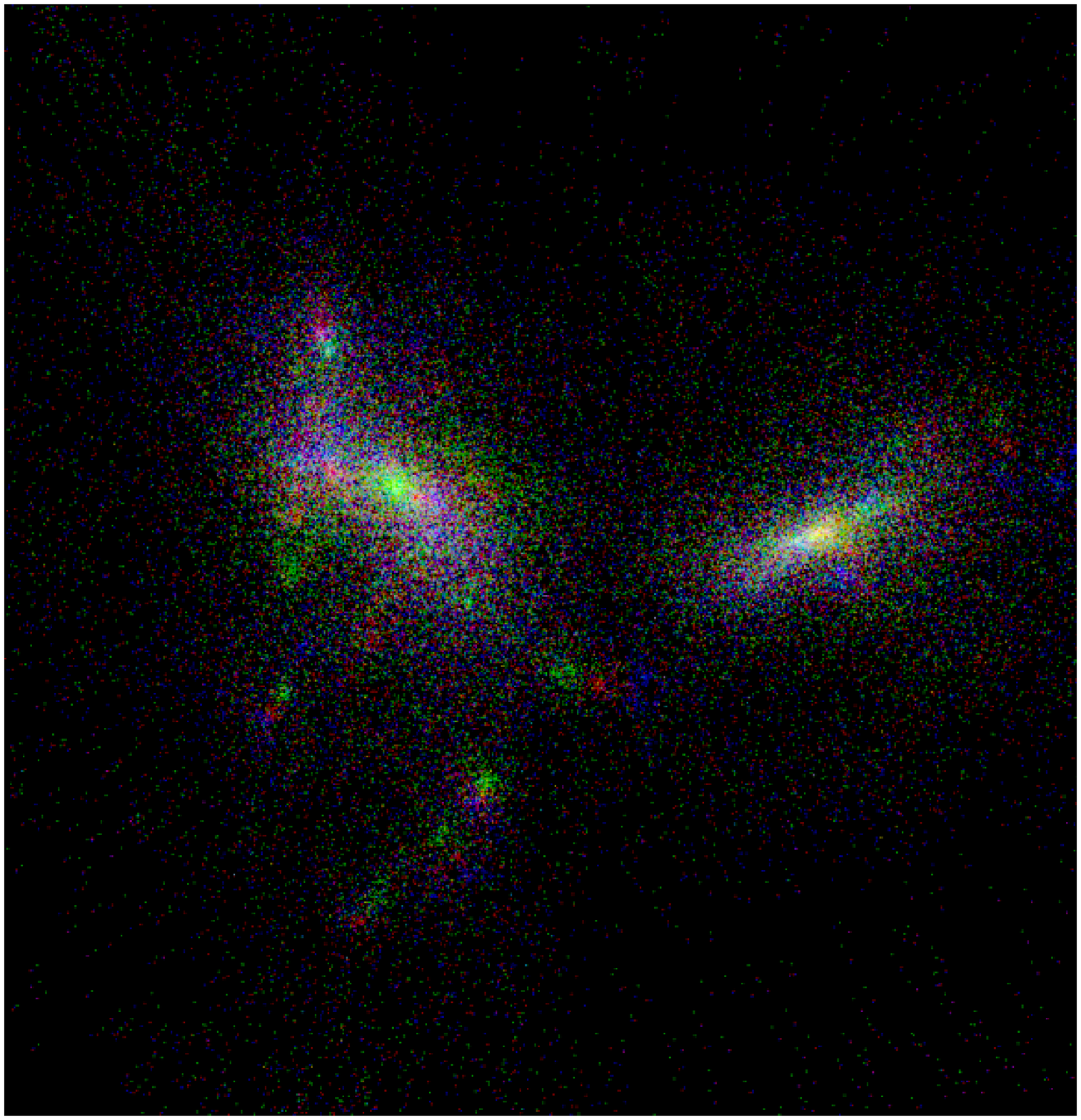}\\

\end{tabular}
\caption{\label{lin_vs_nl} Projected overdensity maps for the $\Lambda$CDM-W5 and L-RPCDM models (top) and $\Lambda$CDM-W5, L-$\Lambda$CDM and SCDM$^*$ 
(bottom) on scale $40$~h$^{-1}Mpc$ (left) and $10$~h$^{-1}Mpc$ (right) respectively at z=0. The views are centered onto two halos and extracted 
from simulations of box length $648$~h$^{-1}$Mpc with the same realisation of the initial conditions and the same $\sigma_8$. Each cosmology 
is represented with different color coding (but with the same intensity level). Top panels: $\Lambda$CDM (red) and L-RPCDM (green). 
Here the distinct green or red regions at small scales are non-linear imprints related to the different equation of state evolution between the two models. 
Bottom panels: $\Lambda$CDM (red), L-$\Lambda$CDM (green) and SCDM* (blue). Here the distinct colored regions on small scales are 
non-linear signature due to the different amount of dark energy of the models.}
\end{center}
\end{figure*}

We now focus on the toy-model simulations. In Fig.~\ref{lin_vs_nl}
we show the projected density maps from the $648~{\rm h^{-1}Mpc}$
simulation box at $z=0$ zoomed on $40{\rm h^{-1}Mpc}$ (left panels) and $10{\rm h^{-1}Mpc}$ 
(right panels) scales respectively. Since the simulations have the same initial phase, in order to
facilitate a visual comparison between different models we plot their density distribution in the same
image with a different color coding for each model. In the top panels are shown the density maps for the 
$\Lambda$CDM-W5 (red) and L-RPCDM (green), while in the lower panel in addition to the
$\Lambda$CDM-W5, we also show the L-$\Lambda$CDM (green) and SCDM$^*$ (blue). 
The differences in the top panels are indicative of the effects related to having a time evolving 
dark energy with $w(z)\ge-1$, while those in the bottom panels are associated to varying the amount
of dark energy density. We can see that in both cases there are no apparent differences in the particle
distribution on the $40{\rm h^{-1}Mpc}$ scale. In contrast differences are 
clearly manifest on the $10{\rm h^{-1}Mpc}$ scale, where the halo concentration seems 
to differ from one model to another, and the outer parts
of halos show different morphologies as well. 
In particular, we may notice that in the L-RPCDM case, since $\Omega_m$ and $P(k)$ are the same of the reference cosmology, differences in the dark matter distribution are unique signature of the non-linear process of structure formation.
In such a case, it is hard to believe that the
mass function should remain unaffected, that is to say universal in dark energy cosmologies.

\begin{figure} 
\includegraphics[width=\hsize]{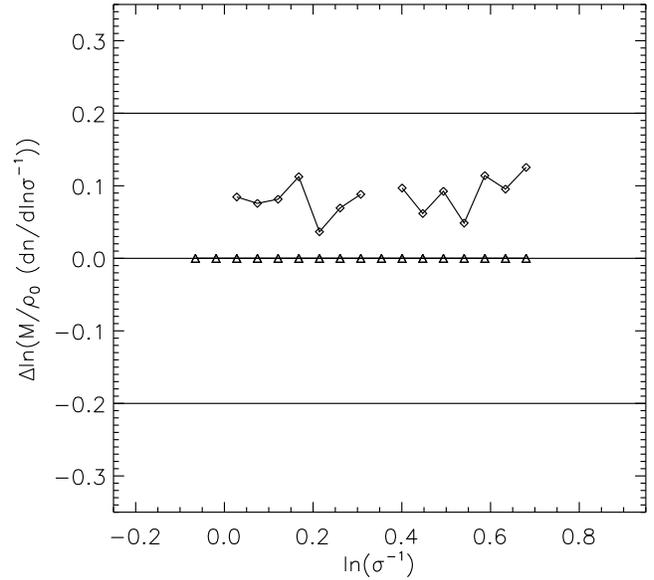}
\caption{\label{univ_break1} The residuals between the measured mass functions of L-RPCDM (diamonds) and the measured mass functions for $\Lambda$CDM-W5 (triangles) 
cosmology at z=0. Deviations from universality are clearly above numerical errors ($<5\%$) and correlated with linear growth history. 
Here the amplitude of the residuals is a signature of the different dark energy equation of state evolution in the L-RPCDM model. }
\end{figure}

\begin{figure} 
\includegraphics[width=\hsize]{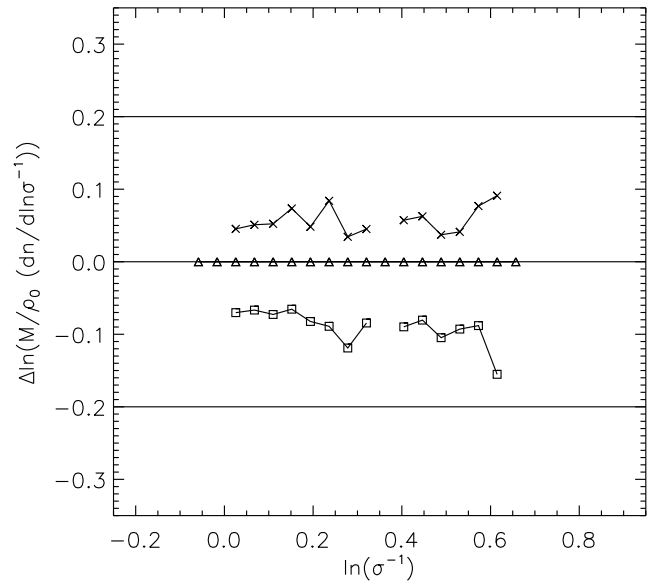}
\caption{\label{univ_break2} The residual between the measured mass functions of L-$\Lambda$CDM (crosses), SCDM$^*$ (squares) and that 
of $\Lambda$CDM-W5 (triangles) cosmology at z=0. Deviations from universality are clearly above numerical errors ($<5\%$) and 
correlated with linear growth history of the models. Here the amplitude of the deviation is related to the different amount of dark energy of the models.}
\end{figure}

For a more quantitative comparison, we plot in Fig.~\ref{univ_break1} the residual of the function 
$f(\sigma)$ measured in L-RPCDM with respect to that of the $\Lambda$CDM-W5. 
As previously mentioned we only plot the mass range
in which halos contain at least 350 particles and where the Poisson noise is within $10\%$, thus allowing us to control
numerical errors to better than about $5\%$. From this plot we can clearly see deviations from universality at the $10\%$ level,
hence well above numerical errors. Similarly in Fig.~\ref{univ_break2} we plot the case of L-$\Lambda$CDM and SCDM$^*$. 
The former lies about $5\%$ above the $\Lambda$CDM-W5 model, while the latter is roughly $10\%$ below, again these
are evidence of departure from universality. However such deviations
are not random, they are
correlated with the linear growth history of the corresponding
model. In fact comparing the evolution of the growth factor
of each model (see Fig.~\ref{phen}) with the corresponding amplitude of the mass function residual, 
we may notice that the greater the deviation in the growth rate history
and the larger is the deviation from universality at z=0.
This is not the case for the $\Lambda$CDM-WMAP cosmologies which share nearly the same linear growth rate 
history and therefore their mass functions are universal to numerical precision.
The physical origin of such deviations is quite intuitive. The cosmic structure formation is more efficient 
in the past when the linear growth rate is greater. This is indeed the case for our toy-models.
Even though by construction they share the same amount of clustering
at the linear level at $z=0$, lots of dense halos were formed earlier and survive at 
lower redshift since they are decoupled from the cosmic expansion on the larger scales.
The same effects are responsible for the imprints which have been shown to affect the non-linear 
matter power spectrum \citet{ma07, alimi09} and halo concentration \citet{dolag04,wechsler02}.
To our knowledge, this is the first time that physical effects leading to a non-universal behaviour of the mass function
have been unambiguously shown.

\section{Non-universality of the mass function and the non-linear collapse}
\label{insight}
In this section we will focus on the non-linear processes which are responsible for
the departure from a universal behaviour of the mass function.
As discussed in Section~\ref{theory}, deviations from universality occur if
and only if the non-linear collapse of dark matter is cosmology and
redshift dependent (or in other words if the relation between the linear and non-linear growth of
structures is cosmology and redshift dependent \citet{francis09b}). 
In the Press-Schechter framework this is parametrized by the dependence of the
mass function on the threshold density $\delta_S$. This suggests that 
we may gain a better insight into the origin of the non-universal
behaviour by explicitly accounting for the non-linear collapse model
in the analysis of the measured mass functions. 
Here we use the spherical collapse model which has been described in
Section~\ref{spherical}. Furthermore we will plot residuals with respect
to the mass function measured from the SCDM$^*$ simulations, rather than 
that of the reference $\Lambda$CDM-W5 cosmology. In fact theoretical arguments (\citet{efstathiou88,blanchard92,lacey94}) 
suggests that for scale-free initial conditions 
(with a power law of slope $-1<n<1$) in the Einstein-de-Sitter
cosmology the structure formation should be universal (or
self-similar) since no length or time scales (other than the one from
the non-linear collapse itself) are involved. Though our power spectrum for
the initial conditions is not a power law, the variations of the
slope as a function of k are very mild such that deviations from
universality in the SCDM case should still be small. Moreover the
spherical collapse model predict constant $\delta_c$ and
$\Delta_{vir}$ values, and all our cosmologies converge 
to a SCDM-like behaviour at high redshift 
since the influence of dark energy becomes negligible.
Therefore residuals with respect to SCDM$^*$ should be more sensitive
to the cosmological and redshift dependent signature of the non-linear collapse
on the halo mass function. The mass function for the reference SCDM$^*$ used here for comparison is measured from the simulations of box lengths 648~Mpc/h and 1296~Mpc/h. As for the $\Lambda$CDM-W5 model we use a fitting formula with the ST form. We find for the SCDM$^*$ model the following parameter values: 
$\tilde{A}=0.350$, $\tilde{a}=0.720$ and $\tilde{p}=0.1$ and $\tilde{\delta_c}=1.686$.

\subsection{The role of the threshold of collapse}


\begin{table} 
\begin{center}
\begin{tabular}{ccccc}
\hline \hline
Cosmology & $\delta_c$ &Deviations&$\Delta_{vir}$ &Deviations \\
&&for $f(1/\sigma)$&&for $f(\delta_c/\sigma)$\\
\hline
SCDM*&1.686&$0\%$&178&$0\%$\\
 \hline
$\Lambda$CDM-W5&1.673&$10.6\%$&368&$7.6\%$\\
\hline
L-$\Lambda$CDM&1.665&$15.8\%$&708&$11.5\%$\\
\hline
L-RPCDM&1.638&$19.9\%$&436&$9.7\%$\\
\hline\hline
\end{tabular}
\caption{\label{deviations} Standard deviations from a universal
  behaviour of the mass function with and without accounting for the
  time of collapse as encoded by $\delta_c$ predicted in the
  spherical collapse model. For comparison we also quote the values of $\delta_c$ and $\Delta_{vir}$ for each model as in Table~\ref{deltac_deltavir}.}
\end{center}
\end{table}

\begin{figure*} 
\begin{tabular}{cc}
\includegraphics[width=0.48\hsize]{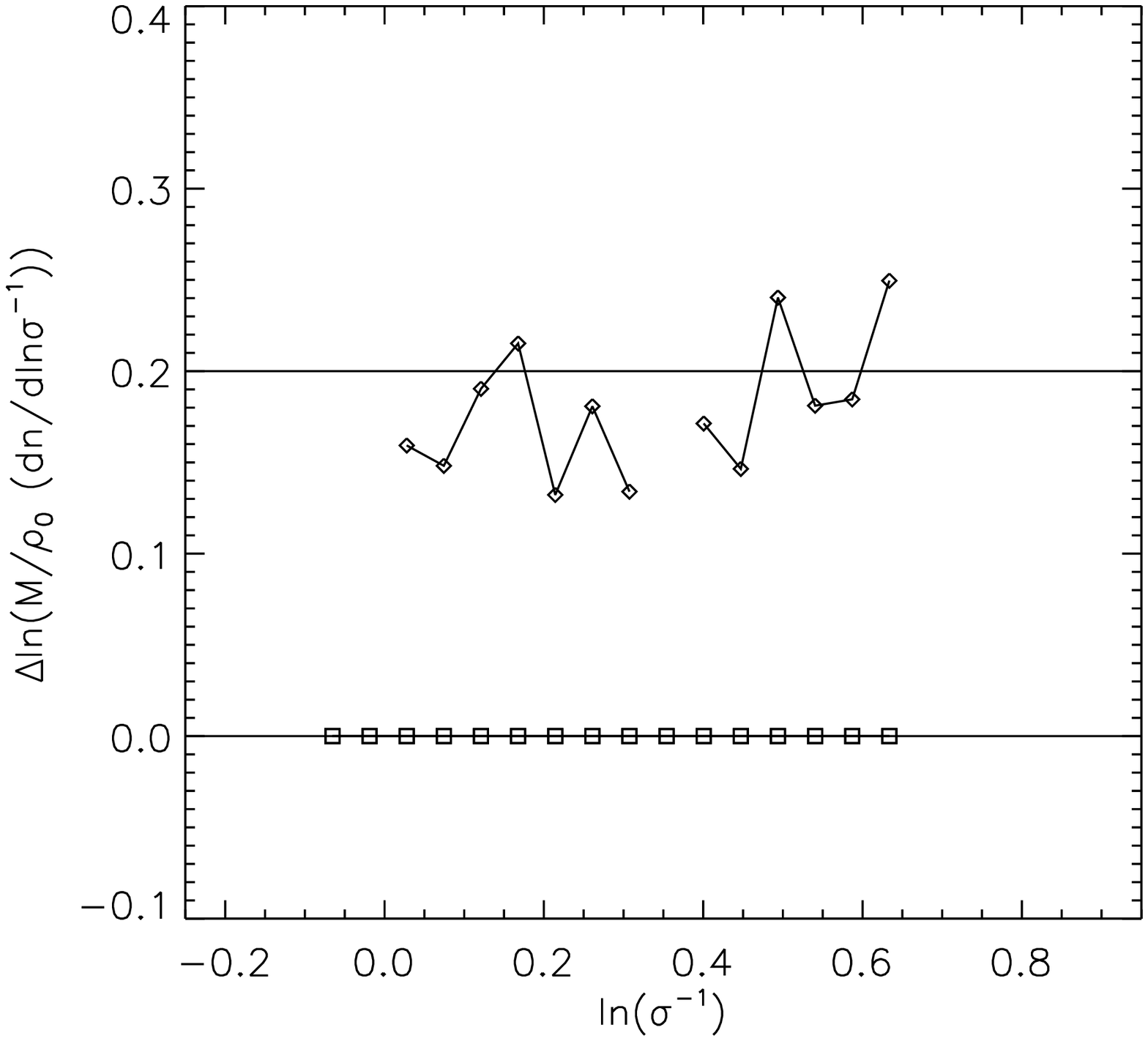}&
\includegraphics[width=0.48\hsize]{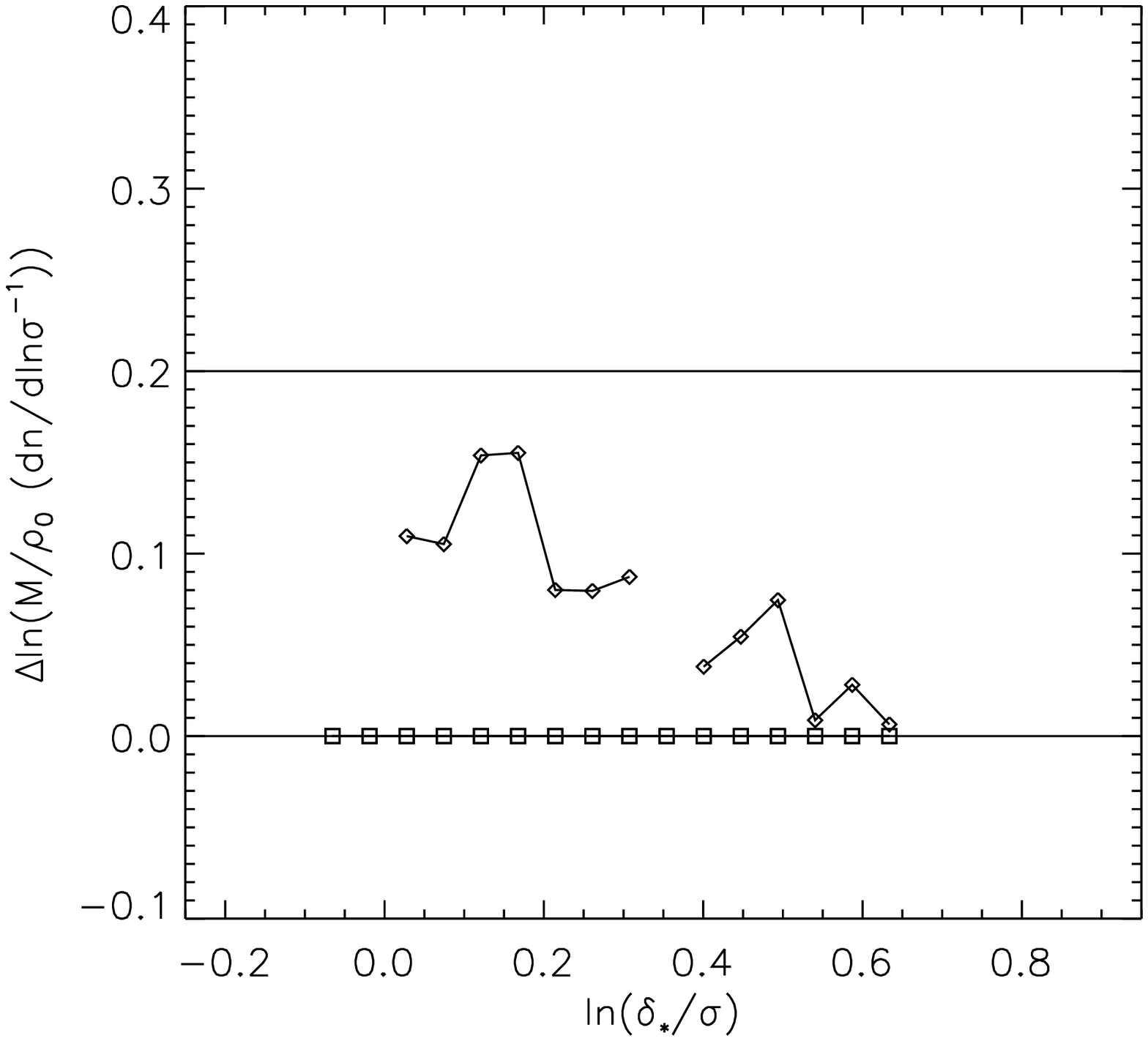}
\end{tabular}
\caption{\label{error_rpl} Residual between the measured mass functions in L-RPCDM (diamonds) and SCDM* (squares) models respectively at z=0. Left: Residual for the mass functions is shown in the $f-\textrm{ln}(1/\sigma)$ plane. Right: Residual for the mass function in the $f-\textrm{ln}(\delta_*/\sigma)$ plane, where $\delta_*=\delta_c/1.686$. Taking into account the different halo formation times through $\delta_c$ strongly decreases deviations from universality in the high-mass end.}
\begin{tabular}{cc}
\includegraphics[width=0.48\hsize]{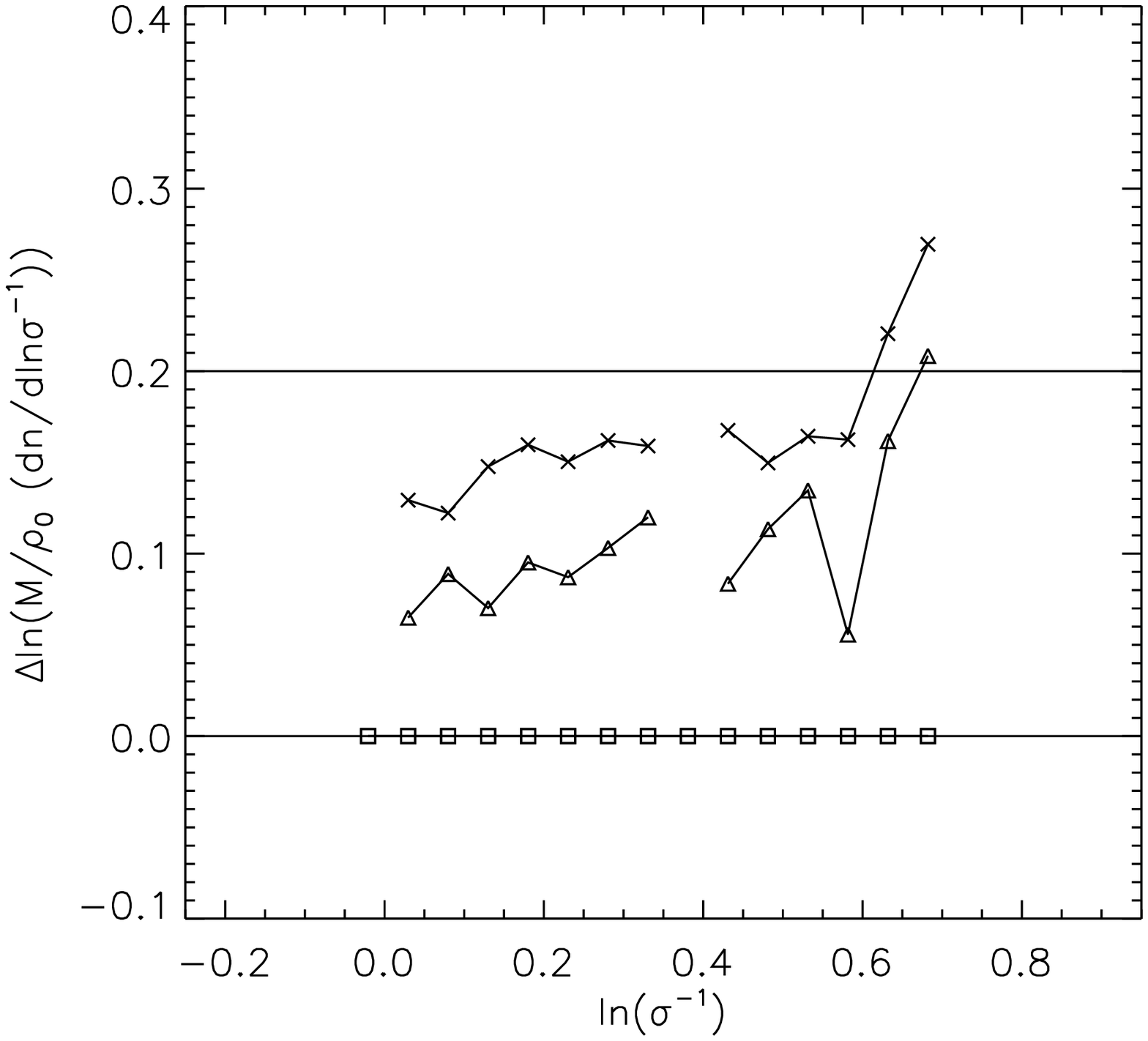}&
\includegraphics[width=0.48\hsize]{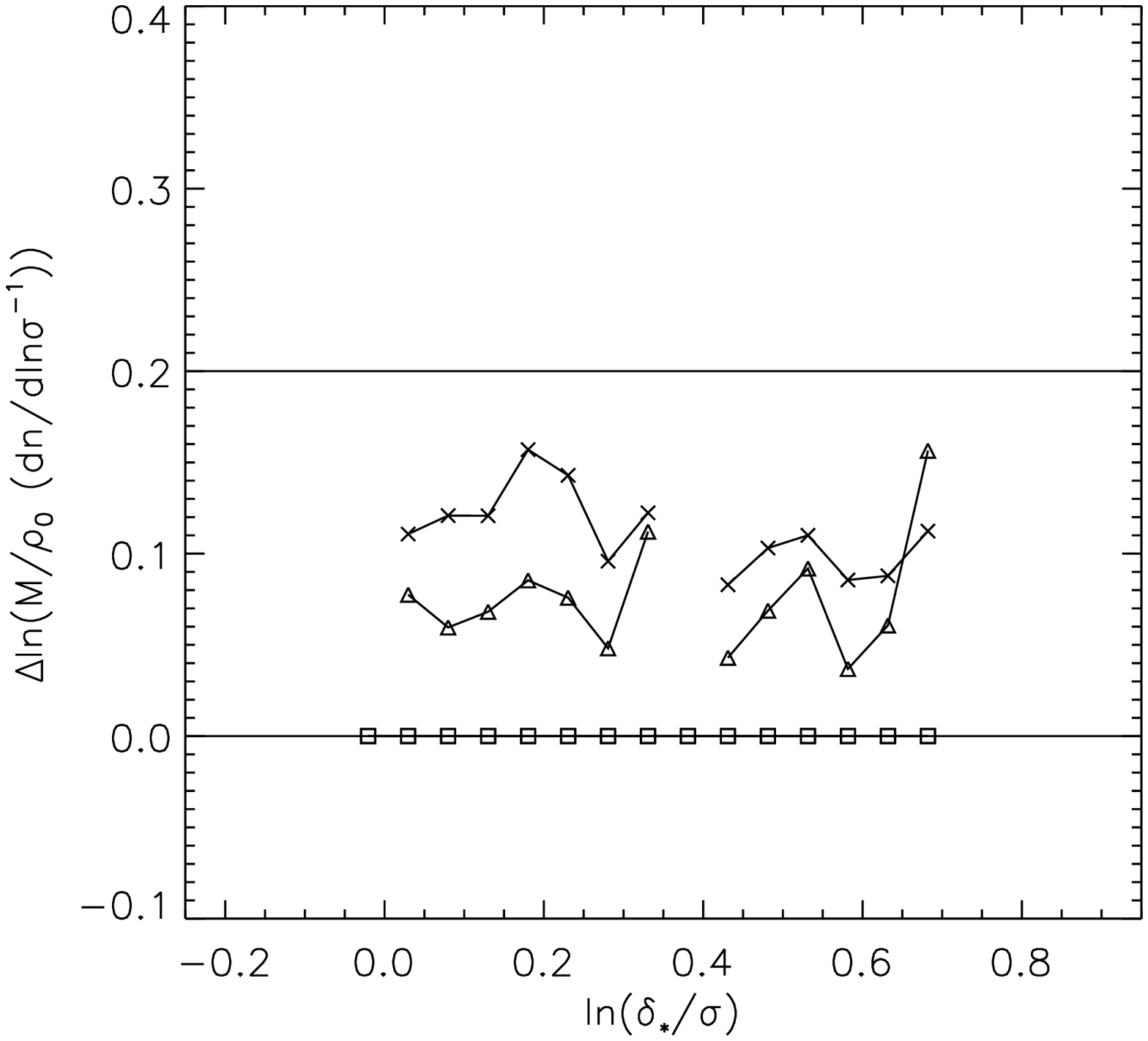}
\end{tabular}
\caption{\label{error_ll} Residual between the measured L-$\Lambda$CDM (crosses), $\Lambda$CDM-W5 (triangles) mass functions and that of SCDM$^*$ (squares) at z=0. Left: Residual of the mass functions in the $f-\textrm{ln}(1/\sigma)$ plane. Right: Residual of the mass functions in the $f-\textrm{ln}(\delta_*/\sigma)$ plane, where $\delta_*=\delta_c/1.686$. Taking into account the different halo formation times through $\delta_c$ decreases deviations from universality in the high-mass end.}
\end{figure*}

Here we consider the effect of the critical density threshold 
(or equivalently the time of collapse) on the halo mass function
as predicted by the spherical collapse model.
In the left panel of Fig.~\ref{error_rpl} we plot the residual 
of the function $f$ for the L-RPCDM model with respect to the SCDM$^*$ case
as a function of $\ln(\sigma^{-1})$. The amplitude of the residual is of 
order $20\%$ over the entire mass range. 
In the right panel of Fig.~\ref{error_rpl} we plot the residual as a function of $\ln(\delta_*/\sigma)$, 
where $\delta_*=\delta_c/1.686$ is the critical density threshold of
the L-RPCDM model rescaled to the SCDM$^*$ value. As we can see the
deviation with respect to the SCDM$^*$ model is strongly 
reduced, more in the high mass end than at low masses. 
A similar trend occurs if we consider the residual of the L-$\Lambda$CDM and $\Lambda$CDM-W5
models, which we plot in Fig.~\ref{error_ll} as a function of 
$\ln(\sigma^{-1})$ (left panel) and $\ln(\delta_*/\sigma)$
  (right panel), with $\delta_*$ the value of the critical density threshold of
each model rescaled to the SCDM$^*$ value. Again accounting for the
collapsing time of halos as encoded in $\delta_c$ reduces the residuals. 
For a more quantitative comparison we quote in Table~\ref{deviations}
the standard deviations\footnote{The standard deviation used to quantify the differences with respect to the SCDM$^*$ mass function is defined as $\sqrt{\sum_i{\Delta \log{f_i}^2}/(N-1)}$ } of the residuals of each cosmology 
as a function of $\sigma^{-1}$ and $\delta_*/\sigma$ respectively. We also quote the values of $\delta_c$
and $\Delta_{vir}$ at $z=0$ already discussed in Section~\ref{spherical}.
We may notice that deviations with respect to the SCDM$^*$
mass function are correlated with the difference in the value of $\delta_c$ specific to 
each model. In particular as $\delta_c$ decreases with respect to the SCDM$^*$ value, the standard deviation
increases up to $\sim20\%$ for the L-RPCDM. Indeed accounting for the spherical
collapse threshold reduces discrepancies between the halo mass functions by a factor
$\sim1.4$ for $\Lambda$CDM-W5 and L-$\Lambda$CDM, and a factor $\sim2$ in L-RPCDM. This clearly show that
the density threshold plays a role in the cosmic
structure formation and a general prescription for the mass function should include its dependence upon cosmology.

From Table~\ref{deviations} we can also notice that the remaining residuals are still relevant, above numerical uncertainties, and not correlated with $\delta_c$. Moreover if we consider the L-RPCDM model, accounting for $\delta_c$
improves only the high mass range and the deviations on the lower 
mass range can be hardly be attributed to $\delta_c$.
This indicates that there is at least one more process
responsible for the departure from universality, which will discuss in the next paragraph.

\subsection{Halo virialization and redshift evolution of the mass function}

\begin{figure*}
\begin{tabular}{cc}
\includegraphics[width=0.48\hsize]{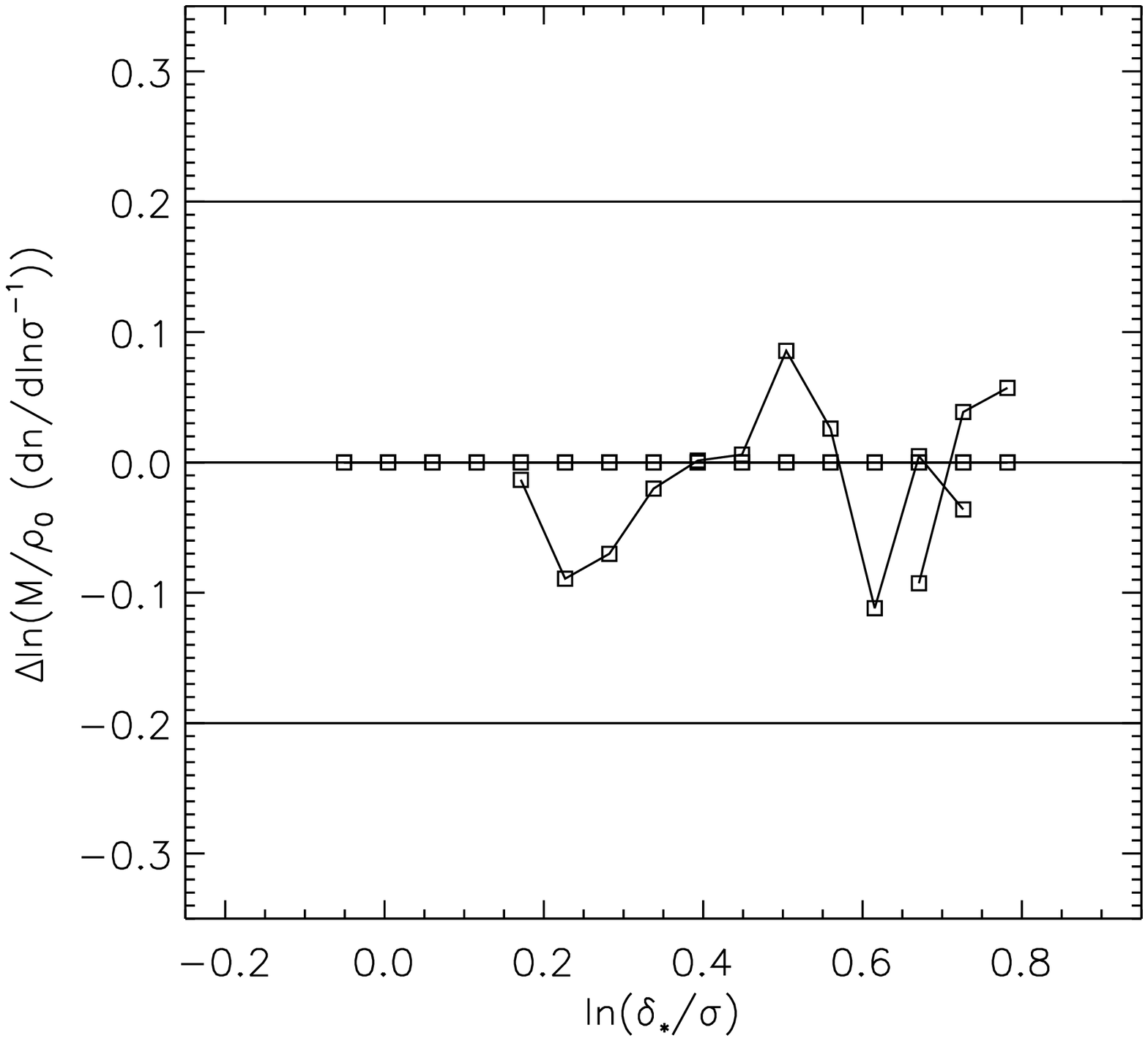}&
\includegraphics[width=0.48\hsize]{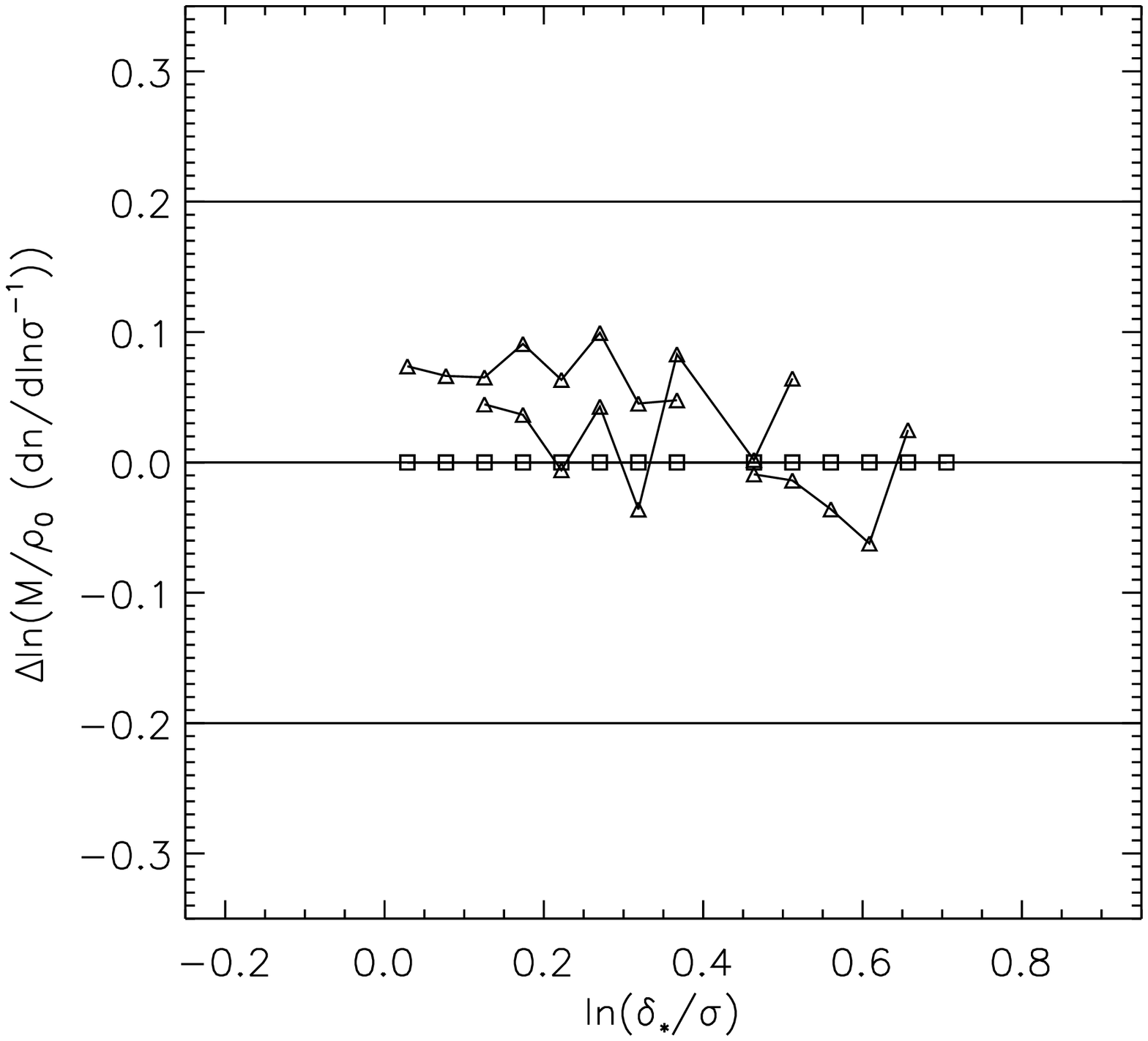}
\\
\includegraphics[width=0.48\hsize]{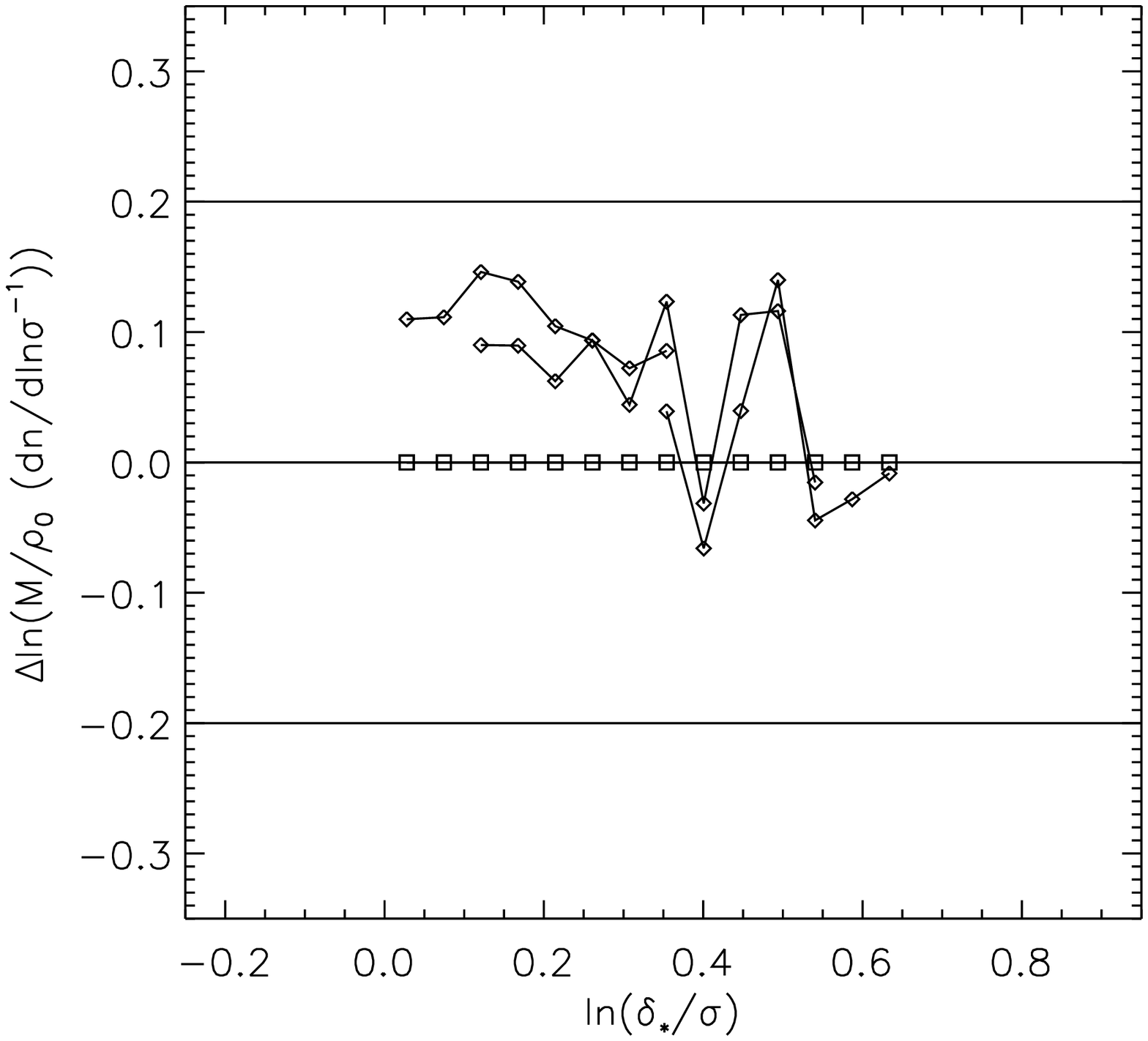}&
\includegraphics[width=0.48\hsize]{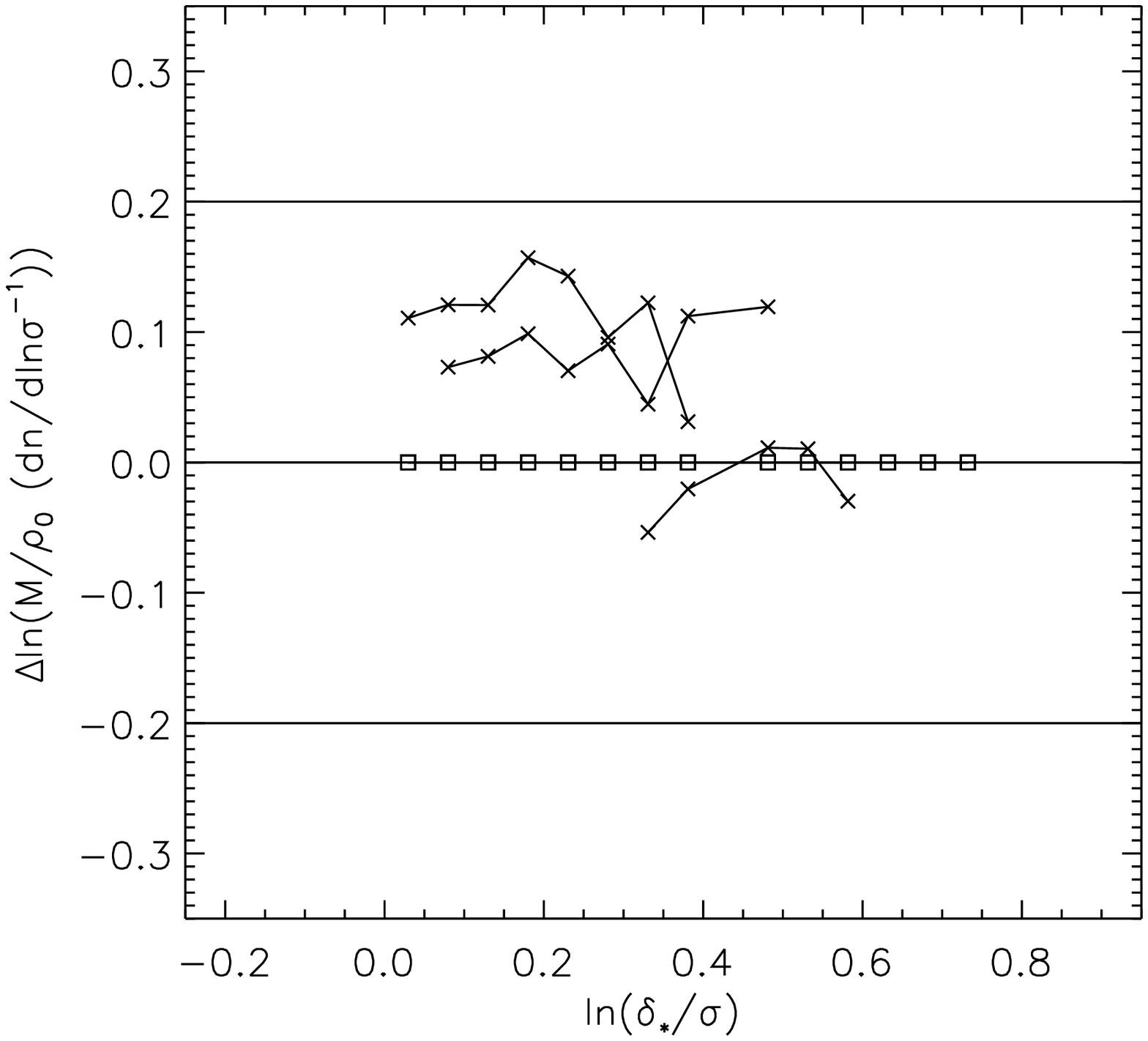}

\end{tabular}
\caption{\label{error_z} 
  Redshift evolution of the deviations from a universal behaviour.  We plot residuals between the measured mass functions for SCDM$^*$ (squares, upper left panel), 
$\Lambda$CDM-W5 (triangles, upper right panel), L-RPCDM (diamonds, lower left panel), L-$\Lambda$CDM (crosses, lower right panel) with respect to the
SCDM$^*$ mass function at z=0 (squares on the central line in the upper left panel).
In each plot the left curve corresponds to $z=0$, the middle curve to $z=0.25$ and the right one to $z=1$. The box length of the
  simulations is $648~h^{-1}Mpc$.  As expected, all mass functions tend towards SCDM$^*$ at high redshift.}
\end{figure*}
Confronting the level of improvement of the mass function residual as a function of $\delta_*/\sigma$
and the difference of the $\Delta_{vir}$ value of each cosmological model with respect to SCDM$^*$ quoted
in Table~\ref{deviations} we notice a strong correlation between the two. 
The larger the difference in the value of $\Delta_{vir}$  the
greater the deviations of the function $f(\delta_*/\sigma)$. Such a
correlation is indicative of the fact that the virialization of halos
does indeed play a role in determining the mass function and since the characteristics of this 
process are dependent upon cosmology and redshift, so are the deviations
from universality. Qualitatively this can be understood as follows.
Let us consider two perturbations, one in SCDM$^*$ and another in L-$\Lambda$CDM, 
both collapsing and virializing at the same moments, then the spherical collapse model suggests that the
resulting halos should be much more compact or ``concentrated'' 
in L-$\Lambda$CDM than in SCDM*. Because we measure mass functions 
with a constant linking length parameter (corresponding to a fixed
overdensity), this would explain why the mass function is greater 
in L-$\Lambda$CDM than in SCDM$^*$.

Now, if we extend this argument to the redshift evolution of
the virial overdensity, since for every cosmology $\Delta_{vir}(z)$ 
converges toward the SCDM$^*$ value at higher
redshifts, then the mass function residuals 
must decrease with redshift
in correlation with the behaviour of
$\Delta_{vir}(z)$ specific to each model. 
In Fig.~\ref{error_z}, we plot the mass function residuals from the  
648~h$^{-1}$Mpc simulation boxes at $z=0,0.25$ and $1$ for the
SCDM$^*$ (upper left panel), $\Lambda$CDM-W5 (upper right panel),
L-RPCDM (lower left panel) and L-$\Lambda$CDM (lower right panel)\footnote{We
  limit this analysis to the 648~h$^{-1}$ box length and $z<1$, 
since higher redshifts or larger box lengths would
  result in too few halos and larger Poisson error bars in the
  reference SCDM$^*$ model.}. 
We can see that the deviations  correlate well with $\Delta_{vir}(z)$
of each model. For the SCDM$^*$ the residual is consistent with a zero
value, this is a non-trivial result which shows that clustering in
SCDM models occurs in a universal manner as a function of
redshift. As shown in Section~\ref{spherical} the evolution of the virial 
overdensity in $\Lambda$CDM-W5 rapidly converges toward the SCDM$^*$ 
value and again in Fig.~\ref{error_z} we can see that
the mass function residual rapidly vanishes as a function of
redshift. However this also shows that the mass function as measured 
with FoF($b=0.2$) for the $\Lambda$CDM-W5 is
not universal for $z<1$, where dark energy starts
dominating the cosmic expansion, which is in agreement with recent
findings by \citet{tinker08,crocce09}. The models with the largest
deviations of $\Delta_{vir}(z)$ are the L-RPCDM which
converges towards the SCDM$^*$ value very slowly at high redshifts and L-$\Lambda$CDM which converges more quickly. In
the lower panels of Fig.~\ref{error_z} we can see that this is indeed the case for 
the mass function residuals.
Although errorbars become very large, we also checked that at higher
redshift (for instance at z=2.3), the mass functions are all compatible with a null
deviation at the same level independently of the cosmological model. 

\begin{figure} 
\includegraphics[width=\hsize]{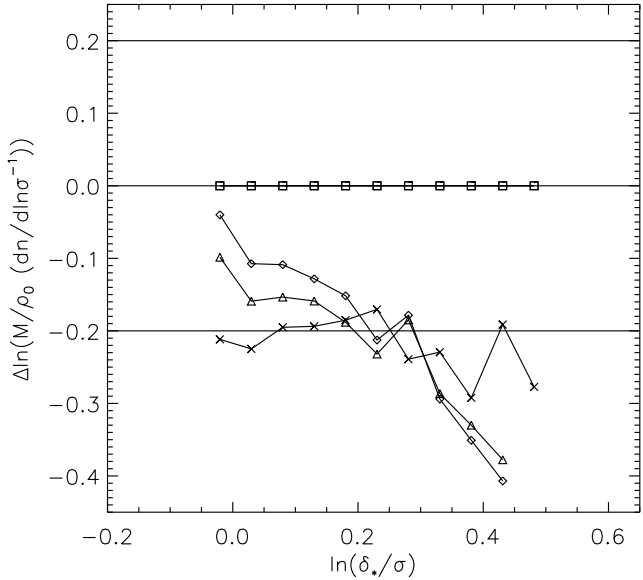}
\caption{\label{b_col} Residuals between the measured mass functions for the considered cosmology at z=0 using FoF with $b=b_{vir}$ (specific for each cosmology) from the spherical collapse and the mass functions for the SCDM* cosmology. Conventions are the same as in previous graphs.  Accounting for the virialization overdensity from the spherical collapse models strongly overcorrects the deviations from universality. Such deviations are larger than those one finds using a constant b for detection (which is consistent with the findings of \citet{jenkins01}).}
\end{figure}

These results clearly demonstrate that the virialization
process also contribute to shaping the halo mass function in a cosmological
and redshift dependent way. At this point we may ask ourselves whether accounting
for the virialization in the measurement of the mass function may reduce the deviations
from universality, similarly to the collapse threshold. After all we have 
detected halos using a constant linking length parameter, $b=0.2$, and using instead a
value of $b$ as predicted by the spherical collapse may further reduce discrepancies of the mass function residuals.
Hence we have run the FoF algorithm on the simulations at $z=0$ with parameter $b_{vir}$ 
given by the conversion $\Delta_{vir}/178\approx(0.2/b_{vir})^3$
\citep{cole96}, with $\Delta_{vir}$ the value predicted by the
spherical collapse model specific to each cosmology. 
As illustrated in Fig.~\ref{b_col} deviations from
universality are even greater than simply using $b=0.2$. 
The conversion formula is not accurate, nevertheless even using the SO halo
finder with density threshold  $\Delta_{vir}$ gives similar results.
From Fig.~\ref{b_col} we can also notice that this type of halo detection 
tends to overcorrect the different mass functions differently for the different
cosmologies. The largest deviation occurs for L-$\Lambda$CDM, then
L-RPCDM and $\Lambda$CDM-W5. This means that other effects may contribute
to the final shaping of the mass function, effects which go beyond the simple
spherical collapse model. Overall using $b_{vir}$ to account for the virialization process is incorrect, for this very reason it will be useful to find an empirical relation which can account for it. This will be discussed in the next paragraph.

\subsection{An empirical relation: $b_{univ}$ vs. $\Delta_{vir}$}

Rather than trying to identify non-linear mechanisms which are not
well modelled by the spherical collapse we may take a different
approach and investigate the following question: which virial density
parameter $\Delta_{univ}$ or alternatively linking length $b_{univ}$ 
is needed to recover a universal behaviour to numerical precision of the simulation? 

\begin{figure} 
\begin{center}
\includegraphics[width=\hsize]{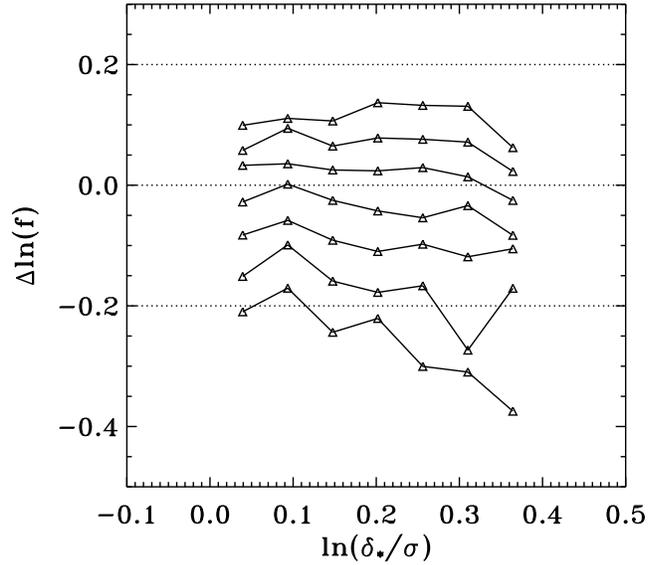}
\end{center}
\caption{Sensitivity of halo mass functions to the FoF linking length. The plot shows the deviations of the mass functions in the plane $f-ln(\delta_*/\sigma)$ (where $\delta_*=\delta_c/1.686$) for linking length ranging from 0.15 (lower curve) to 0.21 (upper curve) by step of 0.01. This plot is only true for $\Lambda CDM$ W5 cosmology at z=0. The deviations from universal behaviour also affect small masses.
\label{deltafsurf_deltavir}}
\end{figure}

\begin{figure} 
\begin{center}
\includegraphics[width=\hsize]{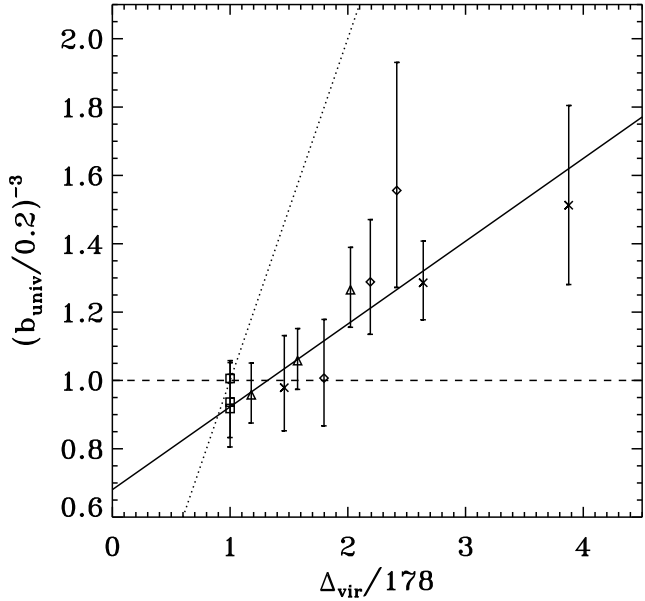}
\caption{\label{deltauniv} FoF linking length $(b_{univ}/0.2)^{-3}$ (or corresponding $\Delta_{univ}/\Delta_{SCDM*}$) required to explain the deviations from universality as a function of virial overdensities from the spherical collapse $\Delta_{vir}/178$ (or corresponding FoF linking length $(b_{vir}/0.2)^{-3}$) . The symbols are the same as for previous graphs.   The horizontal dashed line corresponds to the usually assumed universality which
  is strongly ruled out here. The dotted upper line corresponds to deviations
  exactly explained by the spherical collapse. This is also ruled out.  The
  continuous intermediate line is the linear best fit. This correlation indicates that dark energy effects are indeed encapsulated in $\Delta_{vir}$ (and $\delta_{c}$) but in a non trivial way.}
\end{center}
\end{figure}

One possibility would be to use a halo mass conversion formula such as \citet{hu03}, however this would lead to results which are dependent on the specific form of the halo profile.
Therefore we prefer to adopt a more robust
approach which, although more time consuming, is independent on the
halo profile. We have run a series of 7 FoF halo finders with various
linking lengths ranging from 0.15 to 0.21 (in steps of 0.01) for each
cosmological simulation and redshift. For each measured mass function
we have computed the deviation from universality and through
interpolation we have determined the
linking length parameter which minimizes the residual with respect to
SCDM$^*$. For instance in Fig.~\ref{deltafsurf_deltavir} we plot the
deviations measured for the various linking lengths in $\Lambda
CDM$-W5 simulations at z=0. Notice that deviations are also important
at low masses unlike when varying $\delta_c$. In this specific case the residual is minimal for
$b_{univ}=0.187$ (between the third and fourth curve from the top). 
For each data-point shown in Fig.~\ref{error_z}, this
procedure allows us to determine the value of $b_{univ}$, or better
still $(b_{univ}/0.2)^{-3}$ for which a given data-point of each 
mass function lies on the zero residual axis. This provides us with an ensemble of
$(b_{univ}/0.2)^{-3}$ value for each cosmological
model and at each redshift ($z=0,0.25$ and $1$), for which
we calculate the average and the standard deviation. So in total we
have nine estimations that we plot in Fig.~\ref{deltauniv}
as a function of $\Delta_{vir}(z)/178$ for 
each cosmology and redshift. Assuming the conversion
formula $\Delta/178\approx(0.2/b)^3$ is valid, we can interpret this plot as
$(b_{univ}/0.2)^{-3}$ as a function of $(b_{vir}/0.2)^{-3}$ or alternatively as
a plot of $\Delta_{univ}/\Delta_{SCDM*}$ as a function of $\Delta_{vir}/178$.
Not surprising we find that these points are not randomly distributed but are 
compatible with a linear regression.
We find 
\begin{equation}
  \label{eq_buniv}
  \left(\frac{b_{univ}}{0.2}\right)^{-3}=0.24\times \left[\frac{\Delta_{vir}}{178}-1\right]+0.92.
\end{equation}
which we plot in Fig.~\ref{deltauniv} as solid line. We can see that a fixed
universal value of the linking length parameter $b=0.2$ (short dashed
line) does not exist, as it is ruled out at more than $3\sigma$. Similarly the
spherical collapse model prediction $\Delta_{vir}/178\approx(0.2/b_{univ})^3$
(dotted line) is ruled out, this was expected since the spherical
collapse is a too simplistic model. Nevertheless even in the context of the spherical collapse, accounting for the virialization provide a good qualitative description of the dark matter halos. As an example, in L-$\Lambda$CDM halos are more ``compact'' than in SCDM given the different value of $\Delta_{vir}$, hence requiring a smaller linking-length parameter $b_{univ}$ as shown in Fig.~\ref{deltauniv}. The linear regression best-fit lies between these two extreme cases. However, this relation should not be interpreted as a new form of universality, as indicated by the fact that the slope is neither 0 or 1. Moreover the dispersion of the points around the linear regression is a signature of all the effects which contribute to departure from universality such as non-sphericity, concentration parameters, halo merging rates, etc... which are not modelled by the spherical collapse and which  may also depend on the cosmological model (i.e. the properties and abundance of dark energy).

\begin{figure} 
\begin{center}
\includegraphics[width=\hsize]{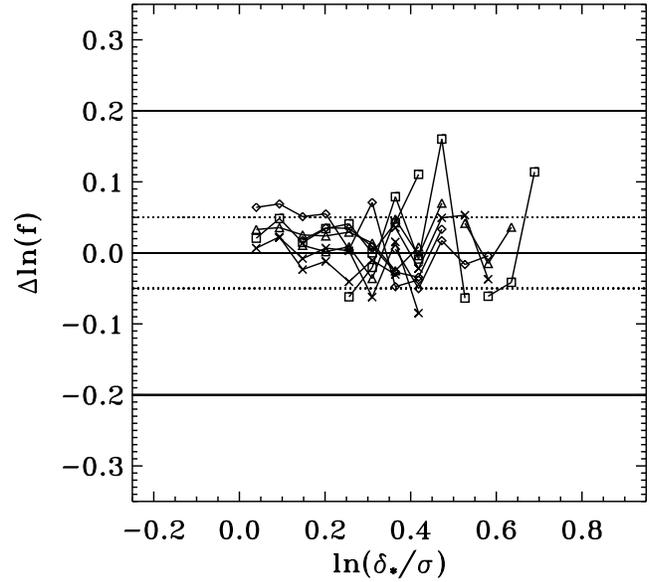}
\caption{\label{plot_mfct_buniv} Deviations from our reference mass functions (SCDM*, FoF, b=0.2) for all cosmologies and redshifts when taking into account the cosmology and redshift dependent time of collapse (by plotting as a function of $\delta_*=\delta_c/1.686$) and virialization processes (by using a linking length $b_{univ}(\Delta_{vir})$ given by Eq.~\ref{eq_buniv} where $\Delta_{vir}$ is given Fig.~\ref{delta_sc_evol}). The convention for the cosmologies are the same as in precedent plots: SCDM* (squares), $\Lambda$CDM (triangles), L-$\Lambda$CDM (crosses) and L-RPCDM (diamonds). The redshifts are z=0, 0.25 and 1 from left to right. The average position of the mass functions for all cosmologies and redshifts are well within the 5 percent deviations band (dotted lines) and are of order of our numerical uncertainties. Overall this shows than one can develop a prescription to go beyond the universality approximation which accounts for dark energy and redshift dependant non-linear effects.}
\end{center}
\end{figure}

Finally, in order to show that Eq.~(\ref{eq_buniv}) explains most of the deviations from a rigorous universal behaviour of the mass function to numerical precision, 
we plot in Fig~\ref{plot_mfct_buniv} the residual of the mass function for
SCDM$^*$, $\Lambda$CDM-W5, L-$\Lambda$CDM and L-RPCDM at $z=0,0.25$ and $1$
as a function of $\delta_*/\sigma$ , with each mass function
determined using a linking length parameter $b_{univ}$ obtained from
Eq.~(\ref{eq_buniv}). The reference cosmology is SCDM$^*$ with FoF($b=0.2$). 
As we can see all deviations reduce to
below the 5 percent level, which is of order of our numerical
precision. 

These results show the importance of taking into account 
non-linear effects in the prescription of the mass function. 
Again we want to stress that we are not claiming to 
have found a new universal behaviour. As already
mentioned other non-linear effects, besides those modelled by the
spherical collapse contributes to the mass function. Future works
with much smaller numerical errors may found departures from
Eq.~(\ref{eq_buniv}) which are correlated with other non-linear collapse
quantities. Here we have shown that taking into account the time of
collapse (as encoded by the density threshold $\delta_c$) and the
virialization process (as described by the virial overdensity
$\Delta_{vir}$) play an important role in determining the halo mass
function and as such these effects should be included in a theoretical formulation, 
also extended to dark energy cosmologies, since they greatly improve the agreement
between theory and simulations.

From a phenomenological point of view, this implies that when constraining dark energy
models through measurements of cluster number counts, rather than using standard
universal fitting formula of the mass function, it will be preferable 
to use a prescription which explicitly depends on $\delta_c(z)$ predicted by a given dark energy model. Here, we have shown that Eq.~(\ref{bestfit}), with SCDM$^*$ parameters (given at the beginning of Section~\ref{insight}) provide an accurate description of the mass function in dark energy cosmologies. In addition, one can account for the effect of the virialization by using Eq.~(\ref{eq_buniv}), with $\Delta_{vir}$ value predicted by a given model. In such case one has to convert the corresponding mass $M_{b_{univ}}$ to the observed one usually defined at a given overdensity $M_{\Delta}$. These conversions can be performed using for instance \citet{hu03} and \citet{lukic09}.

A final point concerns whether the non-universality found here extends to models with different $\sigma_8$ values.
In order to isolate the effects of dark energy we have focused on toy-models for which we have forced $\sigma_8$ to be
that of the reference $\Lambda$CDM-W5 cosmology. Would we observe a non-universality also in ``realistic models'' of dark energy
calibrated on SNIa and CMB data such as those considered in \citep{alimi09}? In this case $\sigma_8$ differs from one cosmology 
to another, but the deviations from universal behavios must be present also for these models.
In fact from the analysis of the mass function in the WMAP cosmologies presented in Section \ref{UNIV_WMAP} 
we have shown that $\sigma_8$ has very little effect on $f(\rm{ln}~\sigma^{-1})$. Despite having very different $\sigma_8$
values the mass function of the WMAP models are universal to numerical precision. This is because what really matters is 
whether cosmological models share the same structure formation history or not. Therefore our conclusions 
on the non-universality of the mass function can be easily extended to cosmologies with different $\sigma_8$. 
For instance, from Fig.~\ref{jenkinsplot} and Fig.~\ref{univ_break1}, it is straighforward to see that the deviations between
L-RPCDM and $\Lambda$CDM-W5 (we share the same $\sigma_8$) still hold if we had confronted  
L-RPCDM to $\Lambda$CDM-W1 (for which $\sigma_8$ is very different). Furthermore, having shown that the non-universality of the mass
function results from the cosmological and redshift dependence of the past structure formation history, it implies that 
our conclusions do not simply restrict to the models considered here, but have more general validity.
 
\section{Conclusion}\label{conclu}

In this work we have studied the cosmology and redshift dependence of the halo mass function through
high resolution cosmological N-body simulations. By comparing the results of $\Lambda$CDM-WMAP calibrated models 
against toy-models characterized by the same distribution of linear density fluctuations today, but with 
 different expansion histories and growth of the linear perturbations, we have been
able to infer a number of new and important results. Previous works have focused on the universality of the halo
mass functions and its deviations as a function of redshifts. We have used a FoF($b=0.2$) halo finder to construct 
 catalogues of halos from our simulations, and we have limited our analysis to mass ranges for which the Poisson noise is below
$10\%$ level and halos contain at least $350$ particles. By focusing on mass function residuals we have further limited
the effect of numerical systematics uncertainties to better than $5\%$. Using such an approach we have been able to clearly show that universality
does not exists in absolute terms, rather it can be verified to numerical precision at $z=0$, 
but only for those cosmologies which share very similar evolution histories at the background and linear level.
This is indeed the case of the $\Lambda$CDM-WMAP models, thus confirming past results \citep{tinker08,crocce09}. In contrast our toy-models show departures from universality above numerical uncertainties, larger than $10\%$ at $z=0$.

Using the spherical collapse model as guiding tool,
we have been able to incontrovertibly identify the non-linear mechanisms responsible for such deviations. Firstly, the spherical collapse threshold which differs from one cosmological model to another is responsible for deviations in the high mass end of the mass function.
Models with values of $\delta_c$ lower than the SCDM$^*$ prediction form structures earlier, thus leading to different imprint on the present halo mass function
consistently with the exponential cut-off expected in the Press-Schechter formalism. Indeed accounting for the collapse threshold reduces the discrepancy
between the mass function of the different models. We have provided a fitting formula, based on the Sheth-Tormen functional form 
calibrated on the $\Lambda$CDM-W5 model for which model dependent deviations from universality are of $\sim10\%$, with an 
explicit dependence on $\delta_c$, thus it can be efficiently used for a robust model parameter inference against halo mass function measurements.

Nevertheless the residuals lie still above numerical errors and are not correlated with the value of $\delta_c$ predicted by each model,
suggesting that at least one additional non-linear mechanism is responsible for departures from universality. We find that the residual deviations at $z<1$ are correlated with the redshift evolution of the 
virial overdensity $\Delta_{vir}$ specific to each model. On the other hand, the mass functions tend to that of the standard cold
dark matter scenario at higher redshifts, which is consistent with the fact that all cosmologies tends to a matter dominated universe at $z>1$.
We find an empirical relation between the linking length parameter of the FoF algorithm necessary to recover a universal form of the mass 
function and the virial overdensity at a given redshift for each cosmology. Such a relation 
lies between the prediction of purely spherical collapse and a fully universal behaviour. It also suggests that other non-linear mechanisms probably exists 
within the numerical precision of our simulations which contributes to further shaping the mass function.

Furthermore the empirical relation we have found between $b_{univ}$ or equivalently $\Delta_{univ}$
and $\Delta_{vir}$ may have important implications for the definition of the physical ``frontier'' of halos. 
As we have seen here, most of the works in the literature which concern the mass function use a mass definition based on a constant detection parameter (either $\Delta$ or $b$). In contrast internal halo profiles are usually defined as a function of $\Delta_{vir}$.
Henceforth, using a value of $\Delta_{univ}$ (or $b_{univ}$) corresponding to that of the assumed cosmology may provide a consistent 
halo definition applicable to both mass function and halo profile. Overall these results point to the fact that the collapse threshold and the virialization process play an important role
in determining the halo mass function, encoding cosmological (dark energy) dependent features which are neglected
in the standard universal fitting formula \citep{jenkins01,warren06}, and which we have shown to be of limited precisions.

These cosmology-dependent effects are coherent with those found on the non-linear matter power spectrum \citep{ma07,alimi09} 
and the halo profile \citep{wechsler02,dolag04}, and which are a direct consequence of the fact that the non-linear
structure formation conserves a record of the linear phase of collapse. Henceforth the halo mass function contains specific
cosmology dependent features which can be tested through observations, provided that predictions from numerical simulations or semi-analytical prescriptions are sufficiently accurate to account for such imprints \citep[see][]{wu10}. The results of this work provide also an understanding of the deviations from universality that are present in realistic dark energy models (calibrated on CMB and SNIa data) as those discussed in \citet{alimi09}, and which will be presented in an upcoming paper. Finally, we can speculate that our findings are relevant 
also in the context of non-minimally coupled inhomogeneous dark energy models, for which the linear growth history is scale dependent,
thus deviating from that of standard LCDM cosmologies. In such a case imprints on the halo mass function should be larger, 
thus needing accurate studies of the non-linear structure formation of these scenarios
beyond those already discussed in the literature.


\section*{Acknowledgements}
This work was granted access to the HPC resources of CCRT under the
allocation 2008-t20080412191 made by GENCI (Grand Equipement National de Calcul Intensif)
 We would like to thank Romain Teyssier, St\'ephane Colombi and Simon Prunet for their valuable advice concerning
the RAMSES, POWMES and MPGRAFIC softwares.  We thank Patrick Valageas for his commentaries. This work was supported by the Horizon Project (www.projet-horizon.fr).
A.F.  is associated researcher at LUTH, Observatory of Paris. V.B. is supported by the Belgian Federal Office for Scientific, Technical and Cultural Affairs through the Interuniversity Attraction Pole P6/11.

\end{document}